\documentclass[prd,aps,showpacs,twocolumn]{revtex4-1}
\usepackage{bm}
\usepackage{amsmath}
\usepackage{amssymb}
\usepackage{graphicx}
\usepackage{slashed}
\usepackage{accents}
\usepackage{mathrsfs}
\usepackage{braket}
\usepackage{units}
\usepackage{upgreek}

\usepackage[breaklinks=true]{hyperref}
\usepackage{breakurl} 

\DeclareSymbolFont{eulerscript}{U}{eur}{m}{n}
\DeclareSymbolFontAlphabet{\matheuler}{eulerscript}
\DeclareMathAlphabet{\boldgreek}{OML}{zplm}{b}{it}

\newcommand{\T}{T}
\newcommand{\Q}{\mathcal{Q}}
\newcommand{\I}{i}
\newcommand{\D}{d}
\newcommand{\nfrac}[2]{{#1}/{#2}}
\newcommand{\one}{\mathbf{1}}
\newcommand{\del}{\partial}
\newcommand{\eps}{\epsilon}
\newcommand{\mc}[1]{\mathcal{#1}}
\newcommand{\spvec}[1]{\boldsymbol{#1}}
\newcommand{\lb}{\left(}
\newcommand{\rb}{\right)}
\newcommand{\rsb}{\right]}
\newcommand{\lsb}{\left[}
\newcommand{\rcb}{\right\}}
\newcommand{\lcb}{\left\{}
\newcommand{\abs}[1]{\left|#1\right|}
\DeclareMathOperator{\tr}{\mathbf{tr}}
\DeclareMathOperator{\diag}{diag}
\DeclareMathOperator{\Ai}{Ai}
\DeclareMathOperator{\Gi}{Gi}
\DeclareMathOperator{\sinc}{sinc}
\newcommand{\spvecgreek}[1]{\boldgreek{#1}}
\newcommand{\lplus}{{{}+}}
\newcommand{\lminus}{{{}-}}
\newcommand{\lone}{{\scalebox{.64}{$\matheuler{I}$}}}
\newcommand{\ltwo}{{\scalebox{.64}{$\matheuler{II}$}}}
\newcommand{\lperp}{\perp}
\newcommand{\s}[1]{\slashed{#1}}
\newcommand{\Lambdainv}{{\Lambda^{\hspace*{-1.8pt}\scalebox{0.65}{$-1$}}}}
\newcommand{\Ftilde}{\mathfrak{F}}

\begin{document}
\title{Polarization operator for plane-wave background fields}
\author{S. \surname{Meuren}}
\author{C. H. \surname{Keitel}}
\author{A. \surname{Di Piazza}}
\email{dipiazza@mpi-hd.mpg.de}
\affiliation{Max-Planck-Institut f\"ur Kernphysik, Saupfercheckweg 1, D-69117 Heidelberg, Germany}
\date{\today}

\begin{abstract}
We derive an alternative representation of the leading-order contribution to the polarization operator in strong-field quantum electrodynamics with a plane-wave electromagnetic background field, which is manifestly symmetric with respect to the external photon momenta. Our derivation is based on a direct evaluation of the corresponding Feynman diagram, using the Volkov representation of the dressed fermion propagator. Furthermore, the validity of the Ward-Takahashi identity is shown for general loop diagrams in an external plane-wave background field.
\end{abstract}
 
\maketitle

\section{Introduction}

The most precise calculations known so far in physics are provided by quantum electrodynamics (QED). The reason for this is the smallness of the fine-structure constant~$\alpha = \nfrac{e^2}{(4\pi\eps_0 \hbar c)} \approx \nfrac{1}{137}$, which allows us to use perturbation theory \footnote{We briefly mention here that the perturbative approach cannot be applied to processes involving particles with extremely high energies 
$\varepsilon$ such that $\alpha\log(\varepsilon/mc^2) \sim 1$, with $m$ being the electron mass \cite{landau_quantum_1981}.} ($e$ is the electron charge). The most prominent example is probably the electron $g$-factor, for which experimental and theoretical results have been matched on the record accuracy level of parts per billion \cite{gabrielse_new_2006}. To achieve this outstanding precision, the corresponding theoretical calculation included Feynman diagrams with up to four loops. 

A quite different situation is encountered for QED with electromagnetic background fields. A source for strong electromagnetic fields are modern laser systems. If spatial focusing effects are sufficiently small, laser fields can be well approximated by plane-wave fields. For a plane-wave field, we obtain, besides the fine-structure constant, a second gauge and Lorentz-invariant parameter $\xi_0=|e|E_0/(mc\omega_0)$, where $E_0$ and $\omega_0$ are the peak electric field strength and central angular frequency of the plane wave, respectively [$m$ is the electron mass; see also Sec. \ref{sec:planewavefields}]. If $\xi_0 \gtrsim 1$ the interaction between electron and positrons with the laser field must be taken into account exactly. For optical lasers (photon energy $\hbar\omega_0 \approx \unit[1]{eV}$), this happens already at intensities of the order of~$\unitfrac[10^{18}]{W}{cm^2}$. More precisely, we can generally still treat the interaction of the electrons and the positrons with the quantized radiation field perturbatively as in vacuum QED (QED without background fields), but must include the dependence on $\xi_0$ to all orders if the threshold $\xi_0 \approx 1$ is exceeded.

Another important scale is the so-called critical field $E_{cr}=m^2c^3/(\hbar|e|)=1.3\times 10^{16}\;\text{V/cm}$, which corresponds to a peak laser intensity of~$I_{\mathrm{cr}}=\eps_0 c E_{\mathrm{cr}}^2 =\unitfrac[4.6\times 10^{29}]{W}{cm^2}$. A constant and uniform electric field of this strength can, in principle, produce electron-positron pairs from the vacuum \cite{sauter_ueber_1931,heisenberg_folgerungen_1936,schwinger_gauge_1951}.

The current laser intensity record (in the optical regime) is given by $\unitfrac[2\times 10^{22}]{W}{cm^2}$ \cite{yanovsky_ultra_2008} and future facilities envisage even intensities of the order of $\unitfrac[10^{24}-10^{25}]{W}{cm^2}$ \cite{ELI,HIPER,XCELS}. Thus, the nonperturbative regime (in $\xi_0$) can be entered with presently available laser systems, and even the critical field can be reached in the rest frame of an ultrarelativistic particle (e.g., a $\sim \unit[1]{GeV}$ electron \cite{leemans_gev_2006}). 

So far only one experiment has been carried out to probe strong-field QED effects using laser fields~\cite{bula_observation_1996,burke_positron_1997}. However, this is expected to change in the near future and, correspondingly, the experimental progress has stimulated many theoretical investigations during the last years \cite{di_piazza_quantum_2010,hu_complete_2010,king_matterless_2010,
mackenroth_determining_2010,dumlu_schwinger_2010,heinzl_beam-shape_2010,heinzl_finite_2010,fedotov_limitations_2010,bulanov_schwinger_2010,sokolov_pair_2010,
meuren_quantum_2011,hu_relativistic_2011,kryuchkyan_bragg_2011,hartin_high_2011,boca_thomson_2011,dumlu_interference_2011,hebenstreit_pair_2011,elkina_qed_2011,nerush_laser_2011,ilderton_pair_2011,ilderton_trident_2011,labun_spectra_2011,monden_enhancement_2011,redondo_light_2011,
seipt_nonlinear_2011,seipt_two-photon_2012,nousch_pair_2012, titov_enhanced_2012,king_photonphoton_2012,dobrich_magnetically_2012,dinu_infrared_2012,harvey_radiation_2012,krajewska_compton_2012,king_pair_2012,mackenroth_nonlinear_2013,king_photon_2013}. For a more detailed overview, the reader is referred to the review \cite{di_piazza_extremely_2012}. 

In contrary to vacuum QED, calculations with a plane-wave background field have not been carried out beyond the one-loop order (for constant-crossed fields higher-order calculations have been performed, see, e.g., \cite{ritus_radiative_1972,ritus_vacuum_1972,ritus_1985}). This can be attributed to the fact that already diagrams with just a few propagators correspond to quite complicated expressions. It is therefore of general interest to investigate new techniques, which have the potential to make also the calculation of complicated diagrams tractable. 

\begin{figure}[t!]
\centering
\includegraphics[height=2.1cm]{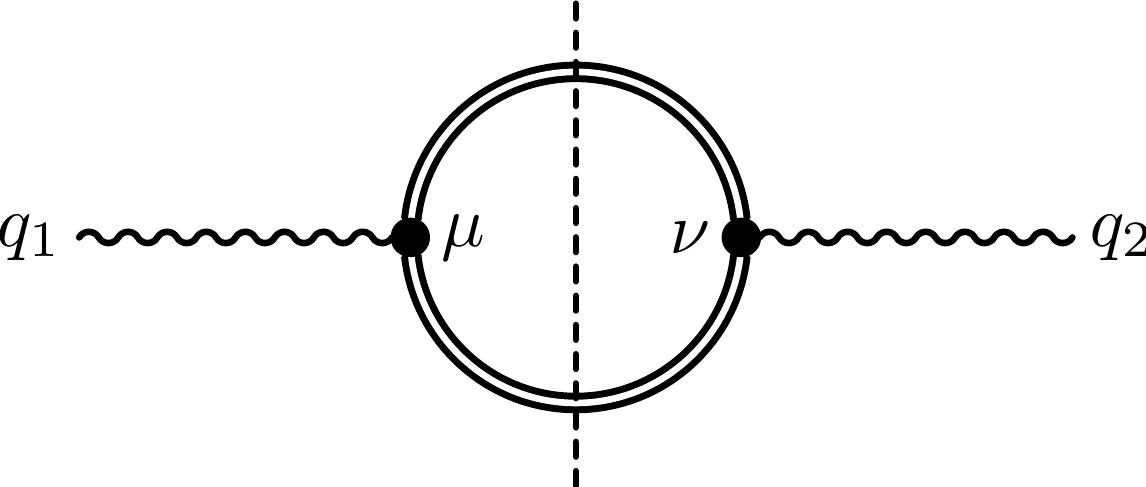}
\caption{\label{fig:polarizationoperator} The Feynman diagram corresponding to the leading-order contribution to the polarization operator~$\mathcal{P}^{\mu\nu}(q_1,q_2)$ in a plane-wave background field. The double lines represent the Volkov propagators for the fermions, which take the external field into account exactly [see Eq.~(\ref{eqn:dressedpropagator})]. The vertical dashed line links the polarization operator to the pair-production diagram due to the unitarity of the $S$-matrix.}
\end{figure}

Here we present a new derivation of the first-order contribution to the polarization operator given in Fig.~\ref{fig:polarizationoperator} \cite{baier_interaction_1975,becker_vacuum_1975}. 
In Ref. \cite{baier_interaction_1975} an operator approach similar to the one introduced by Schwinger~\cite{schwinger_gauge_1951} has been used. We show here how the diagram in Fig.~\ref{fig:polarizationoperator} can be evaluated directly using the Volkov representation of the dressed propagators. This approach has the appealing feature that it is very similar to the calculation techniques used in vacuum QED. Our final result (which is equivalent to the one in Ref. \cite{baier_interaction_1975}) has the interesting property that it is manifestly symmetric with respect to the external photon four-momenta $q_1$ and $q_2$ (see Fig. \ref{fig:polarizationoperator}). Furthermore, we prove the validity of the Ward-Takahashi identity \cite{ward_identity_1950,takahashi_generalized_1957} for general loop diagrams in a plane-wave background field. The calculation techniques employed here are expected to be useful also for other higher-order calculations with strong plane-wave background fields.

The polarization operator itself is of central importance, since it determines the properties of a photon inside the background field via the Schwinger-Dyson equation for the exact photon propagator \cite{baier_interaction_1975,landau_quantum_1981}. As a consequence, a plane-wave field acts as an active medium, e.g. the photon obtains a mass and has a nontrivial dispersion relation. Furthermore, due to the unitarity of the $S$-matrix, the imaginary part of the polarization operator is related to the total photo-production probability for an electron-positron pair (see Fig. \ref{fig:polarizationoperator}) \cite{milstein_polarization-operator_2006,di_piazza_barrier_2009}. The polarization operator is also required for the calculation of radiative corrections to elementary processes like nonlinear Compton scattering or pair production. The significance of the photon polarization tensor can also be inferred from the ongoing effort to calculate it for different field configurations \cite{gies_vacuum_2011,karbstein_optical_2012}.

The paper is divided into three parts. In Sec. \ref{sec:strongfieldqed} the general framework of strong-field QED with plane-wave background fields is described. The actual calculation is then presented in Sec. \ref{sec:polarizationoperator}. Finally, we show how our results are related to those obtained in Ref. \cite{baier_interaction_1975} and discuss the special cases of a constant-crossed field, a linearly polarized field in the quasi-classical approximation, and a circularly polarized, monochromatic field in Sec. \ref{sec:discussion}.

\section{Strong-field QED}
\label{sec:strongfieldqed}

QED is described by the following Lagrangian density \cite{landau_quantum_1981,weinberg_quantum_1995}:
\begin{gather}
\label{eqn:qedlagrangian}
\mathcal{L}_{\mathrm{QED}}
= 
\bar{\psi} \lb \I{}\s{\del} - m\rb \psi  - \frac{1}{4} \mc{F}_{\mu\nu} \mc{F}^{\mu\nu} - e\bar{\psi} \gamma_\mu \psi \mc{A}^\mu,
\end{gather}
where~$\psi$ and~$\mc{A}^\mu$ are the Dirac spinor field and the electromagnetic four-vector potential, respectively, and~$\mc{F}^{\mu\nu} = \del^\mu \mc{A}^\nu - \del^\nu \mc{A}^\mu$ is the electromagnetic field tensor [from now on we will use natural units $\hbar = c = 1$ and Heaviside-Lorentz units for the charge, $\alpha = \nfrac{e^2}{(4\pi)}$, see Appendix \ref{sec:notation} for further details]. The equation of motion for the spinor field $\psi$, which follows from Eq.~(\ref{eqn:qedlagrangian}), is the Dirac equation
\declareslashed{}{/}{.2}{.1}{\mc{A}}
\begin{gather}
\label{eqn:diracequation}
(i\s{\del} - e\s{\mc{A}} - m)\,\psi = 0.
\end{gather}
Correspondingly, we obtain for the photon field $\mc{A}^\mu$ in Lorentz gauge ($\del_\mu \mc{A}^\mu = 0$) the wave equation 
\begin{gather}
\label{eqn:photonwaveequation}
\del^\rho \del_\rho \mc{A}^\mu =  e J^\mu,
\quad
J^\mu = \bar{\psi} \gamma^\mu \psi.
\end{gather}

\subsection{Vacuum QED}

To obtain a quantum theory of electrons, positrons, and photons, both the spinor $\psi$ and the photon field $\mc{A}^\mu$ are promoted to operators with canonical (anti)commutation relations (alternatively, the functional integral formalism can be employed). Once derived in either way, the $S$-matrix element for a given process can be calculated perturbatively using Feynman rules. In vacuum QED, the starting point for the perturbative expansion is a solution of the free Dirac or the free wave equation [Eq.~(\ref{eqn:diracequation}) with~$\mc{A}^\mu=0$ and Eq.~(\ref{eqn:photonwaveequation}) with $J^\mu=0$, respectively]. An electron with a given four-momentum~$p^\mu=(\eps,\spvec{p})$ ($\eps > 0, p^2=m^2$) can then be described by the plane-wave solutions \cite{landau_quantum_1981},
\begin{gather}
\label{eqn:freeplanewave}
\psi_{p} = \frac{1}{\sqrt{2\eps}} e^{-ipx} u_p,
\quad
(\s{p} - m) u_p = 0
\end{gather}
and the corresponding propagator is given by
\begin{gather}
\label{eqn:freepropagator}
iG(x,y) = i\int \frac{d^4p}{(2\pi)^4} \, e^{-ip(x-y)} \frac{\s{p} + m}{p^2 - m^2 + i0}.
\end{gather}
For a photon with polarization four-vector $\eps^\mu$ and four-momentum $k^\mu=(\omega,\spvec{k})$ ($\omega \geq 0$, $k^2=0$), we obtain the following wave-function:
\begin{gather}
\mc{A}^\mu_{k} = \frac{1}{\sqrt{2\omega}} e^{-ikx} \eps^\mu
\end{gather}
and in the Feynman gauge, the photon propagator is given by
\begin{gather}
-iD_{\mu\nu}(x-y) = - i \int \frac{d^4k}{(2\pi)^4} \, e^{-ik(x-y)} \frac{g_{\mu\nu}}{k^2 + i0}.
\end{gather}
Finally, the interaction between electrons, positrons, and photons is represented by the interaction vertex
\begin{gather}
\label{eqn:freevertex}
-ie \int d^4x \, \cdots \gamma^\mu \cdots,
\end{gather}
where the dots indicate that the vertex is always contracted with two fermion and one photon lines. Thus, one can move the  exponential functions from the external lines and propagators to the vertex. After the space-time integrals associated with the vertex are taken, momentum-conserving delta functions are obtained. For a more detailed discussion  see e.g. Refs. \cite{landau_quantum_1981,weinberg_quantum_1995}.

\subsection{QED with background fields}
\label{sec:qedwithbackgroundfields}

Vacuum QED has been tested to a very high precision because, due to the smallness of the fine-structure constant, a perturbative treatment is adequate in most situations. However, for very strong external electromagnetic fields, the (conventional) perturbation series breaks down. Modern laser facilities provide a source of such strong external electromagnetic fields. An intense laser field represents a coherent state of the photon field that can be described by a classical four-potential $A^\mu$. Since only the highly occupied laser modes can be considered as classical, we separate those modes by writing $\mc{A}^\mu = A^\mu_{\mathrm{rad}} + A^\mu$ in the Lagrangian density. Here, $A^\mu$ is treated as a classical field, whereas all other modes, described by $A^\mu_{\mathrm{rad}}$, are properly quantized \cite{fradkin_quantum_1991,glauber_coherent_1963,harvey_signatures_2009} (see Appendix \ref{sec:lasercoherentstate} for further details).

To estimate the laser intensity at which we can start to treat the laser as a classical field, we follow Ref. \cite{landau_quantum_1981}, Sec. 5. The energy density of a laser field with intensity $I_0$ and central angular frequency $\omega_0$ is of the order of $I_0$, and the density of modes is of the order of $\omega_0^3$. Correspondingly, each mode contains $N_\gamma$ photons, where $N_\gamma$ is of the order of $\nfrac{I_0}{\omega_0^4}$. If $N_\gamma \gg 1$, due to the correspondence principle, we can describe the laser modes by a classical field. Thus, we obtain the following condition for the laser intensity:
\begin{gather}
\label{eqn:classicalcondition}
I_0 \gg \omega_0^4 \approx \unitfrac[6 \times 10^5]{W}{cm^2} \, \lb \frac{\omega_0}{\unit[1]{eV}} \rb^4,
\end{gather}
which is well fulfilled at the relativistic intensities ($I_0\gtrsim \unitfrac[10^{18}]{W}{cm^2}$) in the optical regime ($\omega_0 \sim \unit[1]{eV}$) we are interested in here.

Furthermore, we also have to ensure that the depletion of the laser field is sufficiently low, such that the latter can be treated as a given background field. Now, typical available optical petawatt lasers, which are suitable for the investigation of QED processes in a strong laser field, have an energy of the order of 100 J \cite{di_piazza_extremely_2012}, i.e., a total number of about $10^{20}$ photons. In a typical QED process as nonlinear Compton scattering, at an intensity of the order of $\unitfrac[10^{22}]{W}{cm^2}$, about $\xi_0^3\sim 10^6$ photons are absorbed from the laser by each electron \cite{di_piazza_extremely_2012} (such an intensity is, in principle, reachable with a petawatt laser). By assuming a number of electrons in the beam of the order of $10^9$, we obtain that about $10^{15}$ photons are expected to be absorbed from the laser field, which remains then practically unaffected (this number of electrons is typical for current laser-plasma-based electron accelerators \cite{leemans_gev_2006}). In conclusion, we can safely assume here that the background laser field can be treated as a given, classical background field (see also  Refs. \cite{fried_scattering_1964,eberly_electron_1966,berson_electron_1969,bergou_nonlinear_1981,filipowicz_relativistic_1985}).

Working in the so-called Furry picture \cite{furry51} the classical background field $A^\mu$ is taken into account exactly and only the radiation field is treated perturbatively. The starting point for the perturbative expansion of the $S$-matrix is then the solution of the interacting Dirac equation~(\ref{eqn:diracequation}) with the replacement~$\mc{A}^\mu\to A^\mu$. Since photons have no self-interactions at tree level, the photon wave functions and propagators are left unchanged by the background field. A more detailed discussion of strong-field QED can be found in Refs. \cite{battesti_magnetic_2013,di_piazza_extremely_2012,ehlotzky_fundamental_2009,mourou_optics_2006,marklund_nonlinear_2006,dittrich_probingquantum_2000,fradkin_quantum_1991,ritus_1985} and in the references therein.

\subsection{Plane-wave fields}
\label{sec:planewavefields}

In this paper we will consider only plane-wave external fields, i.e., we require that the field tensor~$F^{\mu\nu}$ depends only on the plane-wave phase~$\phi = kx$, where~$k^\mu$ is a momentum four-vector which characterizes the plane wave ($k^2=0$). In the absence of charges and currents, the field tensor must obey the homogeneous Maxwell equations \cite{landau_classical_1987}:
\begin{gather}
\label{eqn:mwefieldtensor}
\del_\mu F^{\mu\nu} = k_\mu F'^{\mu\nu} = 0,
\quad
\del^\mu F^{*}_{\mu\nu} 
=
k^\mu F'^{*}_{\mu\nu} = 0.
\end{gather}
Working in a basis where  $ka_i = a_1 a_2 = 0$, with $i=1,2$ (see Appendix \ref{sec:lccappendix}), we can show that the most general (antisymmetric) field tensor obeying Eq.~(\ref{eqn:mwefieldtensor}) is given by \cite{schwinger_gauge_1951}
\begin{gather}
\label{eqn:fieldtensor}
F^{\mu\nu}(kx) 
= 
f_1^{\mu\nu} \psi_1'(kx)
+
f_2^{\mu\nu} \psi_2'(kx),
\end{gather}
where 
\begin{gather}
f_i^{\mu\nu} = k^\mu a_i^\nu - k^\nu a_i^\mu,\\
f^{\mu}_{i\,\rho} f_{j}^{\rho\nu} = -\delta_{ij} a_i^2\, k^\mu k^\nu,
\quad
k_\mu f_i^{\mu\nu} = 0.
\end{gather}
The scalar functions $\psi_i(kx)$ are arbitrary, restricted only by the physical requirement that the external field is of finite extent [i.e., $\psi_i(\pm\infty)=\psi'_i(\pm \infty)=0$, with $\psi_i(kx)$, $\psi'_i(kx)$ vanishing fast enough at infinity]. Furthermore, we adopt (without restriction) the normalization condition $\abs{\psi_i(kx)},\abs{\psi'_i(kx)} \lesssim 1$. We note that in the Lorentz gauge ($\del_\mu A^\mu = 0$) the four-potential corresponding to the field tensor in Eq.~(\ref{eqn:fieldtensor}) can be chosen in the form
\begin{gather}
\label{eqn:planewavefourpotential}
\begin{aligned}
A^\mu(kx) &= a_1^\mu \psi_1(kx) + a_2^\mu \psi_2(kx),
\\
F^{\mu\nu} &= \del^\mu A^\nu - \del^\nu A^\mu.
\end{aligned}
\end{gather}

Using the tensors $f_i^{\mu\nu}$ and an arbitrary four-momentum vector $q^\mu$ (most commonly the four-momentum of an incoming particle), we can define the following quantum nonlinearity parameters \cite{di_piazza_extremely_2012}:
\begin{gather}
\chi_{i}
=
-\frac{e\sqrt{qf_i^2q}}{m^3}
=
\eta \, \xi_i,
\end{gather}
where
\begin{gather}
\eta =  \frac{\sqrt{(kq)^2}}{m^2},
\end{gather}
and where the quantities
\begin{gather}
\label{eqn:xiparameterdef}
\xi_i 
= 
\frac{1}{m} \sqrt{-a_i^2 e^2}
\end{gather}
are the so-called classical intensity parameters. Since both $\eta$ and $\chi_{i}$ are manifestly gauge and Lorentz invariant, also the parameters $\xi_i$ are gauge and Lorentz invariant. Due to the normalization condition for the shape functions $\psi_i(kx)$, the parameters $\xi_i$ characterize the strength of the plane-wave field. It turns out that the plane-wave field must be taken into account exactly in the calculations if $\xi_i \gtrsim 1$ \cite{di_piazza_extremely_2012}. Modern laser facilities can easily reach this nonperturbative domain, e.g. in Ref. \cite{yanovsky_ultra_2008} $\xi_i \sim 100$ was obtained [we point out that the parameter $\xi_0$, mentioned in the introduction, can be related to the parameters $\xi_i$ by noting that, for a linearly-polarized field with $\psi_2(kx)=0$ and with electric-field amplitude $E_0$ and central angular frequency $\omega_0$, the quantity $\sqrt{-a_1^2}$ can be set equal to $E_0/\omega_0$].

It is also convenient to introduce the integrated field-strength tensor,
\begin{gather}
\Ftilde^{\mu\nu}(kx)
= 
\int^{kx}_{-\infty} d\phi'\, F^{\mu\nu}(\phi'),
\end{gather}
which can be written as
\begin{gather}
\begin{aligned}
\Ftilde^{\mu\nu}(kx)
&=
k^\mu A^\nu(kx) - k^\nu A^\mu(kx)
\\&=
f_1^{\mu\nu} \psi_1(kx)
+
f_2^{\mu\nu} \psi_2(kx)
\end{aligned}
\end{gather}
in the Lorentz gauge [we will use~$\Ftilde^{\mu\nu}_x = \Ftilde^{\mu\nu}(kx)$ interchangeably to denote the argument]. Both~$F^{\mu\nu}$ and~$\Ftilde^{\mu\nu}$ have the important algebraic property that successive contractions of more than two tensors vanish, and their square is proportional to~$k^\mu k^\nu$, e.g.,
\begin{gather}
\Ftilde_x^{\mu\rho} \Ftilde_{y\rho\nu}
=
- k^\mu k_\nu  \sum_{i=1,2} a_i^2 \psi_i(kx) \psi_i(ky).
\end{gather}

If the background field is a plane-wave field, the Dirac equation [Eq.~(\ref{eqn:diracequation}) with $\mc{A}^\mu \to A^\mu$] can be solved analytically \cite{volkov_ueber_1935}. The corresponding so-called Volkov solution with the boundary condition $\Psi_p \to \psi_p$ if $kx \to -\infty$ [see Eq.~(\ref{eqn:freeplanewave})] can be written as \cite{landau_quantum_1981,ritus_radiative_1972,mitter_quantum_1975}
\begin{gather}
\Psi_p = \frac{1}{\sqrt{2\eps}} E_{p,x} u_p,
\quad
E_{p,x} = \lsb\one + \frac{e\s{k}\s{A}(kx)}{2\, kp}\rsb \, e^{iS_p(x)},
\end{gather}
where the phase is given by 
\begin{gather}
S_p(x) 
=
- px - \int_{-\infty}^{kx} d\phi' \, \lsb \frac{e\, pA(\phi')}{\, kp} - \frac{e^2A^2(\phi')}{2\, kp} \rsb.
\end{gather}
Note that Volkov states, although being an exact solution of the Dirac equation and apart from the spin-terms proportional to $\s{k}\s{A}(kx)$, have a quasiclassical structure $\sim \exp[iS_p(x)]$, with $S_p(x)$ being the classical action of an electron inside a plane-wave field \cite{landau_classical_1987}.

The dressed propagator (which is the Green's function of the interacting Dirac equation) is given by
\begin{gather}
\label{eqn:dressedpropagator}
iG(x,y) 
= 
i\int \frac{d^4p}{(2\pi)^4} E_{p,x} \frac{\s{p} + m}{p^2 - m^2 + i0} \bar{E}_{p,y},
\end{gather}
where
\begin{gather}
\bar{E}_{p,x} 
= 
\lsb\one + \frac{e\s{A}(kx)\s{k}}{2\, pk}\rsb \, e^{-iS_p(x)}.
\end{gather}
Thus, in comparison with the vacuum case, the plane waves are replaced by the Ritus $E_p$ functions, which depend nontrivially on the plane-wave phase~$kx$. However, they also form an orthogonal and complete set \cite{ritus_radiative_1972}:
\begin{gather}
\label{eqn:epcompletenessandorthogonality}
\begin{aligned}
\int \frac{\D{}^4p}{(2\pi)^4}\, E_{p,x} \bar{E}_{p,x'} &= \delta^4(x-x'),\\
\int \D{}^4x\, \bar{E}_{p',x} E_{p,x} &= (2\pi)^4 \, \delta^4(p'-p).
\end{aligned}
\end{gather}
The~$E_p$ functions convert the dressed momentum into the free momentum \cite{ritus_radiative_1972}:
\begin{gather}
\label{eqn:Epprojectionproperty}
\begin{aligned}
\phantom{}[i\s{\del}_x - e\s{A}(kx)] E_{p,x} &=  E_{p,x} \s{p},\\
-i \del_x^\mu \bar{E}_{p,x} \gamma_\mu  -e \bar{E}_{p,x} \s{A}(kx) &= \s{p} \bar{E}_{p,x}
\end{aligned}
\end{gather}
(these identities hold only if the derivative acts solely on $E_{p,x}$ and $\bar{E}_{p,x}$, respectively).

\subsection{Dressed vertex}
 
To obtain Feynman rules in momentum space, we can proceed analogously as in the vacuum case and move the $E_{p}$-functions to the vertex \cite{mitter_quantum_1975}. Correspondingly, we define the dressed vertex by 
\begin{gather}
\label{eqn:sfqed_dressedvertex}
\Gamma^\rho(p',q,p) = -ie  \int d^4x \, e^{-iqx}\, \bar{E}_{p',x} \gamma^\rho E_{p,x}.
\end{gather}
Working in momentum space, the only difference between vacuum QED and strong-field QED is the vertex we have to use
[i.e., the free vertex in Eq.~(\ref{eqn:freevertex}) is replaced by the dressed vertex in Eq.~(\ref{eqn:sfqed_dressedvertex})]. Using the relations given in Appendix~\ref{sec:gammamatrixalgebraappendix}, we can write the dressed vertex as
\begin{multline}
\label{eqn:dressedvertexfinal}
\Gamma^\rho(p',q,p)
= 
-ie  \int d^4x \,\big[ \gamma_\mu G^{\mu\rho}(kp',kp;kx) 
\\+ i\gamma_\mu \gamma^5 G_5^{\mu\rho}(kp',kp;kx) \big]  e^{iS_\Gamma(p',q,p;x)},
\end{multline}
where the phase and the coupling tensors are given by
\begin{multline}
S_\Gamma(p',q,p;x)
=
-S_{p'}(x) -qx  + S_p(x)
\\=
(p'-q-p)x
+
\int_{-\infty}^{kx} d\phi'
\,  \Big[ \frac{e p_\mu p'_\nu \Ftilde^{\mu\nu}(\phi')}{(kp)(kp')} 
\\+ \frac{e^2(kp-kp')}{2(kp)^2(kp')^2} p_\mu p'_\nu \Ftilde^{2\mu\nu}(\phi') \Big],
\end{multline}
\begin{gather}
\label{eqn:dressedvertexGtensorsdefinition}
\begin{aligned}
G^{\mu\rho}(kp',kp;kx)
&=
g^{\mu\rho} 
+ 
G_1  \Ftilde^{\mu\rho}_x
+ 
G_2 \Ftilde^{2\mu\rho}_x,\\
G_{5}^{\mu\rho}(kp',kp;kx)
&=
G_3 \Ftilde^{*\mu\rho}_x,
\end{aligned}
\end{gather}
\begin{gather}
\label{eqn:Gidefs}
\begin{gathered}
G_1 = -e\, \frac{kp + kp'}{2 kp \, kp'},
\quad
G_2 = \frac{e^2}{2kp\, kp'},\\
G_3 = -e\, \frac{kp - kp'}{2 kp \, kp'}
\end{gathered}
\end{gather}
(note that $G_1$ and $G_2$ are even in the permutation $kp \leftrightarrow kp'$ while $G_3$ is odd). We point out that the expression given in Eq.~(\ref{eqn:dressedvertexfinal}) is manifestly gauge invariant, since it depends on the external field only through the tensor~$\Ftilde^{\mu\nu}$ \cite{mitter_quantum_1975}. 

In position space the dressed propagator in Eq.~(\ref{eqn:dressedpropagator}) can be interpreted such that the electron (or positron) continuously interacts with the external field during its propagation. Examined in momentum space, we can also visualize the influence of the external field as a modification of the coupling between the photons of the radiation field and the charged particles. From Eq.~(\ref{eqn:dressedvertexfinal}) we see that, besides the modification of the photon  vector current interaction we also obtain a coupling to the axial-vector current inside the plane-wave background. This is possible since the external field provides the pseudotensor $\Ftilde^{*\mu\nu}$. 

Since the external field depends only on the plane-wave phase~$\phi=kx$, it is useful to use light-cone coordinates, which are discussed in Appendix~\ref{sec:lccappendix}. We can then always take the integrals in~$dx^\lplus$ and~$dx^\lperp$ in Eq.~(\ref{eqn:dressedvertexfinal}) and obtain momentum-conserving delta functions in three of four light-cone components,
\begin{gather}
\delta^{(\lminus,\lperp)}(p'-p-q),
\end{gather}
where we used the notation
\begin{gather}
\delta^{(\lminus,\lperp)}(a)
=
\delta(a^\lminus) \delta(a^\lone) \delta(a^\ltwo)
\end{gather}
for a general four-vector $a^\mu$. Thus, the four-momentum is only conserved up to a four-vector proportional to the plane-wave four-momentum $k^\mu$ at each vertex.

\subsection{Ward-Takahashi identity}
\label{sec:wardidentity}

The Ward-Takahashi identity \cite{ward_identity_1950,takahashi_generalized_1957} is a direct consequence of the gauge invariance of QED, which becomes  particularly  transparent in the functional integral approach \cite{collins_renormalization_1984,weinberg_quantum_1995}. Diagrammatically, it is a functional relation for Feynman diagrams (in momentum space), where the polarization four-vector of an external photon leg is replaced by the corresponding momentum four-vector. In Ref. \cite{peskin_introduction_2008} a perturbative proof of the Ward-Takahashi identity in vacuum QED is given. We will show now how this proof can be extended to electron-positron loops inside a plane-wave background field. 

\begin{figure}[b!]
\centering
\includegraphics[height=4.1cm]{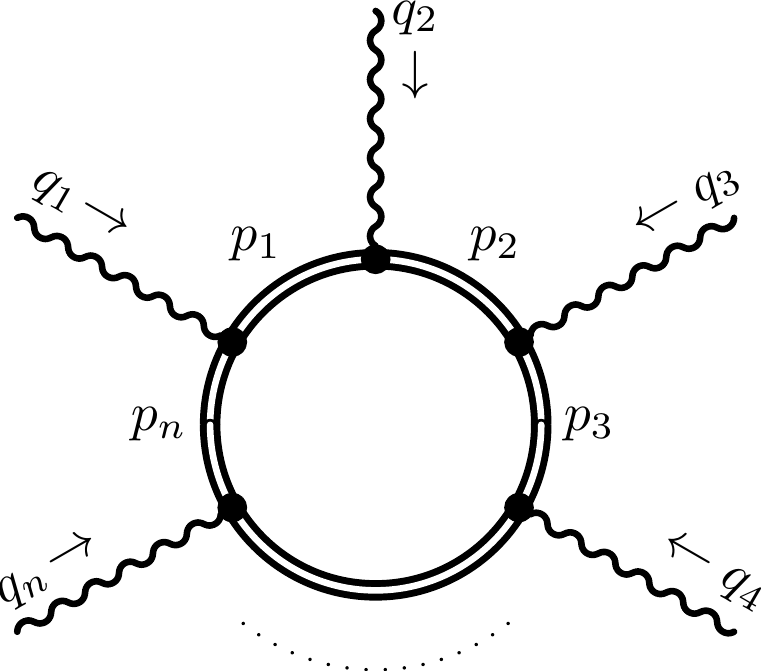}
\caption{\label{fig:wardidentity} Closed electron loop with $n$ dressed vertices and propagators.}
\end{figure}

The starting point is the following algebraic identity for the dressed vertex \cite{mitter_quantum_1975}
\begin{multline}
\label{eqn:algebraicidentitydressedvertex}
q_\rho \Gamma^\rho(p',q,p)
\\=
(\s{p}' - m) I(p',q,p) - I(p',q,p) (\s{p} - m),
\end{multline}
where
\begin{gather}
I(p',q,p) 
=
-ie  \int d^4x \, e^{-iqx}\, \bar{E}_{p',x} E_{p,x}. 
\end{gather}
To verify Eq.~(\ref{eqn:algebraicidentitydressedvertex}), we use Eq.~(\ref{eqn:Epprojectionproperty}) and note that the identity
\begin{gather}
\int d^4x \ i\del_\mu \big[\bar{E}_{p',x} \gamma^\mu e^{-iqx} E_{p,x} \big]
= 0
\end{gather}
holds \cite{mitter_quantum_1975}.

Typically, $(\s{p}' - m)$ and $(\s{p} - m)$ in Eq.~(\ref{eqn:algebraicidentitydressedvertex}) cancel an adjacent propagator, and the associated momentum-integral can be taken using the relations
\begin{gather}
\label{eqn:IGammacontraction}
\begin{aligned}
\int \frac{d^4p''}{(2\pi)^4}
I(p,q',p'') \Gamma^\mu(p'',q,p')
&=
-ie \Gamma^\mu (p,q+q',p'),\\
\int \frac{d^4p''}{(2\pi)^4}
\Gamma^\mu(p,q,p'') I(p'',q',p') 
&=
-ie \Gamma^\mu (p,q+q',p'),
\end{aligned}
\end{gather}
which follow from Eq.~(\ref{eqn:epcompletenessandorthogonality}). Using Eqs. (\ref{eqn:algebraicidentitydressedvertex}) and (\ref{eqn:IGammacontraction}), we can simplify diagrams which contain dressed vertices contracted with the corresponding photon four-momenta. 

As an example, we consider now a closed electron loop which contains~$n$ dressed vertices and electron propagators (see Fig.~\ref{fig:wardidentity}). The $i$th propagator of such a loop together with its adjacent vertices is given by
\begin{gather}
\cdots \Gamma^{\mu_i}(p_{i-1},q_i,p_i) \frac{1}{\s{p}_i - m} \Gamma^{\mu_{i+1}}(p_i,q_{i+1},p_{i+1}) \cdots
\end{gather}
(the electron four-momenta $p_i$ are integrated out). If we insert now a vertex (contracted with its photon four-momentum) at this propagator, we obtain
\begin{multline}
\cdots \Gamma^{\mu_i}(p_{i-1},q_i,p_i) \frac{1}{\s{p}_i - m} 
q_\mu \Gamma^{\mu}(p_i,q,p')  
\\\times \,
\frac{1}{\s{p}'-m}
\Gamma^{\mu_{i+1}}(p',q_{i+1},p_{i+1}) \cdots
\end{multline}
and, by using Eqs.~(\ref{eqn:algebraicidentitydressedvertex}) and (\ref{eqn:IGammacontraction}), we find that this is equivalent to
\begin{gather}
\begin{aligned}
&\phantom{-}\cdots \Gamma^{\mu_i}(p_{i-1},q_i+q,p_i) \frac{1}{\s{p}_i - m} \Gamma^{\mu_{i+1}}(p_i,q_{i+1},p_{i+1}) \cdots
\\
&-
\cdots \Gamma^{\mu_i}(p_{i-1},q_i,p_i) \frac{1}{\s{p}_i - m} \Gamma^{\mu_{i+1}}(p_i,q_{i+1}+q,p_{i+1}) \cdots
\end{aligned}
\end{gather}
(in the first line, we have changed the name of the integration variable from $p'$ to $p_i$). Thus, the insertion splits the diagram into the sum of twice the original diagram with the additional four-momentum $q$ added once at the adjacent vertex before and after the insertion. If we sum now over all possible insertion points of the loop, we obtain zero since all contributions cancel pairwise (as in the vacuum case \cite{peskin_introduction_2008}).

We point out that the above discussion is shortened, since possible issues arising due to the renormalization of the theory were not addressed (in general, the validity of the Ward-Takahashi identity may be spoiled by anomalies \cite{adler_axial-vector_1969}). In this paper, however, we are mainly interested in modifications induced by the background field, which turn out to be finite. Thus, subtleties arising from manipulations of divergent integrals can be addressed as in vacuum QED.

\section{Polarization operator}
\label{sec:polarizationoperator}
\enlargethispage{\baselineskip}

\subsection{General expression}

The leading-order contribution to the polarization operator $\mathcal{P}^{\mu\nu}(q_1,q_2)$ for plane-wave background fields (see Ref. \cite{landau_quantum_1981}, Sec. 104) is determined by the diagram in Fig.~\ref{fig:polarizationoperator}. This diagram corresponds to the following expression: 
\begin{multline}
\label{eqn:polarizationoperator}
\T^{\mu\nu}(q_1,q_2)
=
\int  \frac{d^4p\, d^4p'}{(2\pi)^8} 
\tr 
\, 
\Gamma^\mu(p',q_1,p) 
\\ \times  \frac{(\s{p} + m)}{p^2-m^2+i0} \, 
\Gamma^\nu(p,-q_2,p')
\frac{(\s{p}' + m)}{p'^2-m^2+i0}
\end{multline}
and $\T^{\mu\nu} = i \mathcal{P}^{\mu\nu}$ (see Ref. \cite{landau_quantum_1981}, Sec. 115; \cite{baier_operator_1975}). We note that $\T^{\mu\nu}(q_1,q_2)$ is divergent, but if we write
\begin{multline}
\label{eqn:regularizationpolarizationoperator}
\T^{\mu\nu}(q_1,q_2) 
=
\lsb \T^{\mu\nu}(q_1,q_2) - \T^{\mu\nu}_{\Ftilde=0}(q_1,q_2) \rsb
\\+ \T^{\mu\nu}_{\Ftilde=0}(q_1,q_2),
\end{multline}
the contribution in square brackets is finite \cite{baier_interaction_1975}, and the regularization of the vacuum contribution is well known \cite{landau_quantum_1981,weinberg_quantum_1995}. In the following, we will focus on the tensor in square brackets which contains only the corrections induced by the external background field.

To determine the expression in Eq.~(\ref{eqn:polarizationoperator}), we have to insert the dressed vertex given in Eq.~(\ref{eqn:dressedvertexfinal}) [we will denote the vertex integrals associated with~$\Gamma^\mu(p',q_1,p)$ and~$\Gamma^\nu(p,-q_2,p')$ by~$d^4x$ and~$d^4y$, respectively]. We then obtain for $\T^{\mu\nu}(q_1,q_2)$
\begin{multline}
\label{eqn:polarizationoperatorB}
\T^{\mu\nu}(q_1,q_2)
=
4 \, (-ie)^2 \int  \frac{d^4p\, d^4p'}{(2\pi)^8} \int d^4x d^4y\,
\\ \times \,
\frac{\frac14 \tr \big[ \cdots \big]^{\mu\nu}}{(p^2-m^2+i0)(p'^2-m^2+i0)} e^{iS_\T}
\end{multline}
(the prefactor $\nfrac14$ in front of the trace is included explicitly for later convenience), where the phase reads [see Eq.~(\ref{eqn:dressedvertexfinal})]
\begin{multline}
\label{eqn:sfqed_polarizationoperatorphasestructure}
iS_\T
= 
i(p'-p-q_1)x 
+
i(p-p'+q_2)y
\\+ 
i\int_{ky}^{kx} d\phi'\, \bigg[ \frac{e p_\mu p'_\nu \Ftilde^{\mu\nu}}{(kp)(kp')} 
+ \frac{e^2(kp-kp')}{2(kp)^2(kp')^2} p_\mu p'_\nu \Ftilde^{2\mu\nu} \bigg]
\end{multline}
and $\frac14 \tr \big[ \cdots \big]^{\mu\nu}$ in Eq.~(\ref{eqn:polarizationoperatorB}) can be calculated using the relations given in Appendix~\ref{sec:gammamatrixalgebraappendix}:
\begin{widetext}
\begin{multline}
\label{eqn:sfqed_polarizationoperatortraceA}
\frac14 \tr 
\big[ \gamma_\alpha a^{\alpha\mu} + i\gamma_\alpha \gamma^5 b^{\alpha\mu} \big]
(\s{p} + m) \, 
\big[ \gamma_\beta c^{\beta\nu} + i\gamma_\beta \gamma^5 d^{\beta\nu} \big]
(\s{p}' + m)
\\
=
m^2[(a^{\alpha\mu} c_\alpha^{\phantom{\alpha}\nu})
+(b^{\alpha\mu}d_\alpha^{\phantom{\alpha}\nu})] 
+ (pp')(b^{\alpha\mu}d_\alpha^{\phantom{\alpha}\nu}) 
- (pp')(a^{\alpha\mu} c_\alpha^{\phantom{\alpha}\nu})
+ (p_\alpha a^{\alpha\mu})(p'_\beta c^{\beta\nu}) 
+ (p'_\alpha a^{\alpha\mu})(p_\beta c^{\beta\nu}) 
\\
- (p_\alpha b^{\alpha\mu})(p'_\beta d^{\beta\nu}) 
- (p'_\alpha b^{\alpha\mu})(p_\beta d^{\beta\nu}) 
- \eps_{\rho\sigma\alpha\beta} p^\rho p'^\sigma (a^{\alpha\mu} d^{\beta\nu} + b^{\alpha\mu} c^{\beta\nu}),
\end{multline}
\end{widetext}
where
\begin{gather}
\begin{aligned}
a^{\alpha\mu} 
&=
G^{\alpha\mu}(kp',kp;kx),&
c^{\beta\nu}
&=
G^{\beta\nu}(kp,kp';ky),\\
b^{\alpha\mu}
&= 
G_5^{\alpha\mu}(kp',kp;kx),& 
d^{\beta\nu} 
&=
G_5^{\beta\nu}(kp,kp';ky).
\end{aligned}
\end{gather}

\subsection{Evaluation of the integrals}
\label{sec:polarizationoperatorintegralevaluation}

Working in light-cone coordinates (see Appendix~\ref{sec:lccappendix}) we can take all space-time integrals except of those in $dx^\lminus$ and~$dy^\lminus$ and obtain the momentum-conserving delta functions
\begin{gather}
(2\pi)^6 \delta^{(\lminus,\lperp)}(p'-p-q) \, \delta^{(\lminus,\lperp)}(q_1-q_2).
\end{gather}
Here and in the following, we write $q^\mu$ if $q^\mu_1$ and $q^\mu_2$ can be used interchangeably due to the above delta function. Successively, we can take the integrals in~$dp'^\lminus$ and~$dp'^\lperp$ (for simplicity we will continue writing $p'$ and identify $p'=p+q$ for the components~$\lminus,\lperp$). 

It is now convenient to introduce the two four-vectors:
\begin{gather}
\label{eqn:Lambdavectors}
\Lambda_1^\mu = \frac{f_1^{\mu\nu} q_\nu}{kq \sqrt{-a_1^2}},
\quad
\Lambda_2^\mu = \frac{f_2^{\mu\nu} q_\nu}{kq \sqrt{-a_2^2}},
\end{gather}
which obey~$\Lambda_i \Lambda_j = -\delta_{ij}$, $k\Lambda_i = q_i\Lambda_j = 0$ and
\begin{gather}
f_1^{\mu\nu} \Lambda_{1\nu} = - \frac{m}{e} k^\mu \xi_1,
\quad
f_2^{\mu\nu} \Lambda_{2\nu} = - \frac{m}{e} k^\mu \xi_2.
\end{gather}
They allow us to write the remaining phase as
\begin{multline}
\label{eqn:sfqed_polarizationoperatorphasestructureB}
iS_\T
= 
i(p'-p-q_1)^\lplus x^\lminus 
\\+
i(p-p'+q_2)^\lplus y^\lminus
+
i p\uplambda + i \Uplambda,
\end{multline}
where we defined
\begin{gather}
\label{eqn:sfqed_polarizationoperatorintegraldefinitions}
\begin{aligned}
\uplambda^\mu 
&=
- \frac{m (kq)}{(kp)(kp')} \sum_{i=1,2} \xi_i \Lambda_i^\mu \int_{ky}^{kx} d\phi' \, \psi_i(\phi'),\\
\Uplambda
&=  
-\frac{m^2 (kq)}{2(kp)(kp')} \sum_{i=1,2} \xi^2_i \int_{ky}^{kx} d\phi' \,  \psi_i^2(\phi').
\end{aligned}
\end{gather}
Due to the appearance of $\Lambda_i^\mu$ in $\uplambda^\mu$, it is more convenient to use modified light-cone coordinates from now on [see Eq.~(\ref{eqn:modifiedlcc}); the calculation so far is independent of this choice]. In modified light-cone coordinates, we obtain the convenient relations
\begin{gather}
p\uplambda = - p^\lperp \uplambda^\lperp,
\quad
q^\lperp = 0,
\quad 
p'^\lperp = p^\lperp,
\end{gather}
which simplify the algebra considerably.

If the preexponent would not depend on~$p^\lplus$ and $p'^\lplus$, both integrals could now be taken. We therefore introduce the proper-time representation of the scalar propagators \cite{schwinger_gauge_1951,dittrich_probingquantum_2000}:
\begin{multline}
\label{eqn:sfqed_polarizationoperatorschwingerpropagator}
\frac{1}{p^2-m^2+i0} \frac{1}{p'^2-m^2+i0}
=
(-i)^2 \int_0^\infty ds \, dt 
\\ \times \, \exp \lsb i(p^2-m^2+i0)s + i(p'^2-m^2+i0)t \rsb.
\end{multline}
In the following we will drop the pole prescriptions $i0$ and keep the replacement $m^2 \to m^2 - i0$ in mind. Furthermore, we add the source terms~$i p_\mu j^\mu + i p'_\mu j'^\mu$ to the phase, which allows us to make the replacement 
\begin{gather}
\label{eqn:momentumreplacementtrace}
\s{p} \longrightarrow (-i)\s{\del}_j,
\quad
\s{p}' \longrightarrow (-i)\s{\del}_{j'}
\end{gather} 
in the trace. Now, the preexponent depends only on $p^\lminus$ (through~$kp$ and~$kp'$). Taking the derivatives with respect to the sources out of the integrals, we can take the integrals in $dp^\lplus$, $dp'^\lplus$, which results in the delta functions,
\begin{multline}
\label{eqn:yminuspminusdeltafunctions}
(2\pi) \delta\Big[y^\lminus - x^\lminus -\frac{1}{s+t} (2st q^\lminus - t j^\lminus + s j'^\lminus)\Big]
\\ \times \,
(2\pi) \delta[2p^\lminus (s+t)  + 2 q^\lminus t + j^\lminus + j'^\lminus].
\end{multline}
Successively, these delta functions can be used to take also the integrals in~$dy^\lminus$ and~$dp^\lminus$. To this end we rewrite (since~$s+t\geq 0$)
\begin{multline}
\delta[2p^\lminus (s+t)  + 2 q^\lminus t + j^\lminus + j'^\lminus]
\\=
\frac{1}{2(s+t)} \delta\Big[p^\lminus  + \frac{1}{2(s+t)} (2q^\lminus t + j^\lminus + j'^\lminus)\Big]
\end{multline}
(for simplicity we keep writing $y^\lminus$ and $p^\lminus$). In particular, we obtain the identities
\begin{gather}
\label{eqn:kpkpprimekyidentifications}
\begin{aligned}
kp  &=  -\frac{1}{s+t} \Big[  t kq   + \frac{1}{2}(kj + kj') \Big],\\
kp' &=  +\frac{1}{s+t} \Big[  s kq   - \frac{1}{2}(kj + kj') \Big],\\
ky  &=  kx + \frac{1}{s+t} (2st kq - t kj + s kj'),
\end{aligned}
\end{gather}
which imply for~$j=j'=0$ that
\begin{gather}
\label{eqn:Givssandt}
\begin{aligned}
G_1 &= \frac{e}{2kq} \frac{(s-t)(s+t)}{st} =  \frac{e}{2kq} \frac{v \tau}{\mu},\\
G_2 &= - \frac{e^2}{2(kq)^2} \frac{(s+t)^2}{st} = - \frac{e^2}{2(kq)^2} \frac{\tau}{\mu},\\
G_3 &= - \frac{e}{2kq} \frac{(s+t)^2}{st} = - \frac{e}{2kq} \frac{\tau}{\mu},
\end{aligned}
\end{gather}
where we defined \cite{baier_interaction_1975}
\begin{gather}
\label{eqn:tauvmudefinition}
\tau = s+t,
\quad
v = \frac{s-t}{s+t},
\quad
\mu = \frac{st}{s+t}
= 
\frac14 \tau (1-v^2)
\end{gather}
[the motivation for these definitions becomes clear in Eq.~(\ref{eqn:sfqed_polarizationoperatortraceC})].

The remaining part of the phase structure (including the part coming from the propagators and the sources) is now given by
\begin{multline}
iS'_\T
=
i \Big[ (q_2^\lplus -q_1^\lplus) x^\lminus 
+ 
\frac{st}{s+t} q_2^2 
-
\frac{1}{s+t} (t \, q_2j - s\, q_2j')
\\
\begin{aligned}
&- 
\frac{1}{2(s+t)} (j^\lplus + j'^\lplus) (j^\lminus + j'^\lminus)
\\
&- (p^\lperp p^\lperp + m^2)(s+t)
\end{aligned}
\\-
(j^\lperp + j'^\lperp + \uplambda^\lperp) p^\lperp 
+ 
\Uplambda \Big].
\end{multline}
Taking the Gaussian integrals in~$p^\lone$ and~$p^\ltwo$, we obtain the prefactor $\frac{\pi}{i(s+t)}$, and the final phase is given by
\begin{multline}
\label{eqn:sfqed_polarizationoperatorfinalphase}
iS'_\T
=
i \Big[ (q_2^\lplus -q_1^\lplus) x^\lminus 
- 
m^2 (s+t)
+ 
\frac{st}{s+t} q_2^2 
\\-
\frac{1}{s+t} (t \, q_2j - s\, q_2j')
- 
\frac{1}{4(s+t)} (j+j')^2 
\\-
\frac{1}{2(s+t)} (j+j')\uplambda
-
\frac{1}{4(s+t)} \uplambda^2
+  
\Uplambda \Big],
\end{multline}
which reads for zero sources ($j=j'=0$)
\begin{multline}
\label{eqn:sfqed_polarizationoperatorfinalphasezerosources}
iS'_\T
=
i \Big[ (q_2^\lplus -q_1^\lplus) x^\lminus 
+ 
\mu q_2^2 
\\- 
\tau m^2 + \tau m^2 \sum_{i=1,2} \xi_i^2 (I^2_i-J_i)
\Big],
\end{multline}
where we defined
\begin{gather}
\label{eqn:defIJintegrals}
\begin{aligned}
I_i 
&=
-\frac{1}{2kq \mu} \int_{ky}^{kx} d\phi' \, \psi_i(\phi'),\\
J_i 
&=
-\frac{1}{2kq \mu} \int_{ky}^{kx} d\phi' \, \psi^2_i(\phi')
\end{aligned}
\end{gather}
(the prefactor is chosen such that, for $j=j'=0$ and $\psi_i= 1$, we obtain $I_i=J_i=1$).

Finally, we can write the tensor $\T^{\mu\nu}$ as
\begin{multline}
\label{eqn:polarizationoperatorafterintegralstaken}
\T^{\mu\nu}(q_1,q_2)
= 
-2i\pi e^2\, \delta^{(\lminus,\lperp)}(q_1-q_2) 
\int_0^\infty ds \, dt 
\\ \times\,
\int_{-\infty}^{+\infty} dx^\lminus \,
\frac{1}{(s+t)^2} \frac14 \tr \lsb \ldots \rsb^{\mu\nu}  e^{iS'_\T} \Big|_{j=j'=0},
\end{multline}
where the expression for $\frac14 \tr \lsb \ldots \rsb^{\mu\nu}$ is given in Eq.~(\ref{eqn:sfqed_polarizationoperatortraceA}) with the replacement in Eq.~(\ref{eqn:momentumreplacementtrace}) and where the sources are set to zero after the derivatives are taken.

We point out that the two four-momenta $q_1$ and $q_2$ appear asymmetrically in the final expression [see Eq.~(\ref{eqn:sfqed_polarizationoperatorfinalphasezerosources})]. To remove this asymmetry, we shift the $x^\lminus$ integration by defining 
\begin{gather}
\label{eqn:symmetricintegralshift}
z^\lminus = x^\lminus + \mu q^\lminus. 
\end{gather}
After this shift, the phase contains $q_1q_2$ since
\begin{gather}
\label{eqn:symmetricphaseterms}
(q_2^\lplus -q_1^\lplus) x^\lminus 
+ 
\mu q_2^2 
=
(q_2^\lplus -q_1^\lplus) z^\lminus
+ 
\mu q_1 q_2.
\end{gather}
Furthermore, we obtain (for $j=j'=0$) symmetric representations for the functions in Eq.~(\ref{eqn:defIJintegrals}):
\begin{gather}
\label{eqn:IJsymmetricintegrals}
\begin{aligned}
I_i 
&=
\frac12 \int_{-1}^{+1} d\lambda \, \psi_i(kz - \lambda \mu kq),\\
J_i 
&=
\frac12 \int_{-1}^{+1} d\lambda \, \psi^2_i(kz - \lambda \mu kq),
\end{aligned}
\end{gather}
since
\begin{gather}
\label{eqn:kxkyversuskz}
\begin{aligned}
kx &= kz - \mu kq,\\
ky &= kz + \mu kq + \frac{1}{s+t}(skj'-tkj).
\end{aligned}
\end{gather}

\subsection{Tensor structure}

In principle, the only remaining task is to evaluate the two derivatives with respect to~$j$ and~$j'$ and then set~$j=j'=0$. Despite being elementary, this is the most tedious part of the calculation, since the sources appear in many places in the final expression. The work is considerably reduced if we expand the polarization operator in a convenient basis~\cite{baier_interaction_1975}. To this end we note that
\begin{gather}
q_{1\mu} \T^{\mu\nu}(q_1,q_2) = 0, \quad \T^{\mu\nu}(q_1,q_2) q_{2\nu} = 0
\end{gather}
due to the Ward-Takahashi identity (see Sec. \ref{sec:wardidentity}).

Since the four-vectors $\Lambda_i$ appear  in the phase [see Eq.~(\ref{eqn:sfqed_polarizationoperatorphasestructureB})] and~$q_i\Lambda_j=0$, it is natural to introduce the two complete sets~$q_1$, $\Q_1$, $\Lambda_1$, $\Lambda_2$ and $q_2$, $\Q_2$, $\Lambda_1$, $\Lambda_2$, where
\begin{gather}
\label{eqn:Qidef}
\Q_1^\mu = \frac{k^\mu q_1^2 - q_1^\mu kq}{kq},
\quad
\Q_2^\mu = \frac{k^\mu q_2^2 - q_2^\mu kq}{kq}
\end{gather}
($\Q_1^2 = -q_1^2$,  $\Q_2^2 = -q_2^2$, $\Q_i\Lambda_j=0$, $q_i \Q_i=0$). Using the set including~$q_1$ for the index~$\mu$ and the set including~$q_2$ for the index~$\nu$, seven of 16 coefficients vanish due to the Ward-Takahashi identity, and we can decompose~$\T^{\mu\nu}(q_1,q_2)$ as \cite{baier_interaction_1975}
\begin{multline}
\label{eqn:polarizationoperatordecomposition}
\T^{\mu\nu}
=
c_1 \Lambda_1^\mu \Lambda_2^\nu
+
c_2 \Lambda_2^\mu \Lambda_1^\nu
+
c_3 \Lambda_1^\mu \Lambda_1^\nu
\\+
c_4 \Lambda_2^\mu \Lambda_2^\nu
+
c_5 \Q_1^\mu \Q_2^\nu
+
c_6 \Q_1^\mu \Lambda_1^\nu
\\+
c_7 \Q_1^\mu \Lambda_2^\nu
+
c_8 \Lambda_1^\mu \Q_2^\nu
+
c_9 \Lambda_2^\mu \Q_2^\nu.
\end{multline}
It turns out that also the coefficients~$c_6-c_9$ vanish. If analyzed perturbatively (with respect to the external field coupling) this can be understood from Furry's theorem \cite{baier_interaction_1975,becker_vacuum_1975}. Since a closed fermion loop with an odd number of vertices does not contribute to the final amplitude, only diagrams with an even number of external field couplings ($eA^\mu$) contribute to~$\T^{\mu\nu}$. Due to gauge invariance and the fact that $T^{\mu\nu}$ is a tensor, the external field can enter only as~$\Ftilde^{\mu\nu}$ (which is linear in $A^\mu$). Since it is not possible to construct a scalar linear in $\Ftilde^{\mu\nu}$ using only the four-vectors~$q_1^\mu$, $q_2^\mu$ and~$k^\mu$, the tensor structure cannot involve an odd number of the tensor $\Ftilde^{\mu\nu}$ (note that $q_1\Ftilde q_2 = q\Ftilde q = 0$). As a consequence, the coefficients~$c_6-c_9$ (which are linear in~$\Lambda^\mu_i$ and thus in the external field) should vanish. We will later see that this is indeed the case.

The coefficients $c_i$ in Eq.~(\ref{eqn:polarizationoperatordecomposition}) can be determined by contracting~$\T^{\mu\nu}(q_1,q_2)$ with the appropriate four-vectors. Especially, using again the Ward-Takahashi identity, we obtain
\begin{gather}
\label{eqn:Qcontractionrelations}
\Q_{1\mu} \T^{\mu\nu} 
= 
\frac{q_1^2}{kq} k_\mu \T^{\mu\nu},
\quad
\T^{\mu\nu} \Q_{2\nu}
= 
\frac{q_2^2}{kq} \T^{\mu\nu} k_\nu.
\end{gather}

Thus, effectively, we need to determine the contractions of $\T^{\mu\nu}(q_1,q_2)$ with the four-vectors $k^\mu$ and $\Lambda_i^\mu$ to determine the coefficients $c_i$, i.e. we need to calculate the $(\lminus,\lperp)$-components of $\T^{\mu\nu}(q_1,q_2)$ in modified light-cone coordinates. Since $k^\mu$ has only a $\lplus$-component, the evaluation of the derivatives is now considerably simplified. Leaving the term proportional to~$pp'$ aside, we see that all derivatives which act on~$kj$ or~$kj'$ can be ignored. They would result in the replacement of~$p^\mu$ or~$p'^\mu$ by~$k^\mu$. Since $k_\mu \Ftilde^{\mu\nu} = k_\mu \Ftilde^{2\mu\nu} = k_\mu \Ftilde^{*\mu\nu} = 0$ and~$k^2=k\Lambda_i=0$, we do not need to consider those terms. The derivatives acting on~$kj$ or~$kj'$ are therefore only important to determine the term proportional to $pp'$. However, this is achieved more easily if the calculation presented in Sec. \ref{sec:polarizationoperatorintegralevaluation} is repeated with a scalar source term~$\mc{J} pp'$ in the exponent (see Sec. \ref{sec:scalarterm}).

To calculate the preexponent of the polarization operator, we must now insert the explicit expressions given in Eq.~(\ref{eqn:dressedvertexGtensorsdefinition}) into the trace in Eq.~(\ref{eqn:sfqed_polarizationoperatortraceA}). Many terms of the trace, e.g., the terms proportional to~$\Ftilde^{\mu\nu}$, $\Ftilde^{2\mu\nu}$, $\Ftilde^{2\mu\rho} p_\rho$ vanish, as they are contracted with~$k^\mu$ or~$\Lambda_i^\mu$ from each side. Using the relations in Appendix~\ref{sec:tensorrelationsappendix}, we can show that Eq.~(\ref{eqn:sfqed_polarizationoperatortraceA}) can be substituted by the following expression:
\begin{widetext}
\begin{multline}
\label{eqn:sfqed_polarizationoperatortraceB}
m^2 g^{\mu\nu} + p^\mu p'^\nu + p'^\mu p^\nu 
+ g^{\mu\nu} \big[G_3 p\Ftilde_{y}p' + G_3 p\Ftilde_{x}p' - 2 G_3^2 (p\Ftilde^2_{xy} p')  -(pp') \big]
\\- G_3 \big[ (\Ftilde_y p')^\mu p^\nu - (\Ftilde_y p)^\mu p'^\nu + p^\mu (\Ftilde_x p')^\nu - p'^\mu (\Ftilde_x p)^\nu \big]
- G_1 \big[ p^\mu (\Ftilde_y p')^\nu + p'^\mu (\Ftilde_y p)^\nu + (\Ftilde_x p)^\mu p'^\nu + (\Ftilde_x p')^\mu p^\nu \big]
\\+ G_1^2 \big[ (\Ftilde_x p)^\mu (\Ftilde_y p')^\nu + (\Ftilde_x p')^\mu (\Ftilde_y p)^\nu \big]
- G^2_3 \big[(\Ftilde_y p)^\mu (\Ftilde_x p')^\nu + (\Ftilde_y p')^\mu (\Ftilde_x p)^\nu \big],
\end{multline}
\end{widetext}
where~$\Ftilde^{2\mu\nu}_{xy}=\Ftilde^{\mu\rho}(kx) \Ftilde_{\rho}^{\phantom{\rho}\nu}(ky)=\Ftilde^{\mu\rho}(ky) \Ftilde_{\rho}^{\phantom{\rho}\nu}(kx)$ [here the replacement $p^\mu \longrightarrow (-i) \del_j^\mu$ and $p'^\mu \longrightarrow (-i) \del_{j'}^\mu$ is understood if the trace is inserted in Eq.~(\ref{eqn:polarizationoperatorafterintegralstaken}); see Eq.~(\ref{eqn:momentumreplacementtrace})]. Since the term proportional to~$pp'$ enters as~$g^{\mu\nu}$, it modifies only the diagonal coefficients~$c_3$ and~$c_4$.

\subsection{Evaluation of the derivatives}
\enlargethispage{2\baselineskip}

Leaving the term proportional to~$pp'$ aside, we can ignore derivatives acting on $kj$ and $kj'$ as discussed above [this implies that the derivatives do not act on $kp$, $kp'$, and $ky$; see Eq.~(\ref{eqn:kpkpprimekyidentifications})]. The remaining nontrivial source-dependent part of the phase is given by [see Eq.~(\ref{eqn:sfqed_polarizationoperatorfinalphase})]
\begin{multline}
-
\frac{i}{s+t} \Big[ t \, q_2j - s\, q_2j'
+ 
\frac{1}{4} (j+j')^2 
+
\frac{1}{2} (j+j')\uplambda \Big].
\end{multline}
The squared term contributes only if both derivatives act on it, which results in the replacement
\begin{subequations}
\begin{gather}
\label{eqn:ppprimereplacementA}
p^\alpha p'^\beta
\longrightarrow
(-i)^2 \del^\alpha_j \del^\beta_{j'}
\longrightarrow
\frac{i}{2(s+t)} g^{\alpha\beta}
\end{gather}
and the only nonzero contribution arises from the first three terms in Eq.~(\ref{eqn:sfqed_polarizationoperatortraceB}). If the derivatives act on the other source terms, we obtain the replacement
\begin{multline}
\label{eqn:ppprimereplacementB}
p^\alpha p'^\beta
\longrightarrow
(-i)^2 \del^\alpha_j \del^\beta_{j'}
\\\longrightarrow
-\frac{1}{(s+t)^2} 
\Big(tq_2^\alpha + \frac12 \uplambda^\alpha\Big) 
\Big(sq_2^\beta - \frac12 \uplambda^\beta\Big).
\end{multline}
\end{subequations}
After these replacements are applied to Eq.~(\ref{eqn:sfqed_polarizationoperatortraceB}) and the sources are set to zero, we obtain (effectively) the following expression for Eq.~(\ref{eqn:sfqed_polarizationoperatortraceB}):
\begin{gather}
\label{eqn:sfqed_polarizationoperatortraceC}
\begin{gathered} 
g^{\mu\nu} \Big[ m^2 + \frac{i}{\tau}
- \frac{e}{4kq\, \mu} (q\Ftilde_y\uplambda + q\Ftilde_x\uplambda) \hspace*{3cm}
\\\hspace*{3cm}+ \frac{e^2}{2(kq)^2} \frac{\tau}{\mu} q\Ftilde^2_{xy}q - pp'  \Big]\\
\begin{aligned}
&-  
2\frac{\mu}{\tau}  q_2^\mu q_2^\nu - \frac{v}{2\tau} (q_2^\mu \uplambda^\nu + \uplambda^\mu q_2^\nu) + \frac{1}{2\tau^2} \uplambda^\mu \uplambda^\nu
\\
&+ \frac{e}{kq} v \big[ q_2^\mu (\Ftilde_y q)^\nu + (\Ftilde_x q)^\mu q_2^\nu \big]
\\
&- \frac{e}{4kq} \frac{1}{\mu} 
\big[ (\Ftilde_y q)^\mu \uplambda^\nu + \uplambda^\mu (\Ftilde_x q)^\nu \big]
\\
&+ \frac{e}{4kq} \frac{v^2}{\mu} 
\big[ \uplambda^\mu (\Ftilde_y q)^\nu + (\Ftilde_x q)^\mu \uplambda^\nu \big]
\\
&+ \frac{e^2}{2(kq)^2} \frac{\tau}{\mu} \big[(\Ftilde_y q)^\mu (\Ftilde_x q)^\nu
-
v^2 (\Ftilde_x q)^\mu (\Ftilde_y q)^\nu \big]
\end{aligned}
\end{gathered}
\end{gather}
[note that terms proportional to $(\Ftilde\uplambda)^{\mu}$, $(\Ftilde\uplambda)^{\nu}$ can be omitted]. By changing the proper-time integrations from $s$, $t$ to $\tau$, $v$ \cite{baier_interaction_1975},
\begin{gather}
\label{eqn:stintegraltransform}
\int_0^\infty ds \, dt  \, f(s,t)
=
\frac12 \int_{-1}^{+1} dv \int_0^\infty d\tau \, \tau \tilde{f}(\tau,v)
\end{gather}
we see that the terms linear in $v$ vanish. Those terms determine the coefficients $c_6-c_9$, which are therefore zero (as already anticipated from Furry's theorem).

\subsection{Scalar term}
\label{sec:scalarterm}

To determine the term proportional to $pp'$, we add the scalar source term~$i \mc{J} pp'$ to the phase (instead of~$i p_\mu j^\mu + i p'_\mu j'^\mu$) and repeat the calculation presented in Sec. \ref{sec:polarizationoperatorintegralevaluation}. The propagators are represented in the same way [see Eq.~(\ref{eqn:sfqed_polarizationoperatorschwingerpropagator})], and we replace~$pp'$ by $-i\frac{\del}{\del\mc{J}}$. Then, we take the integrals in $dx^\lplus$, $dx^\lperp$, $dy^\lplus$, $dy^\lperp$, $dp'^\lminus$, $dp'^\lperp$, $dp'^\lplus$, and $dp^\lplus$. Instead of Eq.~(\ref{eqn:yminuspminusdeltafunctions}), we obtain now
\begin{multline}
(2\pi) \delta\Big[y^\lminus - x^\lminus -\frac{4st - \mc{J}^2}{2(s+t+\mc{J})} q^\lminus\Big]
\\ \times \,
(2\pi) \delta[2(s+t+\mc{J}) p^\lminus   + (2t+\mc{J}) q^\lminus ].
\end{multline}
The remaining part of the phase (including the part from the propagator) can be written as
\begin{multline}
iS_\T' = i\big[ q_2^\lplus y^\lminus - q_1^\lplus x^\lminus - p^\lperp p^\lperp \mc{J} \\+ (-p^\lperp p^\lperp - m^2) (s+t)  - p^\lperp \uplambda^\lperp + \Uplambda \big].
\end{multline}
It is now convenient to shift the proper-time integrations 
\begin{gather}
\label{eqn:stintegrationshift}
s \longrightarrow s - \frac12 \mc{J},
\quad
t \longrightarrow t - \frac12 \mc{J}.
\end{gather}
Due to this shift, also the integral boundaries of the proper-time integrations depend on the source. However, if the derivative acts on the integral boundaries, either $s$ or $t$ is set to zero or to infinity. The resulting terms do not depend on the external field since $s=0$ or $t=0$ implies $\mu=0$, $ky=kx$ and thus $\uplambda^\mu = 0$ and $\Uplambda = 0$. On the other hand, the terms at $s\to\infty$ or $t\to\infty$ do not contribute because the field-dependent part of the integral is convergent. Since we want to calculate only the field-dependent part of the polarization operator [see Eq.~(\ref{eqn:polarizationoperatorfinal})], we will ignore the source dependence of the integral boundaries.

After the shift in Eq.~(\ref{eqn:stintegrationshift}), the delta functions read
\begin{multline}
(2\pi) \delta[y^\lminus - x^\lminus - (2\mu-\mc{J})q^\lminus]
\\ \times \,
(2\pi)  \delta[2p^\lminus (s+t)  + 2 q^\lminus t]
\end{multline}
and the phase is given by
\begin{multline}
iS_\T' = i\Big[ (q_2^\lplus - q_1^\lplus) x^\lminus + \Big(\mu - \frac12 \mc{J}\Big) q_2^2 - m^2 (s+t-\mc{J}) \\- p^\lperp p^\lperp (s+t)  - p^\lperp \uplambda^\lperp + \Uplambda \Big].
\end{multline}
We can now use the delta functions to take the integrals in $dy^\lminus$ and $dp^\lminus$ (we keep writing $y^\lminus$ and $p^\lminus$ for convenience). We then obtain the identities
\begin{gather}
\begin{gathered}
kp  =  - \frac{t}{s+t} kq,
\quad
kp' =  \frac{s}{s+t} kq,\\
ky  =  kx + (2\mu -\mc{J}) kq
\end{gathered}
\end{gather}
[for $\mc{J}=0$ this agrees with Eq.~(\ref{eqn:kpkpprimekyidentifications})]. The shift in the proper-time integrals has the advantage that $kp$ and $kp'$ are now independent of $\mc{J}$. We could have proceeded similarly also in the calculation of the other terms. However, since we ignored sources contracted with $k$, this was not necessary.

Taking now the Gaussian integrals in~$dp^\lone$, $dp^\ltwo$, we obtain the prefactor~$\frac{\pi}{i(s+t)}$, and the final phase is given by
\begin{multline}
\label{eqn:sfqed_polarizationoperatorfinalphasezerosourcesscalarterm}
iS_\T' = i\Big[ (q_2^\lplus - q_1^\lplus) x^\lminus + \Big(\mu - \frac12 \mc{J}\Big) q_2^2 - m^2 (\tau-\mc{J}) \\ + \tau m^2 \sum_{i=1,2} \xi_i^2 (I^2_i-J_i) \Big],
\end{multline}
where $I_i$ and $J_i$ are defined in Eq.~(\ref{eqn:defIJintegrals}) [for zero sources Eq.~(\ref{eqn:sfqed_polarizationoperatorfinalphasezerosourcesscalarterm}) agrees with Eq.~(\ref{eqn:sfqed_polarizationoperatorfinalphasezerosources})]. Since $pp'$ in the preexponent is only multiplied by~$g^{\mu\nu}$ [see Eq.~(\ref{eqn:sfqed_polarizationoperatortraceC})], the evaluation of the derivative is not very cumbersome, and we obtain the replacement
\begin{multline}
\label{eqn:ppprimereplacement}
pp' 
\longrightarrow 
(-i) \frac{\del}{\del\mc{J}} 
\longrightarrow
- \frac12 q_2^2 + m^2 \\+ m^2  \frac{\tau}{2\mu} \sum_{i=1,2} \xi^2_i \big[\psi^2_i(ky) - 2 I_i \psi_i(ky) \big]
\end{multline}
after~$\mc{J}$ is set to zero (as explained above, we have ignored the source dependence of the proper-time integral boundaries).

To symmetrize the final expression, we change the $x^\lminus$-integration by defining [see Eq.~(\ref{eqn:symmetricintegralshift})]
\begin{gather}
\label{eqn:symmetricintegralshiftscalarterm}
\tilde{z}^\lminus = x^\lminus + \Big(\mu-\frac12 \mc{J}\Big) \, q^\lminus 
\end{gather}
($\tilde{z}^\lminus=z^\lminus$ for $\mc{J}=0$), which implies
\begin{gather}
\begin{aligned}
kx &= k\tilde{z} - \Big(\mu-\frac12 \mc{J}\Big) kq,&
ky &= k\tilde{z} + \Big(\mu-\frac12 \mc{J}\Big) kq
\end{aligned}
\end{gather}
and
\begin{multline}
(q_2^\lplus - q_1^\lplus) x^\lminus + \Big(\mu - \frac12 \mc{J}\Big) q_2^2
\\=
(q_2^\lplus - q_1^\lplus) \tilde{z}^\lminus + \Big(\mu - \frac12 \mc{J}\Big) q_1q_2.
\end{multline}
Finally, we obtain the symmetric replacement
\begin{multline}
\label{eqn:symmetricppprimereplacement}
pp' 
\longrightarrow 
(-i) \frac{\del}{\del\mc{J}} 
\longrightarrow
- \frac12 q_1q_2+ m^2 
+ 
m^2  \frac{\tau}{2\mu} \sum_{i=1,2} \xi^2_i  
\\\times \, 
\Big[\frac12\psi^2_i(kx) + \frac12\psi^2_i(ky) - I_i \psi_i(kx) - I_i \psi_i(ky) \Big]
\end{multline}
(we assume that at $x^\lminus=\pm\infty$ the external field vanishes, and therefore the derivative does not act on the integral boundaries, which now also depend on the source).

\subsection{Final result}

To determine the nonvanishing coefficients~$c_1-c_5$ of the polarization operator [see Eq.~(\ref{eqn:polarizationoperatordecomposition})], we combine now Eqs. (\ref{eqn:polarizationoperatorafterintegralstaken}), (\ref{eqn:symmetricintegralshift}), (\ref{eqn:sfqed_polarizationoperatortraceC}), and (\ref{eqn:symmetricppprimereplacement}). Furthermore, we define the following functions:
\begin{gather}
\label{eqn:XZdefinition}
\begin{aligned}
X_{ij} 
&= 
[I_i-\psi_i(ky)] \, [I_j - \psi_j(kx)],\\
Z_i 
&= 
\frac12 [\psi_i(kx) - \psi_i(ky)]^2
\end{aligned}
\end{gather}
and note that, for $j=j'=0$,
\begin{align}
\Ftilde^{\mu\nu}_x \Lambda_{i\nu}
&=
- \frac{m}{e} k^\mu \xi_i \psi_i(kx),
\nonumber\displaybreak[0]\\\nonumber
\uplambda^\mu 
&= 
- 2m\tau \sum_{i=1,2} \Lambda_i^\mu \xi_i I_i,
\nonumber\displaybreak[0]\\\nonumber
e\Lambda_{i\mu} \Ftilde^{\mu\nu}_x q_\nu
&= 
m\, kq\, \xi_i\, \psi_i(kx),
\nonumber\displaybreak[0]\\\nonumber
\Lambda_i\uplambda
&=
2m \tau \xi_i I_i,
\nonumber\displaybreak[0]\\\nonumber
e q\Ftilde_{x}\uplambda 
&=
2 kq \,\tau  m^2 \sum_{i=1,2} \xi^2_i \psi_i(kx) I_i,
\displaybreak[0]\nonumber\\
e^2 q\Ftilde^2_{x,y}q 
&= 
m^2 (kq)^2 \sum_{i=1,2} \xi_i^2 \psi_i(kx) \psi_i(ky).
\end{align}

Using these relations, we obtain the following expression for the field-dependent part of the tensor $\T^{\mu\nu}$:
\begin{multline}
\label{eqn:polarizationoperatorfinal}
\T^{\mu\nu}(q_1,q_2)
-
\T^{\mu\nu}_{\Ftilde=0}(q_1,q_2)
= 
-i\pi e^2\, \delta^{(\lminus,\lperp)}(q_1-q_2)
\\ \times\,
\int_{-1}^{+1} dv \int_0^\infty \frac{d\tau}{\tau} \, 
\int_{-\infty}^{+\infty} dz^\lminus \,
\big[
b_1 \Lambda_1^\mu \Lambda_2^\nu
+
b_2 \Lambda_2^\mu \Lambda_1^\nu
\\+
b_3 \Lambda_1^\mu \Lambda_1^\nu
+
b_4 \Lambda_2^\mu \Lambda_2^\nu
+
b_5 \Q_1^\mu \Q_2^\nu
\big]  e^{i\Phi} ,
\end{multline}
where the field-independent phase reads [see Eqs. (\ref{eqn:sfqed_polarizationoperatorfinalphasezerosources}) and (\ref{eqn:symmetricphaseterms})]
\begin{gather}
\label{eqn:polarizationoperator_fieldindependentphaseA}
e^{i\Phi}
=
\exp \lcb i \lsb
(q_2^\lplus -q_1^\lplus) z^\lminus 
+ 
\mu q_1q_2 
- 
\tau m^2 \rsb \rcb
\end{gather}
[$\mu = \frac14 \tau (1-v^2)$; see  Eq.~(\ref{eqn:tauvmudefinition})] and
\begin{align}
b_1 &= 2m^2 \xi_1 \xi_2 \Big( \frac{\tau}{4\mu} X_{12} - \frac{\tau v^2}{4\mu} X_{21}  \Big) e^{i\tau\beta},\nonumber\displaybreak[0]\\\nonumber
b_2 &= 2m^2 \xi_1 \xi_2 \Big( \frac{\tau}{4\mu} X_{21} - \frac{\tau v^2}{4\mu} X_{12}  \Big) e^{i\tau\beta},\nonumber\displaybreak[0]\\\nonumber
b_3 &= - \Big( \frac{i}{\tau} + \frac{q_1q_2}{2} \Big) \lb e^{i\tau\beta} -1 \rb  \\
&\phantom{= }+ 2m^2 \Big[ \frac{\tau}{4\mu} \lb \xi_1^2 Z_1 + \xi_2^2 Z_2 \rb + \xi_1^2 X_{11} \Big] e^{i\tau\beta},
\nonumber\displaybreak[0]\\\nonumber
b_4 &= - \Big( \frac{i}{\tau} + \frac{q_1q_2}{2} \Big) \lb e^{i\tau\beta} -1 \rb \\ 
&\phantom{= }+ 2m^2 \Big[ \frac{\tau}{4\mu} \lb \xi_1^2 Z_1 + \xi_2^2 Z_2 \rb + \xi_2^2 X_{22} \Big] e^{i\tau\beta},
\nonumber\displaybreak[0]\\
b_5 &= - \frac{2\mu}{\tau}  \lb e^{i\tau\beta} -1 \rb.
\end{align}
The field-dependent phase is given by [see Eq.~(\ref{eqn:sfqed_polarizationoperatorfinalphasezerosources})]
\begin{gather}
\label{eqn:polarizationoperator_fielddependentphaseA}
e^{i\tau\beta} = \exp \big[ i \tau m^2 \sum_{i=1,2} \xi_i^2 (I_i^2-J_i) \big],
\end{gather}
where [see Eq.~(\ref{eqn:IJsymmetricintegrals})]
\begin{gather}
\begin{aligned}
I_i 
&=
\frac12 \int_{-1}^{+1} d\lambda \, \psi_i(kz - \lambda \mu kq),\\
J_i 
&=
\frac12 \int_{-1}^{+1} d\lambda \, \psi^2_i(kz - \lambda \mu kq)
\end{aligned}
\end{gather}
and [see Eq.~(\ref{eqn:XZdefinition})]
\begin{gather}
\label{eqn:XYwithkz}
\begin{aligned}
X_{ij} 
&= 
[I_i-\psi_i(kz + \mu kq)] \, [I_j - \psi_j(kz - \mu kq)],\\
Z_i 
&= 
\frac12 [\psi_i(kz - \mu kq) - \psi_i(kz + \mu kq)]^2.
\end{aligned}
\end{gather}

We note that, using the metric tensor $g^{\mu\nu}$, we can construct the following projection tensor \cite{becker_vacuum_1975}:
\begin{gather}
\label{eqn:projectiontensor}
G^{\mu\nu} 
=
q_2^\mu q_1^\nu - q_1q_2 \, g^{\mu\nu}, 
\end{gather}
which obeys
\begin{gather}
q_{1\mu} G^{\mu\nu} = G^{\mu\nu} q_{2\nu} = 0
\end{gather}
and can be decomposed as
\begin{gather}
\label{eqn:projectiontensordecomposition}
G^{\mu\nu}
=
q_1 q_2 \lb \Lambda^\mu_1 \Lambda^\nu_1 + \Lambda^\mu_2 \Lambda^\nu_2 \rb +  \Q_1^\mu \Q^\nu_2.
\end{gather}
This shows that the decomposition given in Eq.~(\ref{eqn:polarizationoperatorfinal}) has the structure claimed in Ref. \cite{becker_vacuum_1975}.

\section{Discussion of the results}
\label{sec:discussion}

\subsection{Comparison with the literature}
\label{sec:litcomparison}

The expression we obtained for the field-dependent part of $\T^{\mu\nu}$ in Eq.~(\ref{eqn:polarizationoperatorfinal}) is manifestly symmetric in $q_1$ and $q_2$. We will now show how the alternative representation found in  Ref. \cite{baier_interaction_1975} can be derived from our calculation. To this end we do not apply the shift in Eqs. (\ref{eqn:symmetricintegralshift}) and (\ref{eqn:symmetricintegralshiftscalarterm}), which means that we have to use the replacement given in Eq.~(\ref{eqn:ppprimereplacement}) [rather than Eq.~(\ref{eqn:symmetricppprimereplacement})] for the $pp'$ term in Eq.~(\ref{eqn:sfqed_polarizationoperatortraceC}). This modifies the coefficients $b_3$ and $b_4$. Furthermore, we introduce the variable
\begin{gather}
\label{eqn:baierintegralshift}
z'^\lminus = x^\lminus + 2\mu q^\lminus = z^\lminus + \mu q^\lminus,
\end{gather}
which allows us to write [see Eq.~(\ref{eqn:symmetricphaseterms})]
\begin{multline}
(q_2^\lplus -q_1^\lplus) x^\lminus 
+ 
\mu q_2^2 
=
(q_2^\lplus -q_1^\lplus) z^\lminus
+ 
\mu q_1 q_2
\\=
(q_2^\lplus -q_1^\lplus) z'^\lminus 
+ 
\mu q_1^2
\end{multline}
and [see Eq.~(\ref{eqn:kpkpprimekyidentifications})]
\begin{gather}
kx = kz' - 2\mu kq, \quad ky = kz'
\end{gather}
(here and in the remaining subsection, we assume that all sources are set to zero). Thus, we obtain the following representation [see Eq.~(\ref{eqn:defIJintegrals})]:
\begin{gather}
\begin{aligned}
I_i 
&=
\int_{0}^{1} d\lambda\, \psi_i(kz' - 2kq \mu \lambda),\\
J_i 
&=
\int_{0}^{1} d\lambda\, \psi^2_i(kz' - 2kq \mu \lambda),
\end{aligned}
\\
I_i^2-J_i
=
\Big[ \int_{0}^{1} d\lambda\, \Delta_i(\mu \lambda) \Big]^2
-
\int_{0}^{1} d\lambda\, \Delta_i^2(\mu \lambda),
\end{gather}
where we introduced \cite{baier_interaction_1975}
\begin{gather}
\Delta_i(r)
=
\psi_i(kz'-2kq r) - \psi_i(kz').
\end{gather}
Furthermore, it is useful to define [compare with Eq.~(\ref{eqn:XZdefinition})]
\begin{gather}
\label{eqn:XYdefinition}
\begin{aligned}
X_{ij} 
&= 
[I_i-\psi_i(ky)] \, [I_j - \psi_j(kx)],\\
Y_i 
&= 
[I_i-\psi_i(ky)] \, [\psi_i(kx) - \psi_i(ky)]
\end{aligned}
\end{gather}
which can be written as
\begin{gather}
\begin{aligned}
X_{ij} 
&=
\lsb \int_0^1 d\lambda \, \Delta_i(\mu\lambda) \rsb \lsb \int_0^1 d\lambda \, \Delta_j(\mu\lambda) - \Delta_j(\mu) \rsb,\\
Y_i 
&=
\lsb \int_0^1 d\lambda\, \Delta_i(\mu\lambda) \rsb \Delta_i(\mu).
\end{aligned}
\end{gather}

Finally, we obtain the following alternative representation for the field-dependent part of $\T^{\mu\nu}$:
\begin{multline}
\label{eqn:polarizationoperatorbaier}
\T^{\mu\nu}(q_1,q_2)
-
\T^{\mu\nu}_{\Ftilde=0}(q_1,q_2)
= 
-i\pi e^2\, \delta^{(\lminus,\lperp)}(q_1-q_2)
\\ \times\,
\int_{-1}^{+1} dv \int_0^\infty \frac{d\tau}{\tau} \, 
\int_{-\infty}^{+\infty} dz'^\lminus \,
\big[
b_1 \Lambda_1^\mu \Lambda_2^\nu
+
b_2 \Lambda_2^\mu \Lambda_1^\nu
\\+
b'_3 \Lambda_1^\mu \Lambda_1^\nu
+
b'_4 \Lambda_2^\mu \Lambda_2^\nu
+
b_5 \Q_1^\mu \Q_2^\nu
\big]  e^{i\Phi}
\end{multline}
with the coefficients
\begin{gather}
\begin{aligned}
b_1 &= 2m^2 \xi_1 \xi_2 \Big( \frac{\tau}{4\mu} X_{12} - \frac{\tau v^2}{4\mu} X_{21}  \Big) e^{i\tau\beta},\\
b_2 &= 2m^2 \xi_1 \xi_2 \Big( \frac{\tau}{4\mu} X_{21} - \frac{\tau v^2}{4\mu} X_{12}  \Big) e^{i\tau\beta},\\
b'_3 &= - \Big( \frac{i}{\tau} + \frac{q_2^2}{2} \Big) \lb e^{i\tau\beta} -1 \rb  \\
&\phantom{= }+ 2m^2 \Big[ \frac{\tau}{4\mu} \lb \xi_1^2 Y_1 + \xi_2^2 Y_2 \rb + \xi_1^2 X_{11} \Big] e^{i\tau\beta},\\
b'_4 &= - \Big( \frac{i}{\tau} + \frac{q_2^2}{2} \Big) \lb e^{i\tau\beta} -1 \rb \\ 
&\phantom{= }+ 2m^2 \Big[ \frac{\tau}{4\mu} \lb \xi_1^2 Y_1 + \xi_2^2 Y_2 \rb + \xi_2^2 X_{22} \Big] e^{i\tau\beta},\\
b_5 &= - \frac{2\mu}{\tau}  \lb e^{i\tau\beta} -1 \rb
\end{aligned}
\end{gather}
and phases
\begin{gather}
\begin{aligned}
e^{i\Phi}
&=
\exp \lcb i \lsb
(q_2^\lplus -q_1^\lplus) z'^\lminus 
+ 
\mu q_1^2 
- 
\tau m^2 \rsb \rcb,
\\
e^{i\tau\beta} &= \exp \big[ i \tau m^2 \sum_{i=1,2} \xi_i^2 (I_i^2-J_i) \big].
\end{aligned}
\end{gather}
This representation coincides with Eq. (2.27) in Ref. \cite{baier_interaction_1975}.

\subsection{Constant-crossed field}

The polarization operator for a constant-crossed field was first obtained in Refs. \cite{narozhnyi_propagation_1968,batalin_preprint_1968} (see also Refs. \cite{batalin_greens_1971,ritus_radiative_1972,ritus_1985}). We show now how this result can be obtained from the expression in Eq.~(\ref{eqn:polarizationoperatorfinal}). 

A constant-crossed field is characterized by
\begin{gather}
\label{eqn:constantcrossedfield}
\psi_1(\phi) = \phi,
\quad 
\psi_2(\phi) = 0
\end{gather}
(the latter condition corresponds to $\xi_2=0$, and we will write $\xi = \xi_1$ in this paragraph). The field tensor and its square are then given by [see Eq.~(\ref{eqn:fieldtensor})]
\begin{gather}
F^{\mu\nu}
= 
f_1^{\mu\nu},
\quad
F^{2\mu\nu}
=
\frac{m^2 \xi^2}{e^2} k^{\mu} k^{\nu}.
\end{gather}
For a constant-crossed field, we obtain
\begin{gather}
\label{eqn:ccfield_IJXZ}
\begin{gathered}
I_1 = kz,
\quad
J_1 = (kz)^2 + \frac13 (\mu kq)^2,
\quad
I_2 = J_2 = 0,
\\
X_{11} = - (\mu kq)^2,
\quad
Z_{1} = 2 (\mu kq)^2,
\\
Z_{2} = X_{12} = X_{21} = X_{22} = 0.
\end{gathered}
\end{gather}
After inserting these expressions into Eq.~(\ref{eqn:polarizationoperatorfinal}), we can take the integral in $dz^\lminus$ and obtain the remaining delta function $2\pi \, \delta^{(\lplus)}(q_1-q_2)$, which implies that the polarization tensor for a constant-crossed field is diagonal in the external photon four-momenta. We define therefore the four-vectors [see Eq.~(\ref{eqn:Qidef})]
\begin{gather}
q^\mu = q_1^\mu = q_2^\mu,
\quad
\Q^\mu = \Q_1^\mu = \Q_2^\mu = \frac{k^\mu q^2 - q^\mu kq}{kq}.
\end{gather}
They obey
\begin{gather}
k\Q=-kq,
\quad
q\Q=0,
\quad 
\Q^2=-q^2.
\end{gather}
The four-vectors $q^\mu$, $\Q^\mu$, $\Lambda_1^\mu$, and $\Lambda_2^\mu$ form a complete set, and we obtain the following representation of the metric tensor:
\begin{gather}
g^{\mu\nu} = \frac{1}{q^2}\lb q^\mu q^\nu - \Q^\mu \Q^\nu \rb - \Lambda_1^\mu \Lambda_1^\nu - \Lambda_2^\mu \Lambda_2^\nu.
\end{gather}
From Eq.~(\ref{eqn:polarizationoperatorfinal})  we obtain now the following representation of the field-dependent part of $\T^{\mu\nu}$ in a constant-crossed field [see Eq.~(\ref{eqn:constantcrossedfield})]:
\begin{multline}
\label{eqn:polarizationoperatorccfieldA}
\T^{\mu\nu}(q_1,q_2)
-
\T^{\mu\nu}_{\Ftilde=0}(q_1,q_2)
= 
-2i\pi^2 e^2\, \delta^{4}(q_1-q_2)
\\ \times\,
\int_{-1}^{+1} dv \int_0^\infty \frac{d\tau}{\tau} \, 
\big[
b_{3} \Lambda_1^\mu \Lambda_1^\nu
+
b_{4} \Lambda_2^\mu \Lambda_2^\nu
+
b_{5} \Q^\mu \Q^\nu
\big]  e^{i\Phi_{}},
\end{multline}
where
\begin{gather}
\begin{aligned}
b_{3} &= - \Big( \frac{i}{\tau} + \frac{q^2}{2} \Big) \lb e^{i\tau\beta_{}} -1 \rb  \\
&\phantom{= }+ m^6 \chi^2 \tau^2 \frac{1}{4}(1-v^2) \Big[ 1 - \frac{1}{2}(1-v^2) \Big] e^{i\tau\beta_{}},\\
b_{4} &= - \Big( \frac{i}{\tau} + \frac{q^2}{2} \Big) \lb e^{i\tau\beta_{}} -1 \rb + m^6 \chi^2 \tau^2 \frac{1}{4}(1-v^2)  e^{i\tau\beta_{}},\\
b_{5} &= - \frac{1}{2}(1-v^2)  \lb e^{i\tau\beta_{}} -1 \rb
\end{aligned}
\end{gather}
and the phases are given by
\begin{gather}
\begin{aligned}
i\Phi_{} &=  -i\tau a,& a &=  m^2 \Big[ 1-\frac{1}{4}(1-v^2) \frac{q^2}{m^2} \Big],\\
i\tau\beta_{} &= - \frac{i}{3} \tau^3 b,& b &= m^6 \chi^2 \Big[ \frac14(1-v^2) \Big]^2
\end{aligned}
\end{gather}
(in the following, we will make the change of variables $\tau \to t$, where $\tau^3 b = t^3$ and $\rho = \nfrac{a}{\sqrt[3]{b}}$). Here we have introduced the quantum nonlinearity parameter
\begin{gather}
\label{eqn:chidefinition}
\chi 
= 
-\frac{e\sqrt{qF^2q}}{m^3}
= 
\xi \frac{\sqrt{(kq)^2}}{m^2}
\end{gather}
($\kappa$ in Refs. \cite{narozhnyi_propagation_1968,ritus_radiative_1972}).

We can rewrite now
\begin{gather}
\label{eqn:LambdaiLambdairewritten}
\begin{aligned}
\Lambda_1^\mu \Lambda_1^\nu
&=
-\frac{(Fq)^\mu (Fq)^\nu}{(Fq)^2},\\
\Lambda_2^\mu \Lambda_2^\nu
&= -\frac{(F^*q)^\mu (F^*q)^\nu}{(F^*q)^2},
\end{aligned}
\end{gather}
where
\begin{gather}
(F^*q)^2 = (Fq)^2 = -\frac{m^2 \xi^2}{e^2} (kq)^2
\end{gather}
and obtain [see Eq.~(\ref{eqn:projectiontensor})]
\begin{multline}
G^{\mu\nu} 
=
q^\mu q^\nu - q^2 \, g^{\mu\nu}
\\=
q^2 \lb \Lambda^\mu_1 \Lambda^\nu_1 + \Lambda^\mu_2 \Lambda^\nu_2 \rb + \Q^\mu \Q^\nu.
\end{multline}
We note the following relations:
\begin{gather}
\begin{aligned}
q_\rho G^{\rho\nu} &= G^{\mu\rho} q_\rho = 0,
\\
k_\rho G^{\rho\mu} &=  G^{\mu\rho} k_\rho = - kq \Q^\mu,
\\
G^{\mu\rho} F^{2}_{\rho\sigma}  G^{\sigma\nu}
&=
\frac{m^2}{e^2} \xi^2 (kq)^2  \Q^\mu \Q^\nu.
\end{aligned}
\end{gather}
To obtain the representation given in Refs. \cite{ritus_radiative_1972,ritus_1985}, we pass over to different basis tensors
\begin{multline}
b_3 \Lambda_1^\mu \Lambda_1^\nu
+
b_4 \Lambda_2^\mu \Lambda_2^\nu 
+
b_5 \Q^\mu \Q^\nu 
=
(q^2 b_5-b_3) \frac{(Fq)^\mu (Fq)^\nu}{(Fq)^2} 
\\+ (q^2 b_5-b_4)
\frac{(F^*q)^\mu (F^*q)^\nu}{(F^*q)^2}
+ b_5 G^{\mu\nu}
\end{multline}
and define the following functions \cite{ritus_radiative_1972,ritus_1985}:
\begin{align}
f(x)
&= 
i\int_0^\infty dt \exp\Big[-i\Big(t x + \frac{1}{3}t^3\Big)\Big] 
\nonumber\\&\hspace*{3cm}=
\pi \Gi(x) + i\pi \Ai(x),\\
f'(x)
&= 
\int_0^\infty t dt \exp\Big[-i\Big(t x + \frac{1}{3}t^3\Big)\Big],\\
f_1(x)
&=
\int_0^\infty \frac{dt}{t} \exp\lb-it x \rb \Big[ \exp\Big(-\frac{i}{3} t^3 \Big) -1 \Big]
\nonumber\\&\hspace*{3cm}=
\int_x^\infty dt \lsb f(t) - \frac{1}{t} \rsb\\
\intertext{and}
f_2(x)
&=
\int_0^\infty \frac{dt}{t^2} \exp\lb-it x \rb \Big[ \exp\Big(-\frac{i}{3} t^3 \Big) -1\Big]
\nonumber\\&\hspace*{3cm}=
-i\lsb x f_1(x) + f'(x) \rsb,
\end{align}
where $\Ai$ and $\Gi$ denote the Airy and Scorer functions, respectively \cite{olver_nist_2010}.  These functions obey the following differential equations:
\begin{gather}
\begin{aligned}
f''(x) &= x f(x) - 1,
\\
f_1'(x) &= \frac{1}{x} - f(x) = -\frac{1}{x} f''(x).
\end{aligned}
\end{gather}
Using the latter, we can replace the function $f_1(x)$ by $f'(x)$ in the following way (if all boundary terms vanish):
\begin{gather}
\int_{-1}^{+1} dv \, g(v) f_1[\rho(v)] 
=
-\int_{-1}^{+1} dv \, \lsb \frac{G(v)}{\rho(v)} \rsb' f'[\rho(v)],
\end{gather}
where $G'(v)=g(v)$.

Using the above notation, we can represent the field-dependent part of the tensor $\T^{\mu\nu}$ for a constant-crossed field given in Eq. (\ref{eqn:polarizationoperatorccfieldA}) by
\begin{multline}
\label{eqn:polarizationoperatorccfieldB}
\T^{\mu\nu}(q_1,q_2)
-
\T^{\mu\nu}_{\Ftilde=0}(q_1,q_2)
= 
i(2\pi)^4 \delta^{4}(q_1-q_2)
\\ \times\,
\Big[
\pi_1 \frac{(Fq)^\mu (Fq)^\nu}{(Fq)^2} 
+
\pi_2 \frac{(F^*q)^\mu (F^*q)^\nu}{(F^*q)^2}
- 
\frac{\pi_3}{q^2}   G^{\mu\nu}
\Big],
\end{multline}
where
\begin{gather}
\label{eqn:ccfield_coefficients}
\begin{aligned}
\pi_1
&=
\phantom{-}\alpha\, \frac{m^2}{3\pi} \int_{-1}^{+1} dv \, (w-1) \Big(\frac{\chi}{w}\Big)^{\nicefrac23} f'(\rho),\\
\pi_2
&=
\phantom{-}\alpha\, \frac{m^2}{3\pi} \int_{-1}^{+1} dv \, (w+2) \Big(\frac{\chi}{w}\Big)^{\nicefrac23} f'(\rho),\\
\pi_3 
&= - \alpha\, \frac{q^2}{\pi} \int_{-1}^{+1} dv \, \frac{f_1(\rho)}{w}
\end{aligned}
\end{gather}
[$\frac{1}{w} = \frac14(1-v^2)$, $\rho = \big(\nfrac{w}{\chi}\big)^{\nicefrac23}(1- \frac{q^2}{m^2} \frac{1}{w})$]. Since all nonvanishing functions are even in $v$, we can now apply the following change of variables:
\begin{gather}
\int_{-1}^{+1} dv = 2 \int_{0}^{1} dv = \int_{4}^{\infty} dw \, \frac{4}{w\sqrt{w(w-4)}},
\end{gather}
which shows that the result in Eq.~(\ref{eqn:polarizationoperatorccfieldB}) is equivalent to the one given in Refs. \cite{ritus_radiative_1972,ritus_1985}.

\subsection{Quasiclassical limit}

We consider now a linearly polarized plane-wave field
\begin{gather}
\label{eqn:linearpolarizationdef}
\psi_1(\phi) = \psi(\phi), 
\quad
\psi_2(\phi) = 0 
\end{gather}
(we will use $\xi = \xi_1$ and $f^{\mu\nu} = f_1^{\mu\nu}$ in this paragraph) in the quasiclassical limit defined by $\xi\to\infty$ while [see Eq.~(\ref{eqn:chidefinition})]
\begin{gather}
\chi 
= 
-\frac{e\sqrt{qf^2q}}{m^3}
= 
\xi \frac{\sqrt{(kq)^2}}{m^2}
\end{gather}
is kept constant. In the optical regime (photon energy $\omega_0 \sim \unit[1]{eV}$), $\chi \gtrsim 1$ requires $\xi \gg 1$ (unless the incoming photon energy exceeds the threshold of about 1 TeV), which means that the quasiclassical limit is sufficient to analyze most of the upcoming strong-field experiments with optical lasers. 

By employing the identity $\abs{kq} = m^2 \nfrac{\chi}{\xi}$, we can expand all functions depending on $\mu kq$
\begin{gather}
\label{eqn:qclimit_IJXZ}
\begin{aligned}
I^2_1 - J_1 &= -(\nfrac13) (\mu kq)^2 \big[\psi'(kz)\big]^2 + \mc{O}(\mu kq)^3,\\
Z_1 &= 2 (\mu kq)^2 \big[\psi'(kz)\big]^2 + \mc{O}(\mu kq)^3,\\
X_{11} &= - (\mu kq)^2 \big[\psi'(kz)\big]^2 + \mc{O}(\mu kq)^3
\end{aligned}
\end{gather}
($X_{12}=X_{21}=X_{22}=Z_2=I_2=J_2=0$ for linear polarization). Thus, if multiplied by $\xi^2$, only the leading-order terms are independent of $\xi$, and all others are suppressed. In the limit $\xi\to \infty$, the expressions in Eq.~(\ref{eqn:qclimit_IJXZ}) correspond to those in Eq.~(\ref{eqn:ccfield_IJXZ}) with the replacement $\chi \to \chi(kz) = \chi \psi'(kz)$. The remaining calculation is therefore similar to the one in the constant-crossed field case, and the final result in Eq.~(\ref{eqn:polarizationoperatorlinearpolqc}) corresponds essentially to Eq.~(\ref{eqn:polarizationoperatorccfieldB}) with the above replacement. Using [see Eq.~(\ref{eqn:LambdaiLambdairewritten})]
\begin{gather}
\begin{aligned}
\Lambda_1^\mu \Lambda_1^\nu
&=
- \frac{(fq)^\mu (fq)^\nu}{(fq)^2},
\\
\Lambda_2^\mu \Lambda_2^\nu
&=
- \frac{(f^*q)^\mu (f^*q)^\nu}{(f^*q)^2}
\end{aligned}
\end{gather}
and Eq.~(\ref{eqn:projectiontensordecomposition}), we obtain for a linearly polarized plane-wave field in the quasiclassical approximation the following representation for the field-dependent part of the tensor $\T^{\mu\nu}$ [see Eq.~(\ref{eqn:polarizationoperatorfinal})]:
\begin{multline}
\label{eqn:polarizationoperatorlinearpolqc}
\T^{\mu\nu}(q_1,q_2)
-
\T^{\mu\nu}_{\Ftilde=0}(q_1,q_2)
= 
i(2\pi)^4\delta^{(\lminus,\lperp)}(q_1-q_2) 
\\ \times\,
\frac{1}{2\pi} \int_{-\infty}^{+\infty} dz^\lminus \, e^{i(q_2^\lplus - q_1^\lplus)z^\lminus} 
\bigg[\pi'_1\, \frac{(fq)^\mu (fq)^\nu}{(fq)^2}
\\+ \pi'_2\, \frac{(f^*q)^\mu (f^*q)^\nu}{(f^*q)^2} - \frac{\pi'_3}{q_1 q_2}\, G^{\mu\nu}  \bigg],
\end{multline}
where [see Eq.~(\ref{eqn:ccfield_coefficients})]
\begin{gather}
\begin{aligned}
\pi'_1
&=
\phantom{-}\alpha\, \frac{m^2}{3\pi} \int_{-1}^{+1} dv \, (w-1) \bigg[\frac{\abs{\chi(kz)}}{w}\bigg]^{\nicefrac23} f'(\rho),\\
\pi'_2
&=
\phantom{-}\alpha\, \frac{m^2}{3\pi} \int_{-1}^{+1} dv \, (w+2) \bigg[\frac{\abs{\chi(kz)}}{w}\bigg]^{\nicefrac23} f'(\rho),\\
\pi'_3 
&= - \alpha\, \frac{q_1 q_2}{\pi} \int_{-1}^{+1} dv \, \frac{f_1(\rho)}{w}
\end{aligned}
\end{gather}
with $\tfrac{1}{w} = \tfrac14(1-v^2)$, $\rho = \big[\nfrac{w}{\abs{\chi(kz)}}\big]^{\nicefrac23}(1- \tfrac{q_1 q_2}{m^2} \tfrac{1}{w})$ and $G^{\mu\nu}  = q_2^\mu q_1^\nu - q_1q_2 \, g^{\mu\nu}$ [see Eq.~(\ref{eqn:projectiontensor})].

\subsection{Circular polarization}

The general result in Eq.~(\ref{eqn:polarizationoperatorfinal}) also simplifies considerably if the plane wave is circularly polarized and monochromatic, 
\begin{gather}
\psi_1(\phi) = \Re{e^{i\phi}},
\quad
\psi_2(\phi) = \Im{e^{i\phi}},
\quad
\xi_1=\xi_2=\xi.
\end{gather}
We then obtain
\begin{gather}
\begin{gathered}
I_{1}
=
\sinc(\mu kq) \Re e^{ikz},
\quad
I_{2}
=
\sinc(\mu kq) \Im e^{ikz},\\
J_{1} + J_{2}
= 1,
\quad
Z_1 + Z_2 = 2\sin^2(\mu kq), 
\end{gathered}
\end{gather}
\begin{gather}
\begin{aligned}
I_{1} - \psi_{1}(kz+\mu kq)
&=
\Re A,\\
I_{2} - \psi_{2}(kz+\mu kq)
&=
\Im A,\\
I_{1} - \psi_{1}(kz-\mu kq)
&=
\Re B,\\
I_{2} - \psi_{2}(kz-\mu kq)
&=
\Im B,
\end{aligned}
\end{gather}
where
\begin{gather}
\begin{aligned}
A &= e^{ikz} \lsb \sinc(\mu kq) - \cos(\mu kq) - i \sin(\mu kq) \rsb,\\
B &= e^{ikz} \lsb \sinc(\mu kq) - \cos(\mu kq) + i \sin(\mu kq) \rsb
\end{aligned}
\end{gather}
[we define $\sinc x = \nfrac{(\sin x)}{x}$]. Thus,
\begin{gather}
\begin{aligned}
X_{12} - X_{21} &= \Im A^*B,&
X_{11} - X_{22} &= \Re AB,\\
X_{12} + X_{21} &= \Im AB,&
X_{11} + X_{22} &= \Re A^*B,
\end{aligned}
\end{gather}
where
\begin{gather}
\begin{aligned}
A^*B &= \sinc^2(\mu kq) + \cos(2\mu kq) - 2\sinc(2\mu kq) \\
&\phantom{=}+ i\lsb -\sin(2\mu kq) + 2 \sinc(\mu kq) \sin(\mu kq) \rsb,\\
AB &= e^{2ikz} \lsb \sinc^2(\mu kq) - 2\sinc(2\mu kq) + 1 \rsb.
\end{aligned}
\end{gather}

Thus, we can write the field-dependent part of the tensor $\T^{\mu\nu}$ for a circularly polarized plane wave as [see Eq.~(\ref{eqn:polarizationoperatorfinal})]
\begin{multline}
\label{eqn:polarizationoperatorcircularpolarization}
\T^{\mu\nu}(q_1,q_2)
-
\T^{\mu\nu}_{\Ftilde=0}(q_1,q_2)
= 
-i\pi e^2\, \delta^{(\lminus,\lperp)}(q_1-q_2)
\\ 
\begin{aligned}
&\times\,
\int_{-1}^{+1} dv \int_0^\infty \frac{d\tau}{\tau} \, 
\int_{-\infty}^{+\infty} dz^\lminus \,
\Big[
b_+ \Lambda_+^\mu \Lambda_+^\nu 
\\&+
b_- \Lambda_-^\mu \Lambda_-^\nu
+
\frac12(b_1-b_2) (\Lambda^\mu_1 \Lambda^\nu_2 - \Lambda^\mu_2 \Lambda^\nu_1)
\\&+
\frac12(b_3+b_4) (\Lambda^\mu_1 \Lambda^\nu_1 + \Lambda^\mu_2 \Lambda^\nu_2)
+
b_5 \Q_1^\mu \Q_2^\nu \Big]  e^{i\Phi},
\end{aligned}
\end{multline}
where we defined
\begin{gather}
\Lambda^\mu_{\pm} = \Lambda_1^\mu \pm i \Lambda_2^\mu
\end{gather}
and the coefficients are given by
\begin{multline}
b_\pm = \frac14\lsb(b_3-b_4)\mp i(b_1+b_2)\rsb  
= \frac12 m^2\xi^2 
\\\times \, 
\lsb \sinc^2(\mu kq) - 2\sinc(2\mu kq) + 1 \rsb \, e^{\mp 2ikz +i\tau\beta},
\end{multline}
\begin{multline}
\frac12(b_1-b_2) 
= m^2\xi^2 \frac{(1+v^2)}{(1-v^2)} \big[ - \sin(2\mu kq)\\
+ 2 \sinc(\mu kq) \sin(\mu kq) \big]\, e^{i\tau\beta}, 
\end{multline}
\begin{multline}
\frac12(b_3+b_4) 
= -\Big( \frac{i}{\tau} + \frac{q_1q_2}{2}\Big) \lb e^{i\tau\beta} - 1\rb \\
\begin{aligned}
&+ m^2 \xi^2 \Big[ 2 \, \frac{(1+v^2)}{(1-v^2)} \,  \sin^2(\mu kq) 
\\&+ \sinc^2(\mu kq)  - 2\sinc(2\mu kq) + 1 \Big] e^{i\tau\beta},
\end{aligned}
\end{multline}
\begin{gather}
b_5 = - \frac{2\mu}{\tau}  \lb e^{i\tau\beta} -1 \rb
\end{gather}
and the phases read
\begin{gather}
\begin{aligned}
i\tau\beta 
&= 
i \tau m^2 \xi^2 \lsb \sinc^2(\mu kq) -1 \rsb,\\
i \Phi
&=
i \lsb (q_2^\lplus -q_1^\lplus) z^\lminus  +  \mu q_1q_2 -  \tau m^2 \rsb
\end{aligned}
\end{gather}
[$\mu = \frac14 \tau (1-v^2)$]. Finally, the integral in $dz^\lminus$ can be taken and  we obtain the following expression for the field-dependent part of $\T^{\mu\nu}(q_1,q_2)$ for a monochromatic, circularly polarized plane-wave laser field
\begin{multline}
\label{eqn:polarizationoperatorcircularpolarizationfinal}
\T^{\mu\nu}(q_1,q_2)
-
\T^{\mu\nu}_{\Ftilde=0}(q_1,q_2)
= 
- \frac{i(2\pi)^4 \,e^2}{8\pi^2}  \,
\int_{-1}^{+1} dv 
\\ \int_0^{\infty} \frac{d\tau}{\tau} \,\, \big[ T^{\mu\nu}_0 \delta^4(q_1-q_2) 
+ T^{\mu\nu}_+ \delta^4(q_1-q_2+2k) \\+ T^{\mu\nu}_- \delta^4(q_1-q_2-2k) \big] 
e^{i\Phi_{\text{cp}}},
\end{multline}
where
\begin{gather}
i\Phi_{\text{cp}}
=
- i\tau m^2 \big\{1 + \xi^2 [1 - \sinc^2(\mu kq)] \big\} + i \mu q_1q_2,
\end{gather}
\begin{multline}
T^{\mu\nu}_0 
=
\tau_1 (\Lambda^\mu_1 \Lambda^\nu_2 - \Lambda^\mu_2 \Lambda^\nu_1)
+
\tau_2 (\Lambda^\mu_1 \Lambda^\nu_1 + \Lambda^\mu_2 \Lambda^\nu_2)
\\+
\tau_3 \Q_1^\mu \Q_2^\nu, 
\end{multline}
\begin{gather}
T^{\mu\nu}_\pm = \frac12 m^2 \xi^2 \big[ \sinc^2(\mu kq) - 2\sinc(2\mu kq) + 1 \big]\, \Lambda^\mu_\pm \Lambda^\nu_\pm
\end{gather}
and
\begin{gather}
\begin{aligned}
\tau_1 
&= 
m^2\xi^2 \frac{(1+v^2)}{(1-v^2)} \big[2\, \nfrac{\sin^2(\mu kq)}{(\mu kq)} -\sin(2\mu kq)\big],
\\
\tau_2
&=
2 m^2 \xi^2 \, \frac{(1+v^2)}{(1-v^2)} \,  \sin^2(\mu kq)
\\&\phantom{=}+ 
\Big[ \Big(\frac{\mu}{\tau} - \frac{1}{2}\Big) q_1q_2  - m^2 \Big] \lb 1 - e^{-i\tau\beta}\rb,
\\
\tau_3 
&= 
- \frac{2\mu}{\tau}  \lb 1 - e^{-i\tau\beta} \rb.
\end{aligned}
\end{gather}
This result agrees with Eq.~(2.34) in Ref. \cite{baier_interaction_1975}. The terms described by $T^{\mu\nu}_\pm$ can be interpreted as describing processes where two photons from the background field are absorbed or emitted, respectively (since the external field is not quantized, this interpretation relies only on the momentum-conserving delta function).

In order to obtain Eq.~(\ref{eqn:polarizationoperatorcircularpolarizationfinal}) from Eq.~(\ref{eqn:polarizationoperatorcircularpolarization}), we have used the identity
\begin{multline}
\int_0^\infty \frac{d\tau}{\tau} e^{i\Phi} \, m^2\xi^2  \big[ \sinc^2(\mu kq)  - 2\sinc(2\mu kq) + 1 \big] e^{i\tau\beta}
\\=
\int_0^\infty \frac{d\tau}{\tau} e^{i\Phi}  \, \Big[\frac{i}{\tau} + \frac{\mu}{\tau}\, q_1q_2 - m^2 \Big]  \big( e^{i\tau\beta} - 1 \big),
\end{multline}
which follows from
\begin{multline}
i\frac{d}{d\tau} \big( e^{i\tau\beta} - 1 \big)
=
i\frac{d}{d\tau} e^{i\tau\beta}
\\=
m^2\xi^2 \big[ \sinc^2(\mu kq)  - 2\sinc(2\mu kq) + 1 \big] e^{i\tau\beta}
\end{multline}
via integration by parts.

\section{Conclusion}

In the present paper, we have proven for the first time the Ward-Takahashi identity for general loop diagrams in a plane-wave background field  (see Sec. \ref{sec:wardidentity}). Moreover, we have presented a new derivation of the leading-order contribution to the polarization operator in a plane-wave background field for arbitrary polarization and dependence on the plane-wave phase (see Sec. \ref{sec:polarizationoperator}). Our calculation relies on a direct evaluation of the space-time integrals without using Schwinger's operator method \cite{schwinger_gauge_1951} that was employed in Ref. \cite{baier_interaction_1975}. An interesting feature of our final representation is the manifest symmetry with respect to the external photon four-momenta~$q_1$ and $q_2$ [see Eq.~(\ref{eqn:polarizationoperatorfinal})]. 

\begin{acknowledgments}
S.M. is grateful to the Studienstiftung des deutschen Volkes for financial support.
\end{acknowledgments}

\appendix

\section{Notation}
\label{sec:notation}

In this paper we use natural units~$\hbar=c=1$ (in some formulas~$\hbar$ and~$c$ are restored for clarity) and the charge is measured in Heaviside-Lorentz units ($\eps_0=1$). The electron mass and charge are denoted by~$m$ and~$e<0$, respectively. Thus, the fine-structure constant is given by $\alpha = \nfrac{e^2}{(4\pi)} \approx \nfrac{1}{137}$. In covariant expressions the space-time metric~$g_{\mu\nu}$ with signature~$(1,-1,-1,-1)$ is used, and $\del_\mu =(\del/\del t,\spvecgreek{\nabla})$ is the four-derivative. This implies $\del_\mu x_\nu = g_{\mu\nu}$, where~$x^\mu =(t,\spvec{x})$ denotes the position four-vector. The unit tensor  is denoted by~$\delta^{\mu}_{\nu} = g^{\mu\rho} g_{\rho\nu} = \diag(1,1,1,1)$ ($\delta_\mu^\mu = 4$), and space-time indices (lowercase Greek letters) are raised and lowered using the metric $a_\mu = g_{\mu\nu} a^\nu$ (summation over all types of repeated indices is understood if they do not appear on both sides of an equation). Greek and Latin indices take the values~(0,1,2,3) and~(1,2,3), respectively.  Contractions of four-vectors are denoted by $a^\mu b_\mu = ab$ and scalar products of three-vectors by~$\spvec{a}^i \spvec{b}^i = \spvec{a}\spvec{b}$. We denote the dual of a second-rank tensor $T^{\mu\nu}$ by $T^{*\mu\nu} = \frac{1}{2} \eps^{\mu\nu\rho\sigma} T_{\rho\sigma}$, where $\eps^{\mu\nu\rho\sigma}$ is the totally antisymmetric tensor in four dimensions with~$\eps^{0123}=-\eps_{0123}=1$. For contractions of second-rank tensors and vectors, a matrix notation is sometimes used, e.g.~$aTb= a_\mu T^{\mu\nu} b_\nu$, $(T_1T_2)^{\mu\nu} = T^{\mu}_{1\rho} T_{2}^{\rho\nu}$, $T^{2\mu\nu} = T^{\mu\rho} T_{\rho}^{\phantom{\rho}\nu}$, $(Ta)^\mu = T^{\mu\nu} a_\nu$. All spinors are Dirac spinors (with four components); spinor indices are usual suppressed. The Dirac gamma matrices are denoted by~$\gamma^\mu$, $\s{a}=a_\mu \gamma^\mu$, $\gamma^5 = -i\gamma^0\gamma^1\gamma^2\gamma^3$ and $2\sigma^{\mu\nu} = \gamma^\mu \gamma^\nu - \gamma^\nu \gamma^\mu$ ($\gamma^\mu \gamma^\nu + \gamma^\nu \gamma^\mu = 2g^{\mu\nu}$). For a spinor $u$, we define $\bar{u}=u^\dagger \gamma^0$, and for a matrix in spinor space  $M$, correspondingly $\bar{M}=\gamma^0 M^\dagger \gamma^0$. A quantization volume $V=1$ is assumed for the normalization of the single-particle electron, positron and photon states. The total derivative of a function with respect to its argument is denoted by a prime $f'(x) = \frac{d}{dx} f(x)$. Integrals without boundaries range from~$-\infty$ to~$+\infty$. We use $i0$ as a short notation for $i\upepsilon$ together with the limit~$\lim_{\upepsilon \to 0^+}$. In general, our notation therefore follows Ref. \cite{landau_quantum_1981} with different units for charge.

\section{Laser field as a coherent state of the electromagnetic field}
\label{sec:lasercoherentstate}

In this appendix we give a detailed justification why a strong laser can be taken into account by applying the shift $\mc{A}^\mu \to A^\mu_{\mathrm{rad}}(x) + A^\mu$ in the Lagrangian density, with $A^\mu$ being treated as a classical field and with  $A^\mu_{\mathrm{rad}}$ describing all other quantized modes  \cite{fradkin_quantum_1991,glauber_coherent_1963,harvey_signatures_2009}. A strong laser field represents a very good experimental realization of a coherent state of the photon field. To be explicit, we consider the four-vector potential operator
\begin{subequations}
\label{eqn:photonfieldoperator}
\begin{gather}
\hat{\mc{A}}^\mu(x)
=
\hat{\mc{A}}_{+}^\mu(x)
+
\hat{\mc{A}}_{+}^{\dagger\mu}(x),
\end{gather}
\begin{gather}
\hat{\mc{A}}_{+}^\mu(x)
=
\sum_{\sigma=1,2} \int \frac{d^3q}{(2\pi)^3}
\frac{1}{\sqrt{2\omega_{\spvec{q}}}} 
\,
\hat{c}_{\spvec{q},\sigma} 
e^{-\I qx} \eps^\mu_{\spvec{q},\sigma},
\end{gather}
\end{subequations}
where $\omega_{\spvec{q}} = \sqrt{\spvec{q}^2}$ and $\eps^\mu_{\spvec{q},\sigma}$ are the orthogonal polarization four-vectors
\begin{gather}
(\eps^{*}_{\spvec{q},\sigma})^\mu (\eps_{\spvec{q},\tau})_{\mu}
=
- \delta_{\sigma\tau}
\end{gather}
[for simplicity we consider here only the two physical degrees of freedom ($\sigma, \tau = 1,2$), see e.g. \cite{weinberg_quantum_1995} for further details]. The photon creation $\hat{c}^\dagger_{\spvec{q},\sigma}$ and annihilation operators $\hat{c}_{\spvec{q},\sigma}$ obey the canonical commutation relations
\begin{gather}
\big[\hat{c}_{\spvec{p},\sigma},\hat{c}^\dagger_{\spvec{q},\tau}\big]
=
(2\pi)^3 \delta^3(\spvec{p}-\spvec{q}) \delta_{\sigma \tau}.
\end{gather}
Using this notation, a coherent state $\ket{A}$ of the photon field can be written as \cite{glauber_coherent_1963}
\begin{gather}
\ket{A} = \hat{D} \, \ket{0},
\end{gather}
where $\ket{0}$ is the vacuum state of the photon Fock space ($\hat{c}_{\spvec{q},\sigma} \ket{0} = 0$ for all $\spvec{q}$ and $\sigma$, and $\braket{0|0} = 1$) and $\hat{D}$ is a unitary displacement operator. If the classical four-potential associated with the coherent state [compare with Eq.~(\ref{eqn:photonfieldoperator})] is
\begin{subequations}
\label{eqn:classicallaserfield}
\begin{gather}
A^\mu(x)=A_{+}^\mu(x)+A_{+}^{*\mu}(x),
\end{gather}
with
\begin{gather}
A_{+}^\mu(x)
=
\sum_{\sigma=1,2} \int \frac{d^3q}{(2\pi)^3}
\frac{1}{\sqrt{2\omega_{\spvec{q}}}} 
\,
C_{\spvec{q},\sigma}
e^{-\I qx} \eps^\mu_{\spvec{q},\sigma},
\end{gather}
\end{subequations}
the displacement operator $\hat{D}$ has the form
\begin{gather}
\label{eqn:displacementoperator}
\hat{D} = \exp \Big[ \sum_{\sigma=1,2} \int \frac{d^3q}{(2\pi)^3} \lb C_{\spvec{q},\sigma} \hat{c}^\dagger_{\spvec{q},\sigma} - C^*_{\spvec{q},\sigma} \hat{c}_{\spvec{q},\sigma} \rb \Big]
\end{gather}
and the properties
\begin{gather}
\label{eqn:displacementoperatorproperties}
\begin{aligned}
\hat{D}^{-1} \hat{c}_{\spvec{q},\sigma} \hat{D} = \hat{c}_{\spvec{q},\sigma} + C_{\spvec{q},\sigma},\\
\hat{D}^{-1} \hat{c}^\dagger_{\spvec{q},\sigma} \hat{D} = \hat{c}^\dagger_{\spvec{q},\sigma} + C^*_{\spvec{q},\sigma}.
\end{aligned}
\end{gather}
By using Eq.~(\ref{eqn:displacementoperatorproperties}), one can show that
\begin{gather}
\hat{\mc{A}}_{+}^\mu(x) \, \ket{A}
=
A_{+}^\mu(x) \, \ket{A}.
\end{gather}
Thus, since
\begin{gather}
\braket{A|\hat{\mc{A}}^\mu(x)|A} = A^\mu(x),
\end{gather}
a coherent state can be seen as the most ``classical'' state of the photon field.

If the coherent part of the photon field is not substantially changed during the interaction, the same coherent state appears on both sides of the $S$-matrix element(s) of an arbitrary QED process,
\begin{gather}
\braket{A^\mu| \cdots | A^\mu} = \braket{0|\hat{D}^{-1} \cdots \hat{D}|0}.
\end{gather}
Physically, this amounts to the assumption that the laser field is not significantly depleted during the interaction, which can be assumed if the laser is sufficiently intense (see also Sec. \ref{sec:qedwithbackgroundfields}). We can then include the coherent part of the photon field nonperturbatively if we adopt the transformation in Eq.~(\ref{eqn:displacementoperatorproperties}). In particular, we obtain [see Eqs.~(\ref{eqn:photonfieldoperator}) and (\ref{eqn:classicallaserfield})]
\begin{gather}
\hat{D}^{-1} \hat{\mc{A}}^\mu(x) \hat{D} 
=
\hat{\mc{A}}^\mu(x) + A^\mu(x).
\end{gather}
Thus, instead of calculating $S$-matrix elements between coherent states, we can apply the shift $\hat{\mc{A}}^\mu(x) \to \hat{A}_{\text{rad}}^\mu(x) + A^\mu(x)$ in the Lagrangian density and consider $S$-matrix elements between vacuum states as usual \cite{fradkin_quantum_1991,harvey_signatures_2009}.

\section{Light-cone coordinates}
\label{sec:lccappendix}

Calculations involving plane-wave background fields become particular transparent if light-cone coordinates are used \cite{dirac_forms_1949,neville_quantum_1971,mitter_quantum_1975}. Since the nontrivial space-time dependence of the momentum-space vertex in Eq.~(\ref{eqn:dressedvertexfinal}) is due to the plane-wave phase $\phi = kx$, it is natural to work in a basis where $k^\mu$ is one of the basis four-vectors. However, since $k^2=0$, this will be a light-cone basis. We introduce now a general light-cone basis by adding three four-vectors $\bar{k}^\mu$, $e_i^\mu$ ($i\in 1,2$) to the set and require the following orthogonality relations:
\begin{subequations}
\label{eqn:lcc_orthogonalityrelations}
\begin{gather}
k^2 = \bar{k}^2 = ke_i = \bar{k}e_i = 0,
\,
k\bar{k} = 1, 
\,
e_i e_j = - \delta_{ij}
\end{gather}
and the orientation
\begin{gather}
\eps_{\mu\nu\rho\sigma} k^\mu \bar{k}^\nu e_1^\rho e_2^\sigma = 1.
\end{gather}
\end{subequations}
To be more specific, we can, in a reference system where the plane wave propagates along the direction~$\spvec{n}$, take the following four-vectors:
\begin{gather}
\label{eqn:canonicallcc}
\begin{gathered}
k^\mu = \omega (1,\spvec{n}),
\quad
\bar{k}^\mu = \frac{1}{2\omega} (1,-\spvec{n}),
\quad
e_i^\mu=(0,\spvec{e}_i) \sim a_i^\mu ,\\
\spvec{n}^2 = 1,
\quad
\spvec{e}_i \spvec{e}_j = \delta_{ij},
\quad
\spvec{n} = \spvec{e}_1 \times \spvec{e}_2
\end{gathered}
\end{gather}
($\spvec{e}_i$ represent the two polarization directions of the plane-wave field, and $\omega$ has the dimension of a frequency).

Due to the relations given in Eq.~(\ref{eqn:lcc_orthogonalityrelations}), we obtain the following decomposition of the metric:
\begin{gather}
g_{\mu\nu} = k_\mu \bar{k}_\nu + \bar{k}_\mu k_\nu - e_{1\mu} e_{1\nu} - e_{2\mu} e_{2\nu}.
\end{gather}
This allows us to define the transformation to light-cone coordinates (primed indices) by
\begin{equation}
a^{\mu'} = \Lambda^{\mu'}_{\phantom{\mu'}\nu} a^\nu\,,
\,
b_{\mu'} = b_\nu \Lambdainv^{\nu}_{\phantom{\nu}\mu'}\,,
\,
\Lambdainv^{\rho}_{\phantom{\rho}\mu'} \Lambda^{\mu'}_{\phantom{\mu'}\sigma} 
=
\delta^{\rho}_{\sigma},
\end{equation}
where the components denote the following scalar products:
\begin{gather}
\begin{aligned}
\Lambda^{\lplus}_{\phantom{\lplus}\mu} &= \bar{k}_\mu,&
\Lambda^{\lone}_{\phantom{\lone}\mu} &= e_{1\mu},\\
\Lambda^{\lminus}_{\phantom{\lminus}\mu} &= k_\mu,&
\Lambda^{\ltwo}_{\phantom{\ltwo}\mu} &= e_{2\mu}
\end{aligned}
\end{gather}
(we label light-cone components by $\lplus$,$\lminus$,$\lone$,$\ltwo$). On the other hand, the inverse transformation is given by
\begin{gather}
\begin{aligned}
\label{eqn:transformationmatrixlcc}
\Lambdainv^{\mu}_{\phantom{\mu}\lplus}  &= k^\mu,&
\Lambdainv^{\mu}_{\phantom{\mu}\lone}  &= -e^\mu_{1},\\
\Lambdainv^{\mu}_{\phantom{\mu}\lminus}  &= \bar{k}^\mu,&
\Lambdainv^{\mu}_{\phantom{\mu}\ltwo}  &= -e^\mu_{2}.
\end{aligned}
\end{gather}
We point out that $k^\mu$ has dimension of momentum and therefore $\bar{k}^\mu$ must have dimension of inverse momentum ($e^\mu_{i}$ are dimensionless). Hence, the dimensions of $v^\lplus$ and $v^\lminus$ differ from those of $v^\mu$ (here $v^\mu$ is an arbitrary Lorentz four-vector). The different dimensions of the light-cone components can be circumvented by defining $k^\mu = \omega n^\mu$ and using the dimensionless quantity $n^\mu$ in place of $k^\mu$. Then, however, $nv$ is not a Lorentz scalar (contrary to $kv = v^\lminus$), and $\omega$ has to appear explicitly in many places.

In light-cone coordinates, the metric is given by
\begin{multline}
g_{\mu'\nu'}
=
g_{\rho\sigma}
\Lambdainv^{\rho}_{\phantom{\rho}\mu'} 
\Lambdainv^{\sigma}_{\phantom{\sigma}\nu'} 
\\=
\delta^\lplus_{\mu'} \delta^\lminus_{\nu'}
+
\delta^\lminus_{\mu'} \delta^\lplus_{\nu'}
-
\delta^\lone_{\mu'} \delta^\lone_{\nu'}
-
\delta^\ltwo_{\mu'} \delta^\ltwo_{\nu'},
\end{multline}
which allows us to write the scalar product of two four-vectors as
\begin{gather}
a_\mu b^\mu = a^\lplus b^\lminus + a^\lminus b^\lplus -  a^\lone b^\lone - a^\ltwo b^\ltwo
\end{gather}
(we also use the short notation $a^\lperp b^\lperp = a^\lone b^\lone + a^\ltwo b^\ltwo$). Due to Eq.~(\ref{eqn:lcc_orthogonalityrelations}), we obtain
\begin{gather}
\abs{\det \Lambda^{\mu'}_{\phantom{\mu}\nu}}
=
\abs{
\Lambda^{\lplus}_{\phantom{\lplus}\mu} \Lambda^{\lminus}_{\phantom{\lminus}\nu} \Lambda^{\lone}_{\phantom{\lone}\rho}\Lambda^{\ltwo}_{\phantom{\ltwo}\sigma} \eps^{\mu\nu\rho\sigma}}
=
1.
\end{gather}
Thus, the four-dimensional integration measure becomes
\begin{gather}
\int d^4a \, = \int da^\lplus da^\lminus da^\lperp,
\quad 
da^\lperp = da^\lone da^\ltwo.
\end{gather}

Since all properties of the light-cone coordinates follow from the relations in Eq.~(\ref{eqn:lcc_orthogonalityrelations}), we are not forced to use the canonical basis in Eq.~(\ref{eqn:canonicallcc}). For the calculation of the polarization operator, it is more convenient to use the two four-vectors [see Eq.~(\ref{eqn:Lambdavectors})],
\begin{subequations}
\label{eqn:modifiedlcc}
\begin{gather}
e_1'^\mu
=
\Lambda_1^\mu
=
\frac{f^{\mu\nu}_1 q_{\nu}}{kq\, \sqrt{-a_1^2}},
\quad
e_2'^\mu
=
\Lambda_2^\mu
=
\frac{f^{\mu\nu}_2 q_{\nu}}{kq\, \sqrt{-a_2^2}}
\end{gather}
together with~$k^\mu$ and
\begin{multline}
\bar{k}'^\mu = \bar{k}^\mu + \frac{a_1q}{a_1^2\, kq} a_1^\mu + \frac{a_2q}{a_2^2\, kq} a_2^\mu \\- \frac{1}{2(kq)^2} \lsb \frac{(a_1q)^2}{a_1^2} + \frac{(a_2q)^2}{a_2^2}\rsb  k^\mu.
\end{multline}
\end{subequations}
The set of four-vectors~$k^\mu$, $\bar{k}'^\mu$, $e_1'^\mu$, $e_2'^\mu$ also obeys the relations in Eq.~(\ref{eqn:lcc_orthogonalityrelations}), and we will call the coordinates, following from this set, modified light-cone coordinates  [the same symbols ($\lplus$, $\lminus$, $\lone$, $\ltwo$) are used to denote the corresponding components].

\section{Gamma matrix algebra}
\label{sec:gammamatrixalgebraappendix}

In this appendix we summarize some general identities, which are useful in calculations involving gamma matrices. The gamma matrices form a complete set in the sense that any matrix in spinor space can be decomposed according to \cite{leader_spin_2001}
\begin{gather}
\label{eqn:decompositionofspinormatrixinfundamentalterms}
\Gamma = c_\one \one + c_5 \gamma^5 + c_\mu \gamma^\mu + c_{5\mu} i\gamma^\mu \gamma^5 + c_{\mu\nu} i \sigma^{\mu\nu},
\end{gather}
where we assume (without restriction) that $c_{\mu\nu} = - c_{\nu\mu}$ and the coefficients can be calculated using
\begin{gather}
\label{eqn:gammamatrixcoefficients}
\begin{gathered}
c_\one =  \frac{1}{4} \tr \one \Gamma,
\quad
c_5 =  \frac{1}{4} \tr \gamma^5 \Gamma,
\quad
c_\mu =  \frac{1}{4} \tr \gamma_\mu \Gamma,\\
c_{5\mu} =  \frac{1}{4} \tr i \gamma_{\mu} \gamma^5 \Gamma,
\quad
c_{\mu\nu} = \frac{1}{8} \tr i \sigma_{\mu\nu} \Gamma.
\end{gathered}
\end{gather}
Due to the cyclic property of the trace, one can recursively calculate traces of arbitrary length without conceptual difficulties by permuting the first gamma matrix to the last position. For completeness we note the following relations
\begin{gather}
\label{eqn:traceformulasuptoordersix}
\begin{aligned} 
&\frac14\,\tr \gamma^\mu \gamma^\nu &
&= g^{\mu\nu},\\
&\frac14\,\tr \gamma^\mu \gamma^\nu \gamma^\rho \gamma^\sigma & 
&=
g^{\mu\sigma}\, g^{\nu\rho} 
- \,g^{\mu\rho}\, g^{\nu\sigma} 
+\,g^{\mu\nu}\, g^{\rho\sigma},\\
&\frac14\,\tr \sigma^{\mu\nu} \gamma^\rho \gamma^\sigma &
&=
g^{\mu\sigma}\, g^{\nu\rho} 
- \,g^{\mu\rho}\, g^{\nu\sigma},\\
&\frac14\, \tr \gamma^\mu\gamma^\nu\gamma^\rho\gamma^\sigma \gamma^5&  
&= 
i\eps^{\mu\nu\rho\sigma}.%
\end{aligned}
\end{gather}
Thus, any identity involving gamma matrices can be proven by calculating the fundamental terms given in Eq.~(\ref{eqn:decompositionofspinormatrixinfundamentalterms}) for both sides of the equation. It is in particular possible to map the gamma matrix algebra to a corresponding tensor algebra once the decomposition of the product of two (arbitrary) gamma matrix expressions is known, 
\begin{gather}
\Gamma_c = \Gamma_a \Gamma_b.
\end{gather}
Here~$\Gamma_x$ is written as in Eq.~(\ref{eqn:decompositionofspinormatrixinfundamentalterms}) with the letter~$c$ replaced by the letter~$x$ appearing in the index. The coefficients of~$\Gamma_c$ are then given by
\begin{gather}
\label{eqn:gammamatrixalgebra_productdecomposition}
\begin{aligned}
c_\one &= a_\one b_\one + a_5 b_5 + a^\mu b_\mu + a_5^\mu b_{5\mu} + 2 a_{\mu\nu} b^{\mu\nu},\\[2ex]
c_5 &=  \lb a_\one b_5 + a_5 b_\one \rb + \lb i a^\mu b_{5\mu} - i a_{5\mu} b^\mu \rb\\
&\phantom{=}- i \eps^{\mu\nu\rho\sigma} a_{\mu\nu} b_{\rho\sigma},\\[2ex]
c_\mu
&= 
\lb a_\one b_\mu + a_\mu b_\one \rb  + \lb i a_{5\mu} b_5 - i a_5 b_{5\mu} \rb\\
&\phantom{=}+ 2 \lb ia_{\mu\nu} b^\nu - i a^\nu b_{\mu\nu} \rb
-i \eps_{\mu\nu\rho\sigma} \lb a^\nu_5 b^{\rho\sigma} + a^{\rho\sigma} b_5^{\nu} \rb,\\[2ex]
c_{5\mu}
&=
\lb a_\one b_{5\mu} + a_{5\mu} b_\one \rb
+ \lb ia_5 b_\mu - i a_\mu b_5 \rb \\
&\phantom{=}+ i\eps_{\mu\nu\rho\sigma} \lb a^\nu b^{\rho\sigma} + a^{\rho\sigma} b^\nu \rb
+ 2\lb ia_{\mu\nu} b^\nu_5 -i a^\nu_5 b_{\mu\nu} \rb,\\[2ex]
c_{\mu\nu}
&= 
\lb a_\one b_{\mu\nu} + a_{\mu\nu} b_\one \rb
-\frac{i}{2} \eps_{\mu\nu\rho\sigma} \lb a^{\rho\sigma} b_5 + a_5 b^{\rho\sigma} \rb\\
&\phantom{=}-\frac{i}{2} \lb a_\mu b_\nu - a_\nu b_\mu \rb
-\frac{i}{2} \eps_{\mu\nu\rho\sigma} \lb a^\rho b_5^\sigma + a_5^\sigma b^\rho \rb\\
&\phantom{=}- \frac{i}{2} \lb a_{5\mu} b_{5\nu} - a_{5\nu} b_{5\mu} \rb
+ 
2i \lb a_{\mu\rho} b^{\rho}_{\phantom{\rho}\nu} - a_{\nu\rho} b^{\rho}_{\phantom{\rho}\mu} \rb.
\end{aligned}
\end{gather}
We point out that taking the trace of the gamma matrix expression~$\Gamma_c$ projects out the coefficient~$c_\one$ [see Eq.~(\ref{eqn:gammamatrixcoefficients})]. Therefore, one can also use Eq.~(\ref{eqn:gammamatrixalgebra_productdecomposition}) in the calculation of traces. 

\section{Tensor relations}
\label{sec:tensorrelationsappendix}

If Eq.~(\ref{eqn:gammamatrixalgebra_productdecomposition}) is used to simplify large gamma matrix expressions, one typically encounters products or contractions of the totally antisymmetric tensor $\eps^{\alpha\beta\gamma\delta}$. They can be simplified using well-known identities stated here for completeness~\cite{landau_classical_1987}:
\begin{gather}
\label{eqn:epsidentities}
\begin{aligned}
\eps^{\alpha\beta\gamma\delta}\eps_{\alpha\beta\gamma\delta} &= -24,\\
\eps^{\alpha\beta\gamma\mu}\eps_{\alpha\beta\gamma\nu} &= -6 \delta^{\mu}_{\nu},\\
\eps^{\alpha\beta\mu\nu}\eps_{\alpha\beta\rho\sigma} &= -2\big(\delta^{\mu}_{\rho}\delta^{\nu}_{\sigma} - \delta^{\mu}_{\sigma}\delta^{\nu}_{\rho} \big),\\
\eps^{\mu\nu\rho\sigma} \eps_{\alpha\beta\gamma\sigma}
& =
-\big(\delta^{\mu}_{\alpha}\delta^{\nu}_{\beta}\delta^{\rho}_{\gamma} 
-
\delta^{\mu}_{\alpha}\delta^{\nu}_{\gamma} \delta^{\rho}_{\beta}
+
\delta^{\mu}_{\gamma}\delta^{\nu}_{\alpha}\delta^{\rho}_{\beta}\\
&\phantom{=}- \delta^{\mu}_{\gamma}\delta^{\nu}_{\beta}\delta^{\rho}_{\alpha}
+
\delta^{\mu}_{\beta}\delta^{\nu}_{\gamma}\delta^{\rho}_{\alpha} 
-
\delta^{\mu}_{\beta}\delta^{\nu}_{\alpha}\delta^{\rho}_{\gamma}
\big),\\
-\eps^{\mu\nu\rho\sigma} \eps_{\alpha\beta\gamma\delta} 
&= 
 \det\left(
 \begin{array}{cccc}
 \delta^{\mu}_{\alpha} 	&	\delta^{\mu}_{\beta}	&	\delta^{\mu}_{\gamma}	& \delta^{\mu}_{\delta}\\
 \delta^{\nu}_{\alpha} 	&	\delta^{\nu}_{\beta} 	&	\delta^{\nu}_{\gamma} 	&	\delta^{\nu}_{\delta}\\
 \delta^{\rho}_{\alpha} 	&	\delta^{\rho}_{\beta} 	&	\delta^{\rho}_{\gamma} 	&	\delta^{\rho}_{\delta}\\
 \delta^{\sigma}_{\alpha} 	&	\delta^{\sigma}_{\beta} 	&	\delta^{\sigma}_{\gamma} 	&	\delta^{\sigma}_{\delta}
 \end{array}\right).
\end{aligned}
\end{gather}
In particular, we note the following formulas for antisymmetric tensors $T^{\mu\nu}$, $T_1^{\mu\nu}$, and $T_2^{\alpha\beta}$:
\begin{gather}
\begin{aligned}
T_1^{*\mu\nu} T_2^{*\alpha\beta}
&=
\frac12 \lb g^{\mu\beta} g^{\nu\alpha} - g^{\mu\alpha} g^{\nu\beta} \rb T_{1\rho\sigma} T_2^{\rho\sigma}
- T_1^{\alpha\beta} T_2^{\mu\nu}
\\&+ g^{\nu\alpha} (T_1 T_2)^{\beta\mu} - g^{\mu\alpha} (T_1 T_2)^{\beta\nu} 
\\&- g^{\nu\beta} (T_1 T_2)^{\alpha\mu} + g^{\mu\beta} (T_1 T_2)^{\alpha\nu},  
\end{aligned}
\\
\begin{aligned}
(T^*_1 T^*_2)^{\mu\nu} &= \frac12 g^{\mu\nu} T_{1\alpha\beta} T_2^{\alpha\beta} + (T_1 T_2)^{\nu\mu},\\
T^*_{1\mu\nu} T_2^{*\mu\nu} &= - T_{1\mu\nu} T_2^{\mu\nu}
\end{aligned}
\end{gather}
and
\begin{gather}
\begin{aligned}
\eps^{\mu\nu\rho\sigma} T^*_{\sigma \alpha} 
&=
\delta^\mu_\alpha T^{\nu\rho} - \delta^\nu_\alpha T^{\mu\rho} + \delta^{\rho}_{\alpha} T^{\mu\nu},
\\
\frac12 \eps_{\mu\nu\rho\sigma} T^{*\rho\sigma}   &= - T_{\mu\nu}.
\end{aligned}
\end{gather}

\begin{thebibliography}{90}%
\makeatletter
\providecommand \@ifxundefined [1]{%
 \@ifx{#1\undefined}
}%
\providecommand \@ifnum [1]{%
 \ifnum #1\expandafter \@firstoftwo
 \else \expandafter \@secondoftwo
 \fi
}%
\providecommand \@ifx [1]{%
 \ifx #1\expandafter \@firstoftwo
 \else \expandafter \@secondoftwo
 \fi
}%
\providecommand \natexlab [1]{#1}%
\providecommand \enquote  [1]{``#1''}%
\providecommand \bibnamefont  [1]{#1}%
\providecommand \bibfnamefont [1]{#1}%
\providecommand \citenamefont [1]{#1}%
\providecommand \href@noop [0]{\@secondoftwo}%
\providecommand \href [0]{\begingroup \@sanitize@url \@href}%
\providecommand \@href[1]{\@@startlink{#1}\@@href}%
\providecommand \@@href[1]{\endgroup#1\@@endlink}%
\providecommand \@sanitize@url [0]{\catcode `\\12\catcode `\$12\catcode
  `\&12\catcode `\#12\catcode `\^12\catcode `\_12\catcode `\%12\relax}%
\providecommand \@@startlink[1]{}%
\providecommand \@@endlink[0]{}%
\providecommand \url  [0]{\begingroup\@sanitize@url \@url }%
\providecommand \@url [1]{\endgroup\@href {#1}{\urlprefix }}%
\providecommand \urlprefix  [0]{URL }%
\providecommand \Eprint [0]{\href }%
\providecommand \doibase [0]{http://dx.doi.org/}%
\providecommand \selectlanguage [0]{\@gobble}%
\providecommand \bibinfo  [0]{\@secondoftwo}%
\providecommand \bibfield  [0]{\@secondoftwo}%
\providecommand \translation [1]{[#1]}%
\providecommand \BibitemOpen [0]{}%
\providecommand \bibitemStop [0]{}%
\providecommand \bibitemNoStop [0]{.\EOS\space}%
\providecommand \EOS [0]{\spacefactor3000\relax}%
\providecommand \BibitemShut  [1]{\csname bibitem#1\endcsname}%
\let\auto@bib@innerbib\@empty
\bibitem [{Note1()}]{Note1}%
  \BibitemOpen
  \bibinfo {note} {We briefly mention here that the perturbative approach
  cannot be applied to processes involving particles with extremely high
  energies $\varepsilon $ such that $\alpha \protect \qopname \relax
  o{log}(\varepsilon /mc^2) \sim 1$, with $m$ being the electron mass \cite
  {landau_quantum_1981}.}\BibitemShut {Stop}%
\bibitem [{\citenamefont {Gabrielse}\ \emph {et~al.}(2006)\citenamefont
  {Gabrielse}, \citenamefont {Hanneke}, \citenamefont {Kinoshita},
  \citenamefont {Nio},\ and\ \citenamefont {Odom}}]{gabrielse_new_2006}%
  \BibitemOpen
  \bibfield  {author} {\bibinfo {author} {\bibfnamefont {G.}~\bibnamefont
  {Gabrielse}}, \bibinfo {author} {\bibfnamefont {D.}~\bibnamefont {Hanneke}},
  \bibinfo {author} {\bibfnamefont {T.}~\bibnamefont {Kinoshita}}, \bibinfo
  {author} {\bibfnamefont {M.}~\bibnamefont {Nio}}, \ and\ \bibinfo {author}
  {\bibfnamefont {B.}~\bibnamefont {Odom}},\ }\href {\doibase
  10.1103/PhysRevLett.97.030802} {\bibfield  {journal} {\bibinfo  {journal}
  {Phys. Rev. Lett.}\ }\textbf {\bibinfo {volume} {97}},\ \bibinfo {pages}
  {030802} (\bibinfo {year} {2006})}\BibitemShut {NoStop}%
\bibitem [{\citenamefont {Sauter}(1931)}]{sauter_ueber_1931}%
  \BibitemOpen
  \bibfield  {author} {\bibinfo {author} {\bibfnamefont {F.}~\bibnamefont
  {Sauter}},\ }\href@noop {} {\bibfield  {journal} {\bibinfo  {journal} {Z.
  Phys.}\ }\textbf {\bibinfo {volume} {69}},\ \bibinfo {pages} {742} (\bibinfo
  {year} {1931})}\BibitemShut {NoStop}%
\bibitem [{\citenamefont {Heisenberg}\ and\ \citenamefont
  {Euler}(1936)}]{heisenberg_folgerungen_1936}%
  \BibitemOpen
  \bibfield  {author} {\bibinfo {author} {\bibfnamefont {W.}~\bibnamefont
  {Heisenberg}}\ and\ \bibinfo {author} {\bibfnamefont {H.}~\bibnamefont
  {Euler}},\ }\href {\doibase 10.1007/BF01343663} {\bibfield  {journal}
  {\bibinfo  {journal} {Z. Phys.}\ }\textbf {\bibinfo {volume} {98}},\ \bibinfo
  {pages} {714} (\bibinfo {year} {1936})}\BibitemShut {NoStop}%
\bibitem [{\citenamefont {Schwinger}(1951)}]{schwinger_gauge_1951}%
  \BibitemOpen
  \bibfield  {author} {\bibinfo {author} {\bibfnamefont {J.}~\bibnamefont
  {Schwinger}},\ }\href {\doibase 10.1103/PhysRev.82.664} {\bibfield  {journal}
  {\bibinfo  {journal} {Phys. Rev.}\ }\textbf {\bibinfo {volume} {82}},\
  \bibinfo {pages} {664} (\bibinfo {year} {1951})}\BibitemShut {NoStop}%
\bibitem [{\citenamefont {Yanovsky}\ \emph {et~al.}(2008)\citenamefont
  {Yanovsky}, \citenamefont {Chvykov}, \citenamefont {Kalinchenko},
  \citenamefont {Rousseau}, \citenamefont {Planchon}, \citenamefont {Matsuoka},
  \citenamefont {Maksimchuk}, \citenamefont {Nees}, \citenamefont {Cheriaux},
  \citenamefont {Mourou} \emph {et~al.}}]{yanovsky_ultra_2008}%
  \BibitemOpen
  \bibfield  {author} {\bibinfo {author} {\bibfnamefont {V.}~\bibnamefont
  {Yanovsky}}, \bibinfo {author} {\bibfnamefont {V.}~\bibnamefont {Chvykov}},
  \bibinfo {author} {\bibfnamefont {G.}~\bibnamefont {Kalinchenko}}, \bibinfo
  {author} {\bibfnamefont {P.}~\bibnamefont {Rousseau}}, \bibinfo {author}
  {\bibfnamefont {T.}~\bibnamefont {Planchon}}, \bibinfo {author}
  {\bibfnamefont {T.}~\bibnamefont {Matsuoka}}, \bibinfo {author}
  {\bibfnamefont {A.}~\bibnamefont {Maksimchuk}}, \bibinfo {author}
  {\bibfnamefont {J.}~\bibnamefont {Nees}}, \bibinfo {author} {\bibfnamefont
  {G.}~\bibnamefont {Cheriaux}}, \bibinfo {author} {\bibfnamefont
  {G.}~\bibnamefont {Mourou}},  \emph {et~al.},\ }\href {\doibase
  10.1364/OE.16.002109} {\bibfield  {journal} {\bibinfo  {journal} {Opt.
  Express}\ }\textbf {\bibinfo {volume} {16}},\ \bibinfo {pages} {2109}
  (\bibinfo {year} {2008})}\BibitemShut {NoStop}%
\bibitem [{ELI()}]{ELI}%
  \BibitemOpen
  \href@noop {} {\enquote {\bibinfo {title} {{E}xtreme {L}ight {I}nfrastructure
  ({E}{L}{I})},}\ }\bibinfo {note}
  {\textit{http://www.eli-beams.eu/}}\BibitemShut {NoStop}%
\bibitem [{HIP()}]{HIPER}%
  \BibitemOpen
  \href@noop {} {\enquote {\bibinfo {title} {{E}uropean {H}igh {P}ower laser
  {E}nergy {R}esearch facility ({H}i{P}{E}{R})},}\ }\bibinfo {note}
  {\textit{http://www.hiperlaser.org}}\BibitemShut {NoStop}%
\bibitem [{XCE()}]{XCELS}%
  \BibitemOpen
  \href@noop {} {\enquote {\bibinfo {title} {{E}xawatt {C}enter for {E}xtreme
  {L}ight {S}tudies {({X}{C}{E}{L}{S})}},}\ }\bibinfo {note}
  {\textit{http://www.xcels.iapras.ru/img/XCELS-Project-english-version.pdf}}\BibitemShut
  {NoStop}%
\bibitem [{\citenamefont {Leemans}\ \emph {et~al.}(2006)\citenamefont
  {Leemans}, \citenamefont {Nagler}, \citenamefont {Gonsalves}, \citenamefont
  {Toth}, \citenamefont {Nakamura}, \citenamefont {Geddes}, \citenamefont
  {Esarey}, \citenamefont {Schroeder},\ and\ \citenamefont
  {Hooker}}]{leemans_gev_2006}%
  \BibitemOpen
  \bibfield  {author} {\bibinfo {author} {\bibfnamefont {W.~P.}\ \bibnamefont
  {Leemans}}, \bibinfo {author} {\bibfnamefont {B.}~\bibnamefont {Nagler}},
  \bibinfo {author} {\bibfnamefont {A.~J.}\ \bibnamefont {Gonsalves}}, \bibinfo
  {author} {\bibfnamefont {C.}~\bibnamefont {Toth}}, \bibinfo {author}
  {\bibfnamefont {K.}~\bibnamefont {Nakamura}}, \bibinfo {author}
  {\bibfnamefont {C.~G.~R.}\ \bibnamefont {Geddes}}, \bibinfo {author}
  {\bibfnamefont {E.}~\bibnamefont {Esarey}}, \bibinfo {author} {\bibfnamefont
  {C.~B.}\ \bibnamefont {Schroeder}}, \ and\ \bibinfo {author} {\bibfnamefont
  {S.~M.}\ \bibnamefont {Hooker}},\ }\href {\doibase 10.1038/nphys418}
  {\bibfield  {journal} {\bibinfo  {journal} {Nature Phys.}\ }\textbf {\bibinfo
  {volume} {2}},\ \bibinfo {pages} {696} (\bibinfo {year} {2006})}\BibitemShut
  {NoStop}%
\bibitem [{\citenamefont {Bula}\ \emph {et~al.}(1996)\citenamefont {Bula},
  \citenamefont {McDonald}, \citenamefont {Prebys}, \citenamefont {Bamber},
  \citenamefont {Boege}, \citenamefont {Kotseroglou}, \citenamefont
  {Melissinos}, \citenamefont {Meyerhofer}, \citenamefont {Ragg}, \citenamefont
  {Burke} \emph {et~al.}}]{bula_observation_1996}%
  \BibitemOpen
  \bibfield  {author} {\bibinfo {author} {\bibfnamefont {C.}~\bibnamefont
  {Bula}}, \bibinfo {author} {\bibfnamefont {K.~T.}\ \bibnamefont {McDonald}},
  \bibinfo {author} {\bibfnamefont {E.~J.}\ \bibnamefont {Prebys}}, \bibinfo
  {author} {\bibfnamefont {C.}~\bibnamefont {Bamber}}, \bibinfo {author}
  {\bibfnamefont {S.}~\bibnamefont {Boege}}, \bibinfo {author} {\bibfnamefont
  {T.}~\bibnamefont {Kotseroglou}}, \bibinfo {author} {\bibfnamefont {A.~C.}\
  \bibnamefont {Melissinos}}, \bibinfo {author} {\bibfnamefont {D.~D.}\
  \bibnamefont {Meyerhofer}}, \bibinfo {author} {\bibfnamefont
  {W.}~\bibnamefont {Ragg}}, \bibinfo {author} {\bibfnamefont {D.~L.}\
  \bibnamefont {Burke}},  \emph {et~al.},\ }\href {\doibase
  10.1103/PhysRevLett.76.3116} {\bibfield  {journal} {\bibinfo  {journal}
  {Phys. Rev. Lett.}\ }\textbf {\bibinfo {volume} {76}},\ \bibinfo {pages}
  {3116} (\bibinfo {year} {1996})}\BibitemShut {NoStop}%
\bibitem [{\citenamefont {Burke}\ \emph {et~al.}(1997)\citenamefont {Burke},
  \citenamefont {Field}, \citenamefont {Horton-Smith}, \citenamefont {Spencer},
  \citenamefont {Walz}, \citenamefont {Berridge}, \citenamefont {Bugg},
  \citenamefont {Shmakov}, \citenamefont {Weidemann}, \citenamefont {Bula}
  \emph {et~al.}}]{burke_positron_1997}%
  \BibitemOpen
  \bibfield  {author} {\bibinfo {author} {\bibfnamefont {D.~L.}\ \bibnamefont
  {Burke}}, \bibinfo {author} {\bibfnamefont {R.~C.}\ \bibnamefont {Field}},
  \bibinfo {author} {\bibfnamefont {G.}~\bibnamefont {Horton-Smith}}, \bibinfo
  {author} {\bibfnamefont {J.~E.}\ \bibnamefont {Spencer}}, \bibinfo {author}
  {\bibfnamefont {D.}~\bibnamefont {Walz}}, \bibinfo {author} {\bibfnamefont
  {S.~C.}\ \bibnamefont {Berridge}}, \bibinfo {author} {\bibfnamefont {W.~M.}\
  \bibnamefont {Bugg}}, \bibinfo {author} {\bibfnamefont {K.}~\bibnamefont
  {Shmakov}}, \bibinfo {author} {\bibfnamefont {A.~W.}\ \bibnamefont
  {Weidemann}}, \bibinfo {author} {\bibfnamefont {C.}~\bibnamefont {Bula}},
  \emph {et~al.},\ }\href {\doibase 10.1103/PhysRevLett.79.1626} {\bibfield
  {journal} {\bibinfo  {journal} {Phys. Rev. Lett.}\ }\textbf {\bibinfo
  {volume} {79}},\ \bibinfo {pages} {1626} (\bibinfo {year}
  {1997})}\BibitemShut {NoStop}%
\bibitem [{\citenamefont {Di~Piazza}\ \emph {et~al.}(2010)\citenamefont
  {Di~Piazza}, \citenamefont {Hatsagortsyan},\ and\ \citenamefont
  {Keitel}}]{di_piazza_quantum_2010}%
  \BibitemOpen
  \bibfield  {author} {\bibinfo {author} {\bibfnamefont {A.}~\bibnamefont
  {Di~Piazza}}, \bibinfo {author} {\bibfnamefont {K.~Z.}\ \bibnamefont
  {Hatsagortsyan}}, \ and\ \bibinfo {author} {\bibfnamefont {C.~H.}\
  \bibnamefont {Keitel}},\ }\href {\doibase 10.1103/PhysRevLett.105.220403}
  {\bibfield  {journal} {\bibinfo  {journal} {Phys. Rev. Lett.}\ }\textbf
  {\bibinfo {volume} {105}},\ \bibinfo {pages} {220403} (\bibinfo {year}
  {2010})}\BibitemShut {NoStop}%
\bibitem [{\citenamefont {Hu}\ \emph {et~al.}(2010)\citenamefont {Hu},
  \citenamefont {M{\"u}ller},\ and\ \citenamefont {Keitel}}]{hu_complete_2010}%
  \BibitemOpen
  \bibfield  {author} {\bibinfo {author} {\bibfnamefont {H.}~\bibnamefont
  {Hu}}, \bibinfo {author} {\bibfnamefont {C.}~\bibnamefont {M{\"u}ller}}, \
  and\ \bibinfo {author} {\bibfnamefont {C.~H.}\ \bibnamefont {Keitel}},\
  }\href {\doibase 10.1103/PhysRevLett.105.080401} {\bibfield  {journal}
  {\bibinfo  {journal} {Phys. Rev. Lett.}\ }\textbf {\bibinfo {volume} {105}},\
  \bibinfo {pages} {080401} (\bibinfo {year} {2010})}\BibitemShut {NoStop}%
\bibitem [{\citenamefont {King}\ \emph {et~al.}(2010)\citenamefont {King},
  \citenamefont {Di~Piazza},\ and\ \citenamefont
  {Keitel}}]{king_matterless_2010}%
  \BibitemOpen
  \bibfield  {author} {\bibinfo {author} {\bibfnamefont {B.}~\bibnamefont
  {King}}, \bibinfo {author} {\bibfnamefont {A.}~\bibnamefont {Di~Piazza}}, \
  and\ \bibinfo {author} {\bibfnamefont {C.~H.}\ \bibnamefont {Keitel}},\
  }\href {\doibase 10.1038/nphoton.2009.261} {\bibfield  {journal} {\bibinfo
  {journal} {Nature Photon.}\ }\textbf {\bibinfo {volume} {4}},\ \bibinfo
  {pages} {92} (\bibinfo {year} {2010})}\BibitemShut {NoStop}%
\bibitem [{\citenamefont {Mackenroth}\ \emph {et~al.}(2010)\citenamefont
  {Mackenroth}, \citenamefont {Di~Piazza},\ and\ \citenamefont
  {Keitel}}]{mackenroth_determining_2010}%
  \BibitemOpen
  \bibfield  {author} {\bibinfo {author} {\bibfnamefont {F.}~\bibnamefont
  {Mackenroth}}, \bibinfo {author} {\bibfnamefont {A.}~\bibnamefont
  {Di~Piazza}}, \ and\ \bibinfo {author} {\bibfnamefont {C.~H.}\ \bibnamefont
  {Keitel}},\ }\href {\doibase 10.1103/PhysRevLett.105.063903} {\bibfield
  {journal} {\bibinfo  {journal} {Phys. Rev. Lett.}\ }\textbf {\bibinfo
  {volume} {105}},\ \bibinfo {pages} {063903} (\bibinfo {year}
  {2010})}\BibitemShut {NoStop}%
\bibitem [{\citenamefont {Dumlu}(2010)}]{dumlu_schwinger_2010}%
  \BibitemOpen
  \bibfield  {author} {\bibinfo {author} {\bibfnamefont {C.~K.}\ \bibnamefont
  {Dumlu}},\ }\href {\doibase 10.1103/PhysRevD.82.045007} {\bibfield  {journal}
  {\bibinfo  {journal} {Phys. Rev. D}\ }\textbf {\bibinfo {volume} {82}},\
  \bibinfo {pages} {045007} (\bibinfo {year} {2010})}\BibitemShut {NoStop}%
\bibitem [{\citenamefont {Heinzl}\ \emph
  {et~al.}(2010{\natexlab{a}})\citenamefont {Heinzl}, \citenamefont {Seipt},\
  and\ \citenamefont {K\"ampfer}}]{heinzl_beam-shape_2010}%
  \BibitemOpen
  \bibfield  {author} {\bibinfo {author} {\bibfnamefont {T.}~\bibnamefont
  {Heinzl}}, \bibinfo {author} {\bibfnamefont {D.}~\bibnamefont {Seipt}}, \
  and\ \bibinfo {author} {\bibfnamefont {B.}~\bibnamefont {K\"ampfer}},\ }\href
  {\doibase 10.1103/PhysRevA.81.022125} {\bibfield  {journal} {\bibinfo
  {journal} {Phys. Rev. A}\ }\textbf {\bibinfo {volume} {81}},\ \bibinfo
  {pages} {022125} (\bibinfo {year} {2010}{\natexlab{a}})}\BibitemShut
  {NoStop}%
\bibitem [{\citenamefont {Heinzl}\ \emph
  {et~al.}(2010{\natexlab{b}})\citenamefont {Heinzl}, \citenamefont
  {Ilderton},\ and\ \citenamefont {Marklund}}]{heinzl_finite_2010}%
  \BibitemOpen
  \bibfield  {author} {\bibinfo {author} {\bibfnamefont {T.}~\bibnamefont
  {Heinzl}}, \bibinfo {author} {\bibfnamefont {A.}~\bibnamefont {Ilderton}}, \
  and\ \bibinfo {author} {\bibfnamefont {M.}~\bibnamefont {Marklund}},\ }\href
  {\doibase 10.1016/j.physletb.2010.07.044} {\bibfield  {journal} {\bibinfo
  {journal} {Phys. Lett. B}\ }\textbf {\bibinfo {volume} {692}},\ \bibinfo
  {pages} {250} (\bibinfo {year} {2010}{\natexlab{b}})}\BibitemShut {NoStop}%
\bibitem [{\citenamefont {Fedotov}\ \emph {et~al.}(2010)\citenamefont
  {Fedotov}, \citenamefont {Narozhny}, \citenamefont {Mourou},\ and\
  \citenamefont {Korn}}]{fedotov_limitations_2010}%
  \BibitemOpen
  \bibfield  {author} {\bibinfo {author} {\bibfnamefont {A.~M.}\ \bibnamefont
  {Fedotov}}, \bibinfo {author} {\bibfnamefont {N.~B.}\ \bibnamefont
  {Narozhny}}, \bibinfo {author} {\bibfnamefont {G.}~\bibnamefont {Mourou}}, \
  and\ \bibinfo {author} {\bibfnamefont {G.}~\bibnamefont {Korn}},\ }\href
  {\doibase 10.1103/PhysRevLett.105.080402} {\bibfield  {journal} {\bibinfo
  {journal} {Phys. Rev. Lett.}\ }\textbf {\bibinfo {volume} {105}},\ \bibinfo
  {pages} {080402} (\bibinfo {year} {2010})}\BibitemShut {NoStop}%
\bibitem [{\citenamefont {Bulanov}\ \emph {et~al.}(2010)\citenamefont
  {Bulanov}, \citenamefont {Esirkepov}, \citenamefont {Thomas}, \citenamefont
  {Koga},\ and\ \citenamefont {Bulanov}}]{bulanov_schwinger_2010}%
  \BibitemOpen
  \bibfield  {author} {\bibinfo {author} {\bibfnamefont {S.~S.}\ \bibnamefont
  {Bulanov}}, \bibinfo {author} {\bibfnamefont {T.~Z.}\ \bibnamefont
  {Esirkepov}}, \bibinfo {author} {\bibfnamefont {A.~G.~R.}\ \bibnamefont
  {Thomas}}, \bibinfo {author} {\bibfnamefont {J.~K.}\ \bibnamefont {Koga}}, \
  and\ \bibinfo {author} {\bibfnamefont {S.~V.}\ \bibnamefont {Bulanov}},\
  }\href {\doibase 10.1103/PhysRevLett.105.220407} {\bibfield  {journal}
  {\bibinfo  {journal} {Phys. Rev. Lett.}\ }\textbf {\bibinfo {volume} {105}},\
  \bibinfo {pages} {220407} (\bibinfo {year} {2010})}\BibitemShut {NoStop}%
\bibitem [{\citenamefont {Sokolov}\ \emph {et~al.}(2010)\citenamefont
  {Sokolov}, \citenamefont {Naumova}, \citenamefont {Nees},\ and\ \citenamefont
  {Mourou}}]{sokolov_pair_2010}%
  \BibitemOpen
  \bibfield  {author} {\bibinfo {author} {\bibfnamefont {I.~V.}\ \bibnamefont
  {Sokolov}}, \bibinfo {author} {\bibfnamefont {N.~M.}\ \bibnamefont
  {Naumova}}, \bibinfo {author} {\bibfnamefont {J.~A.}\ \bibnamefont {Nees}}, \
  and\ \bibinfo {author} {\bibfnamefont {G.~A.}\ \bibnamefont {Mourou}},\
  }\href {\doibase 10.1103/PhysRevLett.105.195005} {\bibfield  {journal}
  {\bibinfo  {journal} {Phys. Rev. Lett.}\ }\textbf {\bibinfo {volume} {105}},\
  \bibinfo {pages} {195005} (\bibinfo {year} {2010})}\BibitemShut {NoStop}%
\bibitem [{\citenamefont {Meuren}\ and\ \citenamefont
  {Di~Piazza}(2011)}]{meuren_quantum_2011}%
  \BibitemOpen
  \bibfield  {author} {\bibinfo {author} {\bibfnamefont {S.}~\bibnamefont
  {Meuren}}\ and\ \bibinfo {author} {\bibfnamefont {A.}~\bibnamefont
  {Di~Piazza}},\ }\href {\doibase 10.1103/PhysRevLett.107.260401} {\bibfield
  {journal} {\bibinfo  {journal} {Phys. Rev. Lett.}\ }\textbf {\bibinfo
  {volume} {107}},\ \bibinfo {pages} {260401} (\bibinfo {year}
  {2011})}\BibitemShut {NoStop}%
\bibitem [{\citenamefont {Hu}\ and\ \citenamefont
  {M{\"u}ller}(2011)}]{hu_relativistic_2011}%
  \BibitemOpen
  \bibfield  {author} {\bibinfo {author} {\bibfnamefont {H.}~\bibnamefont
  {Hu}}\ and\ \bibinfo {author} {\bibfnamefont {C.}~\bibnamefont
  {M{\"u}ller}},\ }\href {\doibase 10.1103/PhysRevLett.107.090402} {\bibfield
  {journal} {\bibinfo  {journal} {Phys. Rev. Lett.}\ }\textbf {\bibinfo
  {volume} {107}},\ \bibinfo {pages} {090402} (\bibinfo {year}
  {2011})}\BibitemShut {NoStop}%
\bibitem [{\citenamefont {Kryuchkyan}\ and\ \citenamefont
  {Hatsagortsyan}(2011)}]{kryuchkyan_bragg_2011}%
  \BibitemOpen
  \bibfield  {author} {\bibinfo {author} {\bibfnamefont {G.~Y.}\ \bibnamefont
  {Kryuchkyan}}\ and\ \bibinfo {author} {\bibfnamefont {K.~Z.}\ \bibnamefont
  {Hatsagortsyan}},\ }\href {\doibase 10.1103/PhysRevLett.107.053604}
  {\bibfield  {journal} {\bibinfo  {journal} {Phys. Rev. Lett.}\ }\textbf
  {\bibinfo {volume} {107}},\ \bibinfo {pages} {053604} (\bibinfo {year}
  {2011})}\BibitemShut {NoStop}%
\bibitem [{\citenamefont {Hartin}\ and\ \citenamefont
  {Moortgat-Pick}(2011)}]{hartin_high_2011}%
  \BibitemOpen
  \bibfield  {author} {\bibinfo {author} {\bibfnamefont {A.}~\bibnamefont
  {Hartin}}\ and\ \bibinfo {author} {\bibfnamefont {G.}~\bibnamefont
  {Moortgat-Pick}},\ }\href {\doibase 10.1140/epjc/s10052-011-1729-8}
  {\bibfield  {journal} {\bibinfo  {journal} {Eur. Phys. J. C}\ }\textbf
  {\bibinfo {volume} {71}},\ \bibinfo {pages} {1729} (\bibinfo {year}
  {2011})}\BibitemShut {NoStop}%
\bibitem [{\citenamefont {Boca}\ and\ \citenamefont
  {Florescu}(2011)}]{boca_thomson_2011}%
  \BibitemOpen
  \bibfield  {author} {\bibinfo {author} {\bibfnamefont {M.}~\bibnamefont
  {Boca}}\ and\ \bibinfo {author} {\bibfnamefont {V.}~\bibnamefont
  {Florescu}},\ }\href {\doibase 10.1140/epjd/e2010-10429-y} {\bibfield
  {journal} {\bibinfo  {journal} {Eur. Phys. J. D}\ }\textbf {\bibinfo {volume}
  {61}},\ \bibinfo {pages} {449} (\bibinfo {year} {2011})}\BibitemShut
  {NoStop}%
\bibitem [{\citenamefont {Dumlu}\ and\ \citenamefont
  {Dunne}(2011)}]{dumlu_interference_2011}%
  \BibitemOpen
  \bibfield  {author} {\bibinfo {author} {\bibfnamefont {C.~K.}\ \bibnamefont
  {Dumlu}}\ and\ \bibinfo {author} {\bibfnamefont {G.~V.}\ \bibnamefont
  {Dunne}},\ }\href {\doibase 10.1103/PhysRevD.83.065028} {\bibfield  {journal}
  {\bibinfo  {journal} {Phys. Rev. D}\ }\textbf {\bibinfo {volume} {83}},\
  \bibinfo {pages} {065028} (\bibinfo {year} {2011})}\BibitemShut {NoStop}%
\bibitem [{\citenamefont {Hebenstreit}\ \emph {et~al.}(2011)\citenamefont
  {Hebenstreit}, \citenamefont {Ilderton},\ and\ \citenamefont
  {Marklund}}]{hebenstreit_pair_2011}%
  \BibitemOpen
  \bibfield  {author} {\bibinfo {author} {\bibfnamefont {F.}~\bibnamefont
  {Hebenstreit}}, \bibinfo {author} {\bibfnamefont {A.}~\bibnamefont
  {Ilderton}}, \ and\ \bibinfo {author} {\bibfnamefont {M.}~\bibnamefont
  {Marklund}},\ }\href {\doibase 10.1103/PhysRevD.84.125022} {\bibfield
  {journal} {\bibinfo  {journal} {Phys. Rev. D}\ }\textbf {\bibinfo {volume}
  {84}},\ \bibinfo {pages} {125022} (\bibinfo {year} {2011})}\BibitemShut
  {NoStop}%
\bibitem [{\citenamefont {Elkina}\ \emph {et~al.}(2011)\citenamefont {Elkina},
  \citenamefont {Fedotov}, \citenamefont {Kostyukov}, \citenamefont {Legkov},
  \citenamefont {Narozhny}, \citenamefont {Nerush},\ and\ \citenamefont
  {Ruhl}}]{elkina_qed_2011}%
  \BibitemOpen
  \bibfield  {author} {\bibinfo {author} {\bibfnamefont {N.~V.}\ \bibnamefont
  {Elkina}}, \bibinfo {author} {\bibfnamefont {A.~M.}\ \bibnamefont {Fedotov}},
  \bibinfo {author} {\bibfnamefont {I.~Y.}\ \bibnamefont {Kostyukov}}, \bibinfo
  {author} {\bibfnamefont {M.~V.}\ \bibnamefont {Legkov}}, \bibinfo {author}
  {\bibfnamefont {N.~B.}\ \bibnamefont {Narozhny}}, \bibinfo {author}
  {\bibfnamefont {E.~N.}\ \bibnamefont {Nerush}}, \ and\ \bibinfo {author}
  {\bibfnamefont {H.}~\bibnamefont {Ruhl}},\ }\href {\doibase
  10.1103/PhysRevSTAB.14.054401} {\bibfield  {journal} {\bibinfo  {journal}
  {Phys. Rev. {ST} Accel. Beams}\ }\textbf {\bibinfo {volume} {14}},\ \bibinfo
  {pages} {054401} (\bibinfo {year} {2011})}\BibitemShut {NoStop}%
\bibitem [{\citenamefont {Nerush}\ \emph {et~al.}(2011)\citenamefont {Nerush},
  \citenamefont {Kostyukov}, \citenamefont {Fedotov}, \citenamefont {Narozhny},
  \citenamefont {Elkina},\ and\ \citenamefont {Ruhl}}]{nerush_laser_2011}%
  \BibitemOpen
  \bibfield  {author} {\bibinfo {author} {\bibfnamefont {E.~N.}\ \bibnamefont
  {Nerush}}, \bibinfo {author} {\bibfnamefont {I.~Y.}\ \bibnamefont
  {Kostyukov}}, \bibinfo {author} {\bibfnamefont {A.~M.}\ \bibnamefont
  {Fedotov}}, \bibinfo {author} {\bibfnamefont {N.~B.}\ \bibnamefont
  {Narozhny}}, \bibinfo {author} {\bibfnamefont {N.~V.}\ \bibnamefont
  {Elkina}}, \ and\ \bibinfo {author} {\bibfnamefont {H.}~\bibnamefont
  {Ruhl}},\ }\href {\doibase 10.1103/PhysRevLett.106.035001} {\bibfield
  {journal} {\bibinfo  {journal} {Phys. Rev. Lett.}\ }\textbf {\bibinfo
  {volume} {106}},\ \bibinfo {pages} {035001} (\bibinfo {year}
  {2011})}\BibitemShut {NoStop}%
\bibitem [{\citenamefont {Ilderton}\ \emph {et~al.}(2011)\citenamefont
  {Ilderton}, \citenamefont {Johansson},\ and\ \citenamefont
  {Marklund}}]{ilderton_pair_2011}%
  \BibitemOpen
  \bibfield  {author} {\bibinfo {author} {\bibfnamefont {A.}~\bibnamefont
  {Ilderton}}, \bibinfo {author} {\bibfnamefont {P.}~\bibnamefont {Johansson}},
  \ and\ \bibinfo {author} {\bibfnamefont {M.}~\bibnamefont {Marklund}},\
  }\href {\doibase 10.1103/PhysRevA.84.032119} {\bibfield  {journal} {\bibinfo
  {journal} {Phys. Rev. A}\ }\textbf {\bibinfo {volume} {84}},\ \bibinfo
  {pages} {032119} (\bibinfo {year} {2011})}\BibitemShut {NoStop}%
\bibitem [{\citenamefont {Ilderton}(2011)}]{ilderton_trident_2011}%
  \BibitemOpen
  \bibfield  {author} {\bibinfo {author} {\bibfnamefont {A.}~\bibnamefont
  {Ilderton}},\ }\href {\doibase 10.1103/PhysRevLett.106.020404} {\bibfield
  {journal} {\bibinfo  {journal} {Phys. Rev. Lett.}\ }\textbf {\bibinfo
  {volume} {106}},\ \bibinfo {pages} {020404} (\bibinfo {year}
  {2011})}\BibitemShut {NoStop}%
\bibitem [{\citenamefont {Labun}\ and\ \citenamefont
  {Rafelski}(2011)}]{labun_spectra_2011}%
  \BibitemOpen
  \bibfield  {author} {\bibinfo {author} {\bibfnamefont {L.}~\bibnamefont
  {Labun}}\ and\ \bibinfo {author} {\bibfnamefont {J.}~\bibnamefont
  {Rafelski}},\ }\href {\doibase 10.1103/PhysRevD.84.033003} {\bibfield
  {journal} {\bibinfo  {journal} {Phys. Rev. D}\ }\textbf {\bibinfo {volume}
  {84}},\ \bibinfo {pages} {033003} (\bibinfo {year} {2011})}\BibitemShut
  {NoStop}%
\bibitem [{\citenamefont {Monden}\ and\ \citenamefont
  {Kodama}(2011)}]{monden_enhancement_2011}%
  \BibitemOpen
  \bibfield  {author} {\bibinfo {author} {\bibfnamefont {Y.}~\bibnamefont
  {Monden}}\ and\ \bibinfo {author} {\bibfnamefont {R.}~\bibnamefont
  {Kodama}},\ }\href {\doibase 10.1103/PhysRevLett.107.073602} {\bibfield
  {journal} {\bibinfo  {journal} {Phys. Rev. Lett.}\ }\textbf {\bibinfo
  {volume} {107}},\ \bibinfo {pages} {073602} (\bibinfo {year}
  {2011})}\BibitemShut {NoStop}%
\bibitem [{\citenamefont {Redondo}\ and\ \citenamefont
  {Ringwald}(2011)}]{redondo_light_2011}%
  \BibitemOpen
  \bibfield  {author} {\bibinfo {author} {\bibfnamefont {J.}~\bibnamefont
  {Redondo}}\ and\ \bibinfo {author} {\bibfnamefont {A.}~\bibnamefont
  {Ringwald}},\ }\href {\doibase 10.1080/00107514.2011.563516} {\bibfield
  {journal} {\bibinfo  {journal} {Contemp. Phys.}\ }\textbf {\bibinfo {volume}
  {52}},\ \bibinfo {pages} {211} (\bibinfo {year} {2011})}\BibitemShut
  {NoStop}%
\bibitem [{\citenamefont {Seipt}\ and\ \citenamefont
  {K\"ampfer}(2011)}]{seipt_nonlinear_2011}%
  \BibitemOpen
  \bibfield  {author} {\bibinfo {author} {\bibfnamefont {D.}~\bibnamefont
  {Seipt}}\ and\ \bibinfo {author} {\bibfnamefont {B.}~\bibnamefont
  {K\"ampfer}},\ }\href {\doibase 10.1103/PhysRevA.83.022101} {\bibfield
  {journal} {\bibinfo  {journal} {Phys. Rev. A}\ }\textbf {\bibinfo {volume}
  {83}},\ \bibinfo {pages} {022101} (\bibinfo {year} {2011})}\BibitemShut
  {NoStop}%
\bibitem [{\citenamefont {Seipt}\ and\ \citenamefont
  {K\"ampfer}(2012)}]{seipt_two-photon_2012}%
  \BibitemOpen
  \bibfield  {author} {\bibinfo {author} {\bibfnamefont {D.}~\bibnamefont
  {Seipt}}\ and\ \bibinfo {author} {\bibfnamefont {B.}~\bibnamefont
  {K\"ampfer}},\ }\href {\doibase 10.1103/PhysRevD.85.101701} {\bibfield
  {journal} {\bibinfo  {journal} {Phys. Rev. D}\ }\textbf {\bibinfo {volume}
  {85}},\ \bibinfo {pages} {101701} (\bibinfo {year} {2012})}\BibitemShut
  {NoStop}%
\bibitem [{\citenamefont {Nousch}\ \emph {et~al.}(2012)\citenamefont {Nousch},
  \citenamefont {Seipt}, \citenamefont {K{\"a}mpfer},\ and\ \citenamefont
  {Titov}}]{nousch_pair_2012}%
  \BibitemOpen
  \bibfield  {author} {\bibinfo {author} {\bibfnamefont {T.}~\bibnamefont
  {Nousch}}, \bibinfo {author} {\bibfnamefont {D.}~\bibnamefont {Seipt}},
  \bibinfo {author} {\bibfnamefont {B.}~\bibnamefont {K{\"a}mpfer}}, \ and\
  \bibinfo {author} {\bibfnamefont {A.~I.}\ \bibnamefont {Titov}},\ }\href
  {\doibase 10.1016/j.physletb.2012.07.040} {\bibfield  {journal} {\bibinfo
  {journal} {Phys. Lett. B}\ }\textbf {\bibinfo {volume} {715}},\ \bibinfo
  {pages} {246} (\bibinfo {year} {2012})}\BibitemShut {NoStop}%
\bibitem [{\citenamefont {Titov}\ \emph {et~al.}(2012)\citenamefont {Titov},
  \citenamefont {Takabe}, \citenamefont {K\"ampfer},\ and\ \citenamefont
  {Hosaka}}]{titov_enhanced_2012}%
  \BibitemOpen
  \bibfield  {author} {\bibinfo {author} {\bibfnamefont {A.~I.}\ \bibnamefont
  {Titov}}, \bibinfo {author} {\bibfnamefont {H.}~\bibnamefont {Takabe}},
  \bibinfo {author} {\bibfnamefont {B.}~\bibnamefont {K\"ampfer}}, \ and\
  \bibinfo {author} {\bibfnamefont {A.}~\bibnamefont {Hosaka}},\ }\href
  {\doibase 10.1103/PhysRevLett.108.240406} {\bibfield  {journal} {\bibinfo
  {journal} {Phys. Rev. Lett.}\ }\textbf {\bibinfo {volume} {108}},\ \bibinfo
  {pages} {240406} (\bibinfo {year} {2012})}\BibitemShut {NoStop}%
\bibitem [{\citenamefont {King}\ and\ \citenamefont
  {Keitel}(2012)}]{king_photonphoton_2012}%
  \BibitemOpen
  \bibfield  {author} {\bibinfo {author} {\bibfnamefont {B.}~\bibnamefont
  {King}}\ and\ \bibinfo {author} {\bibfnamefont {C.~H.}\ \bibnamefont
  {Keitel}},\ }\href {\doibase 10.1088/1367-2630/14/10/103002} {\bibfield
  {journal} {\bibinfo  {journal} {New J. Phys.}\ }\textbf {\bibinfo {volume}
  {14}},\ \bibinfo {pages} {103002} (\bibinfo {year} {2012})}\BibitemShut
  {NoStop}%
\bibitem [{\citenamefont {D{\"o}brich}\ \emph {et~al.}(2012)\citenamefont
  {D{\"o}brich}, \citenamefont {Gies}, \citenamefont {Neitz},\ and\
  \citenamefont {Karbstein}}]{dobrich_magnetically_2012}%
  \BibitemOpen
  \bibfield  {author} {\bibinfo {author} {\bibfnamefont {B.}~\bibnamefont
  {D{\"o}brich}}, \bibinfo {author} {\bibfnamefont {H.}~\bibnamefont {Gies}},
  \bibinfo {author} {\bibfnamefont {N.}~\bibnamefont {Neitz}}, \ and\ \bibinfo
  {author} {\bibfnamefont {F.}~\bibnamefont {Karbstein}},\ }\href {\doibase
  10.1103/PhysRevLett.109.131802} {\bibfield  {journal} {\bibinfo  {journal}
  {Phys. Rev. Lett.}\ }\textbf {\bibinfo {volume} {109}},\ \bibinfo {pages}
  {131802} (\bibinfo {year} {2012})}\BibitemShut {NoStop}%
\bibitem [{\citenamefont {Dinu}\ \emph {et~al.}(2012)\citenamefont {Dinu},
  \citenamefont {Heinzl},\ and\ \citenamefont {Ilderton}}]{dinu_infrared_2012}%
  \BibitemOpen
  \bibfield  {author} {\bibinfo {author} {\bibfnamefont {V.}~\bibnamefont
  {Dinu}}, \bibinfo {author} {\bibfnamefont {T.}~\bibnamefont {Heinzl}}, \ and\
  \bibinfo {author} {\bibfnamefont {A.}~\bibnamefont {Ilderton}},\ }\href
  {\doibase 10.1103/PhysRevD.86.085037} {\bibfield  {journal} {\bibinfo
  {journal} {Phys. Rev. D}\ }\textbf {\bibinfo {volume} {86}},\ \bibinfo
  {pages} {085037} (\bibinfo {year} {2012})}\BibitemShut {NoStop}%
\bibitem [{\citenamefont {Harvey}\ and\ \citenamefont
  {Marklund}(2012)}]{harvey_radiation_2012}%
  \BibitemOpen
  \bibfield  {author} {\bibinfo {author} {\bibfnamefont {C.}~\bibnamefont
  {Harvey}}\ and\ \bibinfo {author} {\bibfnamefont {M.}~\bibnamefont
  {Marklund}},\ }\href {\doibase 10.1103/PhysRevA.85.013412} {\bibfield
  {journal} {\bibinfo  {journal} {Phys. Rev. A}\ }\textbf {\bibinfo {volume}
  {85}},\ \bibinfo {pages} {013412} (\bibinfo {year} {2012})}\BibitemShut
  {NoStop}%
\bibitem [{\citenamefont {Krajewska}\ and\ \citenamefont
  {Kami{\'n}ski}(2012)}]{krajewska_compton_2012}%
  \BibitemOpen
  \bibfield  {author} {\bibinfo {author} {\bibfnamefont {K.}~\bibnamefont
  {Krajewska}}\ and\ \bibinfo {author} {\bibfnamefont {J.~Z.}\ \bibnamefont
  {Kami{\'n}ski}},\ }\href {\doibase 10.1103/PhysRevA.85.062102} {\bibfield
  {journal} {\bibinfo  {journal} {Phys. Rev. A}\ }\textbf {\bibinfo {volume}
  {85}},\ \bibinfo {pages} {062102} (\bibinfo {year} {2012})}\BibitemShut
  {NoStop}%
\bibitem [{\citenamefont {King}\ \emph {et~al.}(2012)\citenamefont {King},
  \citenamefont {Gies},\ and\ \citenamefont {Di~Piazza}}]{king_pair_2012}%
  \BibitemOpen
  \bibfield  {author} {\bibinfo {author} {\bibfnamefont {B.}~\bibnamefont
  {King}}, \bibinfo {author} {\bibfnamefont {H.}~\bibnamefont {Gies}}, \ and\
  \bibinfo {author} {\bibfnamefont {A.}~\bibnamefont {Di~Piazza}},\ }\href
  {\doibase 10.1103/PhysRevD.86.125007} {\bibfield  {journal} {\bibinfo
  {journal} {Phys. Rev. D}\ }\textbf {\bibinfo {volume} {86}},\ \bibinfo
  {pages} {125007} (\bibinfo {year} {2012})}\BibitemShut {NoStop}%
\bibitem [{\citenamefont {Mackenroth}\ and\ \citenamefont
  {Di~Piazza}(2013)}]{mackenroth_nonlinear_2013}%
  \BibitemOpen
  \bibfield  {author} {\bibinfo {author} {\bibfnamefont {F.}~\bibnamefont
  {Mackenroth}}\ and\ \bibinfo {author} {\bibfnamefont {A.}~\bibnamefont
  {Di~Piazza}},\ }\href {\doibase 10.1103/PhysRevLett.110.070402} {\bibfield
  {journal} {\bibinfo  {journal} {Phys. Rev. Lett.}\ }\textbf {\bibinfo
  {volume} {110}},\ \bibinfo {pages} {070402} (\bibinfo {year}
  {2013})}\BibitemShut {NoStop}%
\bibitem [{\citenamefont {King}\ \emph {et~al.}(2013)\citenamefont {King},
  \citenamefont {Elkina},\ and\ \citenamefont {Ruhl}}]{king_photon_2013}%
  \BibitemOpen
  \bibfield  {author} {\bibinfo {author} {\bibfnamefont {B.}~\bibnamefont
  {King}}, \bibinfo {author} {\bibfnamefont {N.}~\bibnamefont {Elkina}}, \ and\
  \bibinfo {author} {\bibfnamefont {H.}~\bibnamefont {Ruhl}},\ }\href {\doibase
  10.1103/PhysRevA.87.042117} {\bibfield  {journal} {\bibinfo  {journal} {Phys.
  Rev. A}\ }\textbf {\bibinfo {volume} {87}},\ \bibinfo {pages} {042117}
  (\bibinfo {year} {2013})}\BibitemShut {NoStop}%
\bibitem [{\citenamefont {Di~Piazza}\ \emph {et~al.}(2012)\citenamefont
  {Di~Piazza}, \citenamefont {M\"uller}, \citenamefont {Hatsagortsyan},\ and\
  \citenamefont {Keitel}}]{di_piazza_extremely_2012}%
  \BibitemOpen
  \bibfield  {author} {\bibinfo {author} {\bibfnamefont {A.}~\bibnamefont
  {Di~Piazza}}, \bibinfo {author} {\bibfnamefont {C.}~\bibnamefont {M\"uller}},
  \bibinfo {author} {\bibfnamefont {K.~Z.}\ \bibnamefont {Hatsagortsyan}}, \
  and\ \bibinfo {author} {\bibfnamefont {C.~H.}\ \bibnamefont {Keitel}},\
  }\href {\doibase 10.1103/RevModPhys.84.1177} {\bibfield  {journal} {\bibinfo
  {journal} {Rev. Mod. Phys.}\ }\textbf {\bibinfo {volume} {84}},\ \bibinfo
  {pages} {1177} (\bibinfo {year} {2012})}\BibitemShut {NoStop}%
\bibitem [{\citenamefont {Ritus}(1972{\natexlab{a}})}]{ritus_radiative_1972}%
  \BibitemOpen
  \bibfield  {author} {\bibinfo {author} {\bibfnamefont {V.~I.}\ \bibnamefont
  {Ritus}},\ }\href {\doibase 10.1016/0003-4916(72)90191-1} {\bibfield
  {journal} {\bibinfo  {journal} {Ann. Phys.}\ }\textbf {\bibinfo {volume}
  {69}},\ \bibinfo {pages} {555} (\bibinfo {year}
  {1972}{\natexlab{a}})}\BibitemShut {NoStop}%
\bibitem [{\citenamefont {Ritus}(1972{\natexlab{b}})}]{ritus_vacuum_1972}%
  \BibitemOpen
  \bibfield  {author} {\bibinfo {author} {\bibfnamefont {V.~I.}\ \bibnamefont
  {Ritus}},\ }\href {\doibase 10.1016/0550-3213(72)90282-9} {\bibfield
  {journal} {\bibinfo  {journal} {Nucl. Phys. B}\ }\textbf {\bibinfo {volume}
  {44}},\ \bibinfo {pages} {236} (\bibinfo {year}
  {1972}{\natexlab{b}})}\BibitemShut {NoStop}%
\bibitem [{\citenamefont {Ritus}(1985)}]{ritus_1985}%
  \BibitemOpen
  \bibfield  {author} {\bibinfo {author} {\bibfnamefont {V.~I.}\ \bibnamefont
  {Ritus}},\ }\href {\doibase 10.1007/BF01120220} {\bibfield  {journal}
  {\bibinfo  {journal} {J. Sov. Laser Res.}\ }\textbf {\bibinfo {volume} {6}},\
  \bibinfo {pages} {497} (\bibinfo {year} {1985})}\BibitemShut {NoStop}%
\bibitem [{\citenamefont {Baier}\ \emph
  {et~al.}(1975{\natexlab{a}})\citenamefont {Baier}, \citenamefont
  {Mil'shtein},\ and\ \citenamefont {Strakhovenko}}]{baier_interaction_1975}%
  \BibitemOpen
  \bibfield  {author} {\bibinfo {author} {\bibfnamefont {V.~N.}\ \bibnamefont
  {Baier}}, \bibinfo {author} {\bibfnamefont {A.~I.}\ \bibnamefont
  {Mil'shtein}}, \ and\ \bibinfo {author} {\bibfnamefont {V.~M.}\ \bibnamefont
  {Strakhovenko}},\ }\href@noop {} {\bibfield  {journal} {\bibinfo  {journal}
  {Sov. Phys. JETP}\ }\textbf {\bibinfo {volume} {42}},\ \bibinfo {pages} {961}
  (\bibinfo {year} {1975}{\natexlab{a}})}\BibitemShut {NoStop}%
\bibitem [{\citenamefont {Becker}\ and\ \citenamefont
  {Mitter}(1975)}]{becker_vacuum_1975}%
  \BibitemOpen
  \bibfield  {author} {\bibinfo {author} {\bibfnamefont {W.}~\bibnamefont
  {Becker}}\ and\ \bibinfo {author} {\bibfnamefont {H.}~\bibnamefont
  {Mitter}},\ }\href {\doibase 10.1088/0305-4470/8/10/017} {\bibfield
  {journal} {\bibinfo  {journal} {J. Phys. A}\ }\textbf {\bibinfo {volume}
  {8}},\ \bibinfo {pages} {1638} (\bibinfo {year} {1975})}\BibitemShut
  {NoStop}%
\bibitem [{\citenamefont {Ward}(1950)}]{ward_identity_1950}%
  \BibitemOpen
  \bibfield  {author} {\bibinfo {author} {\bibfnamefont {J.~C.}\ \bibnamefont
  {Ward}},\ }\href {\doibase 10.1103/PhysRev.78.182} {\bibfield  {journal}
  {\bibinfo  {journal} {Phys. Rev.}\ }\textbf {\bibinfo {volume} {78}},\
  \bibinfo {pages} {182} (\bibinfo {year} {1950})}\BibitemShut {NoStop}%
\bibitem [{\citenamefont {Takahashi}(1957)}]{takahashi_generalized_1957}%
  \BibitemOpen
  \bibfield  {author} {\bibinfo {author} {\bibfnamefont {Y.}~\bibnamefont
  {Takahashi}},\ }\href {\doibase 10.1007/BF02832514} {\bibfield  {journal}
  {\bibinfo  {journal} {Il Nuovo Cimento}\ }\textbf {\bibinfo {volume} {6}},\
  \bibinfo {pages} {371} (\bibinfo {year} {1957})}\BibitemShut {NoStop}%
\bibitem [{\citenamefont {Landau}\ and\ \citenamefont
  {Lifshitz}(1982)}]{landau_quantum_1981}%
  \BibitemOpen
  \bibfield  {author} {\bibinfo {author} {\bibfnamefont {L.~D.}\ \bibnamefont
  {Landau}}\ and\ \bibinfo {author} {\bibfnamefont {E.~M.}\ \bibnamefont
  {Lifshitz}},\ }\href@noop {} {\emph {\bibinfo {title} {{Q}uantum
  {E}lectrodynamics}}},\ \bibinfo {edition} {2nd}\ ed.\ (\bibinfo  {publisher}
  {Butterworth},\ \bibinfo {year} {1982})\BibitemShut {NoStop}%
\bibitem [{\citenamefont {Milstein}\ \emph {et~al.}(2006)\citenamefont
  {Milstein}, \citenamefont {M\"uller}, \citenamefont {Hatsagortsyan},
  \citenamefont {Jentschura},\ and\ \citenamefont
  {Keitel}}]{milstein_polarization-operator_2006}%
  \BibitemOpen
  \bibfield  {author} {\bibinfo {author} {\bibfnamefont {A.~I.}\ \bibnamefont
  {Milstein}}, \bibinfo {author} {\bibfnamefont {C.}~\bibnamefont {M\"uller}},
  \bibinfo {author} {\bibfnamefont {K.~Z.}\ \bibnamefont {Hatsagortsyan}},
  \bibinfo {author} {\bibfnamefont {U.~D.}\ \bibnamefont {Jentschura}}, \ and\
  \bibinfo {author} {\bibfnamefont {C.~H.}\ \bibnamefont {Keitel}},\ }\href
  {\doibase 10.1103/PhysRevA.73.062106} {\bibfield  {journal} {\bibinfo
  {journal} {Phys. Rev. A}\ }\textbf {\bibinfo {volume} {73}},\ \bibinfo
  {pages} {062106} (\bibinfo {year} {2006})}\BibitemShut {NoStop}%
\bibitem [{\citenamefont {Di~Piazza}\ \emph {et~al.}(2009)\citenamefont
  {Di~Piazza}, \citenamefont {L\"otstedt}, \citenamefont {Milstein},\ and\
  \citenamefont {Keitel}}]{di_piazza_barrier_2009}%
  \BibitemOpen
  \bibfield  {author} {\bibinfo {author} {\bibfnamefont {A.}~\bibnamefont
  {Di~Piazza}}, \bibinfo {author} {\bibfnamefont {E.}~\bibnamefont
  {L\"otstedt}}, \bibinfo {author} {\bibfnamefont {A.~I.}\ \bibnamefont
  {Milstein}}, \ and\ \bibinfo {author} {\bibfnamefont {C.~H.}\ \bibnamefont
  {Keitel}},\ }\href {\doibase 10.1103/PhysRevLett.103.170403} {\bibfield
  {journal} {\bibinfo  {journal} {Phys. Rev. Lett.}\ }\textbf {\bibinfo
  {volume} {103}},\ \bibinfo {pages} {170403} (\bibinfo {year}
  {2009})}\BibitemShut {NoStop}%
\bibitem [{\citenamefont {Gies}\ and\ \citenamefont
  {Roessler}(2011)}]{gies_vacuum_2011}%
  \BibitemOpen
  \bibfield  {author} {\bibinfo {author} {\bibfnamefont {H.}~\bibnamefont
  {Gies}}\ and\ \bibinfo {author} {\bibfnamefont {L.}~\bibnamefont
  {Roessler}},\ }\href {\doibase 10.1103/PhysRevD.84.065035} {\bibfield
  {journal} {\bibinfo  {journal} {Phys. Rev. D}\ }\textbf {\bibinfo {volume}
  {84}},\ \bibinfo {pages} {065035} (\bibinfo {year} {2011})}\BibitemShut
  {NoStop}%
\bibitem [{\citenamefont {Karbstein}\ \emph {et~al.}(2012)\citenamefont
  {Karbstein}, \citenamefont {Roessler}, \citenamefont {D{\"o}brich},\ and\
  \citenamefont {Gies}}]{karbstein_optical_2012}%
  \BibitemOpen
  \bibfield  {author} {\bibinfo {author} {\bibfnamefont {F.}~\bibnamefont
  {Karbstein}}, \bibinfo {author} {\bibfnamefont {L.}~\bibnamefont {Roessler}},
  \bibinfo {author} {\bibfnamefont {B.}~\bibnamefont {D{\"o}brich}}, \ and\
  \bibinfo {author} {\bibfnamefont {H.}~\bibnamefont {Gies}},\ }\href {\doibase
  10.1142/S2010194512007520} {\bibfield  {journal} {\bibinfo  {journal} {Int.
  J. Mod. Phys. Conf. Ser.}\ }\textbf {\bibinfo {volume} {14}},\ \bibinfo
  {pages} {403} (\bibinfo {year} {2012})}\BibitemShut {NoStop}%
\bibitem [{\citenamefont {Weinberg}(1995)}]{weinberg_quantum_1995}%
  \BibitemOpen
  \bibfield  {author} {\bibinfo {author} {\bibfnamefont {S.}~\bibnamefont
  {Weinberg}},\ }\href@noop {} {\emph {\bibinfo {title} {{T}he {Q}uantum
  {T}heory of {F}ields {I}}}}\ (\bibinfo  {publisher} {Cambridge University
  Press},\ \bibinfo {year} {1995})\BibitemShut {NoStop}%
\bibitem [{\citenamefont {Fradkin}\ \emph {et~al.}(1991)\citenamefont
  {Fradkin}, \citenamefont {Gitman},\ and\ \citenamefont
  {Shvartsman}}]{fradkin_quantum_1991}%
  \BibitemOpen
  \bibfield  {author} {\bibinfo {author} {\bibfnamefont {E.~S.}\ \bibnamefont
  {Fradkin}}, \bibinfo {author} {\bibfnamefont {D.~M.}\ \bibnamefont {Gitman}},
  \ and\ \bibinfo {author} {\bibfnamefont {S.~M.}\ \bibnamefont {Shvartsman}},\
  }\href@noop {} {\emph {\bibinfo {title} {{Q}uantum electrodynamics: with
  unstable vacuum}}}\ (\bibinfo  {publisher} {Springer},\ \bibinfo {year}
  {1991})\BibitemShut {NoStop}%
\bibitem [{\citenamefont {Glauber}(1963)}]{glauber_coherent_1963}%
  \BibitemOpen
  \bibfield  {author} {\bibinfo {author} {\bibfnamefont {R.~J.}\ \bibnamefont
  {Glauber}},\ }\href {\doibase 10.1103/PhysRev.131.2766} {\bibfield  {journal}
  {\bibinfo  {journal} {Phys. Rev.}\ }\textbf {\bibinfo {volume} {131}},\
  \bibinfo {pages} {2766} (\bibinfo {year} {1963})}\BibitemShut {NoStop}%
\bibitem [{\citenamefont {Harvey}\ \emph {et~al.}(2009)\citenamefont {Harvey},
  \citenamefont {Heinzl},\ and\ \citenamefont
  {Ilderton}}]{harvey_signatures_2009}%
  \BibitemOpen
  \bibfield  {author} {\bibinfo {author} {\bibfnamefont {C.}~\bibnamefont
  {Harvey}}, \bibinfo {author} {\bibfnamefont {T.}~\bibnamefont {Heinzl}}, \
  and\ \bibinfo {author} {\bibfnamefont {A.}~\bibnamefont {Ilderton}},\ }\href
  {\doibase 10.1103/PhysRevA.79.063407} {\bibfield  {journal} {\bibinfo
  {journal} {Phys. Rev. A}\ }\textbf {\bibinfo {volume} {79}},\ \bibinfo
  {pages} {063407} (\bibinfo {year} {2009})}\BibitemShut {NoStop}%
\bibitem [{\citenamefont {Fried}\ and\ \citenamefont
  {Eberly}(1964)}]{fried_scattering_1964}%
  \BibitemOpen
  \bibfield  {author} {\bibinfo {author} {\bibfnamefont {Z.}~\bibnamefont
  {Fried}}\ and\ \bibinfo {author} {\bibfnamefont {J.~H.}\ \bibnamefont
  {Eberly}},\ }\href {\doibase 10.1103/PhysRev.136.B871} {\bibfield  {journal}
  {\bibinfo  {journal} {Phys. Rev.}\ }\textbf {\bibinfo {volume} {136}},\
  \bibinfo {pages} {B871} (\bibinfo {year} {1964})}\BibitemShut {NoStop}%
\bibitem [{\citenamefont {Eberly}\ and\ \citenamefont
  {Reiss}(1966)}]{eberly_electron_1966}%
  \BibitemOpen
  \bibfield  {author} {\bibinfo {author} {\bibfnamefont {J.~H.}\ \bibnamefont
  {Eberly}}\ and\ \bibinfo {author} {\bibfnamefont {H.~R.}\ \bibnamefont
  {Reiss}},\ }\href {\doibase 10.1103/PhysRev.145.1035} {\bibfield  {journal}
  {\bibinfo  {journal} {Phys. Rev.}\ }\textbf {\bibinfo {volume} {145}},\
  \bibinfo {pages} {1035} (\bibinfo {year} {1966})}\BibitemShut {NoStop}%
\bibitem [{\citenamefont {Berson}(1969)}]{berson_electron_1969}%
  \BibitemOpen
  \bibfield  {author} {\bibinfo {author} {\bibfnamefont {I.}~\bibnamefont
  {Berson}},\ }\href@noop {} {\bibfield  {journal} {\bibinfo  {journal} {Sov.
  Phys. {JETP}}\ }\textbf {\bibinfo {volume} {29}},\ \bibinfo {pages} {871}
  (\bibinfo {year} {1969})}\BibitemShut {NoStop}%
\bibitem [{\citenamefont {Bergou}\ and\ \citenamefont
  {Varro}(1981)}]{bergou_nonlinear_1981}%
  \BibitemOpen
  \bibfield  {author} {\bibinfo {author} {\bibfnamefont {J.}~\bibnamefont
  {Bergou}}\ and\ \bibinfo {author} {\bibfnamefont {S.}~\bibnamefont {Varro}},\
  }\href {\doibase 10.1088/0305-4470/14/9/023} {\bibfield  {journal} {\bibinfo
  {journal} {J. Phys. A}\ }\textbf {\bibinfo {volume} {14}},\ \bibinfo {pages}
  {2281} (\bibinfo {year} {1981})}\BibitemShut {NoStop}%
\bibitem [{\citenamefont {Filipowicz}(1985)}]{filipowicz_relativistic_1985}%
  \BibitemOpen
  \bibfield  {author} {\bibinfo {author} {\bibfnamefont {P.}~\bibnamefont
  {Filipowicz}},\ }\href {\doibase 10.1088/0305-4470/18/10/022} {\bibfield
  {journal} {\bibinfo  {journal} {J. Phys. A}\ }\textbf {\bibinfo {volume}
  {18}},\ \bibinfo {pages} {1675} (\bibinfo {year} {1985})}\BibitemShut
  {NoStop}%
\bibitem [{\citenamefont {Furry}(1951)}]{furry51}%
  \BibitemOpen
  \bibfield  {author} {\bibinfo {author} {\bibfnamefont {W.~H.}\ \bibnamefont
  {Furry}},\ }\href {\doibase 10.1103/PhysRev.81.115} {\bibfield  {journal}
  {\bibinfo  {journal} {Phys. Rev.}\ }\textbf {\bibinfo {volume} {81}},\
  \bibinfo {pages} {115} (\bibinfo {year} {1951})}\BibitemShut {NoStop}%
\bibitem [{\citenamefont {Battesti}\ and\ \citenamefont
  {Rizzo}(2013)}]{battesti_magnetic_2013}%
  \BibitemOpen
  \bibfield  {author} {\bibinfo {author} {\bibfnamefont {R.}~\bibnamefont
  {Battesti}}\ and\ \bibinfo {author} {\bibfnamefont {C.}~\bibnamefont
  {Rizzo}},\ }\href {\doibase 10.1088/0034-4885/76/1/016401} {\bibfield
  {journal} {\bibinfo  {journal} {Rep. Prog. Phys.}\ }\textbf {\bibinfo
  {volume} {76}},\ \bibinfo {pages} {016401} (\bibinfo {year}
  {2013})}\BibitemShut {NoStop}%
\bibitem [{\citenamefont {Ehlotzky}\ \emph {et~al.}(2009)\citenamefont
  {Ehlotzky}, \citenamefont {Krajewska},\ and\ \citenamefont
  {Kami{\'n}ski}}]{ehlotzky_fundamental_2009}%
  \BibitemOpen
  \bibfield  {author} {\bibinfo {author} {\bibfnamefont {F.}~\bibnamefont
  {Ehlotzky}}, \bibinfo {author} {\bibfnamefont {K.}~\bibnamefont {Krajewska}},
  \ and\ \bibinfo {author} {\bibfnamefont {J.~Z.}\ \bibnamefont
  {Kami{\'n}ski}},\ }\href {\doibase 10.1088/0034-4885/72/4/046401} {\bibfield
  {journal} {\bibinfo  {journal} {Rep. Prog. Phys.}\ }\textbf {\bibinfo
  {volume} {72}},\ \bibinfo {pages} {046401} (\bibinfo {year}
  {2009})}\BibitemShut {NoStop}%
\bibitem [{\citenamefont {Mourou}\ \emph {et~al.}(2006)\citenamefont {Mourou},
  \citenamefont {Tajima},\ and\ \citenamefont {Bulanov}}]{mourou_optics_2006}%
  \BibitemOpen
  \bibfield  {author} {\bibinfo {author} {\bibfnamefont {G.~A.}\ \bibnamefont
  {Mourou}}, \bibinfo {author} {\bibfnamefont {T.}~\bibnamefont {Tajima}}, \
  and\ \bibinfo {author} {\bibfnamefont {S.~V.}\ \bibnamefont {Bulanov}},\
  }\href {\doibase 10.1103/RevModPhys.78.309} {\bibfield  {journal} {\bibinfo
  {journal} {Rev. Mod. Phys.}\ }\textbf {\bibinfo {volume} {78}},\ \bibinfo
  {pages} {309} (\bibinfo {year} {2006})}\BibitemShut {NoStop}%
\bibitem [{\citenamefont {Marklund}\ and\ \citenamefont
  {Shukla}(2006)}]{marklund_nonlinear_2006}%
  \BibitemOpen
  \bibfield  {author} {\bibinfo {author} {\bibfnamefont {M.}~\bibnamefont
  {Marklund}}\ and\ \bibinfo {author} {\bibfnamefont {P.~K.}\ \bibnamefont
  {Shukla}},\ }\href {\doibase 10.1103/RevModPhys.78.591} {\bibfield  {journal}
  {\bibinfo  {journal} {Rev. Mod. Phys.}\ }\textbf {\bibinfo {volume} {78}},\
  \bibinfo {pages} {591} (\bibinfo {year} {2006})}\BibitemShut {NoStop}%
\bibitem [{\citenamefont {Dittrich}\ and\ \citenamefont
  {Gies}(2000)}]{dittrich_probingquantum_2000}%
  \BibitemOpen
  \bibfield  {author} {\bibinfo {author} {\bibfnamefont {W.}~\bibnamefont
  {Dittrich}}\ and\ \bibinfo {author} {\bibfnamefont {H.}~\bibnamefont
  {Gies}},\ }\href@noop {} {\emph {\bibinfo {title} {{P}robing the {Q}uantum
  {V}acuum}}}\ (\bibinfo  {publisher} {Springer},\ \bibinfo {address}
  {Berlin},\ \bibinfo {year} {2000})\BibitemShut {NoStop}%
\bibitem [{\citenamefont {Landau}\ and\ \citenamefont
  {Lifshitz}(1987)}]{landau_classical_1987}%
  \BibitemOpen
  \bibfield  {author} {\bibinfo {author} {\bibfnamefont {L.~D.}\ \bibnamefont
  {Landau}}\ and\ \bibinfo {author} {\bibfnamefont {E.~M.}\ \bibnamefont
  {Lifshitz}},\ }\href@noop {} {\emph {\bibinfo {title} {{T}he {C}lassical
  {T}heory of {F}ields}}},\ \bibinfo {edition} {4th}\ ed.\ (\bibinfo
  {publisher} {Butterworth},\ \bibinfo {year} {1987})\BibitemShut {NoStop}%
\bibitem [{\citenamefont {Volkov}(1935)}]{volkov_ueber_1935}%
  \BibitemOpen
  \bibfield  {author} {\bibinfo {author} {\bibfnamefont {D.~M.}\ \bibnamefont
  {Volkov}},\ }\href {\doibase 10.1007/BF01331022} {\bibfield  {journal}
  {\bibinfo  {journal} {Z. Phys.}\ }\textbf {\bibinfo {volume} {94}},\ \bibinfo
  {pages} {250} (\bibinfo {year} {1935})}\BibitemShut {NoStop}%
\bibitem [{\citenamefont {Mitter}(1975)}]{mitter_quantum_1975}%
  \BibitemOpen
  \bibfield  {author} {\bibinfo {author} {\bibfnamefont {H.}~\bibnamefont
  {Mitter}},\ }\href {\doibase 10.1007/978-3-7091-8424-0_7} {\bibfield
  {journal} {\bibinfo  {journal} {Acta Phys. Austriaca, Suppl.}\ }\textbf
  {\bibinfo {volume} {XIV}},\ \bibinfo {pages} {397} (\bibinfo {year}
  {1975})}\BibitemShut {NoStop}%
\bibitem [{\citenamefont {Collins}(1984)}]{collins_renormalization_1984}%
  \BibitemOpen
  \bibfield  {author} {\bibinfo {author} {\bibfnamefont {J.~C.}\ \bibnamefont
  {Collins}},\ }\href@noop {} {\emph {\bibinfo {title} {{R}enormalization}}}\
  (\bibinfo  {publisher} {Cambridge University Press},\ \bibinfo {year}
  {1984})\BibitemShut {NoStop}%
\bibitem [{\citenamefont {Peskin}\ and\ \citenamefont
  {Schroeder}(1995)}]{peskin_introduction_2008}%
  \BibitemOpen
  \bibfield  {author} {\bibinfo {author} {\bibfnamefont {M.~E.}\ \bibnamefont
  {Peskin}}\ and\ \bibinfo {author} {\bibfnamefont {D.~V.}\ \bibnamefont
  {Schroeder}},\ }\href@noop {} {\emph {\bibinfo {title} {{A}n {I}ntroduction
  to {Q}uantum {F}ield {T}heory}}}\ (\bibinfo  {publisher} {Addison-Wesley},\
  \bibinfo {year} {1995})\BibitemShut {NoStop}%
\bibitem [{\citenamefont {Adler}(1969)}]{adler_axial-vector_1969}%
  \BibitemOpen
  \bibfield  {author} {\bibinfo {author} {\bibfnamefont {S.~L.}\ \bibnamefont
  {Adler}},\ }\href {\doibase 10.1103/PhysRev.177.2426} {\bibfield  {journal}
  {\bibinfo  {journal} {Phys. Rev.}\ }\textbf {\bibinfo {volume} {177}},\
  \bibinfo {pages} {2426} (\bibinfo {year} {1969})}\BibitemShut {NoStop}%
\bibitem [{\citenamefont {Baier}\ \emph
  {et~al.}(1975{\natexlab{b}})\citenamefont {Baier}, \citenamefont {Katkov},\
  and\ \citenamefont {Strakhovenko}}]{baier_operator_1975}%
  \BibitemOpen
  \bibfield  {author} {\bibinfo {author} {\bibfnamefont {V.~N.}\ \bibnamefont
  {Baier}}, \bibinfo {author} {\bibfnamefont {V.~M.}\ \bibnamefont {Katkov}}, \
  and\ \bibinfo {author} {\bibfnamefont {V.~M.}\ \bibnamefont {Strakhovenko}},\
  }\href@noop {} {\bibfield  {journal} {\bibinfo  {journal} {Sov. Phys. JETP}\
  }\textbf {\bibinfo {volume} {40}},\ \bibinfo {pages} {225} (\bibinfo {year}
  {1975}{\natexlab{b}})}\BibitemShut {NoStop}%
\bibitem [{\citenamefont {Narozhnyi}(1968)}]{narozhnyi_propagation_1968}%
  \BibitemOpen
  \bibfield  {author} {\bibinfo {author} {\bibfnamefont {N.~B.}\ \bibnamefont
  {Narozhnyi}},\ }\href@noop {} {\bibfield  {journal} {\bibinfo  {journal}
  {Sov. Phys. {JETP}}\ }\textbf {\bibinfo {volume} {28}},\ \bibinfo {pages}
  {371} (\bibinfo {year} {1968})}\BibitemShut {NoStop}%
\bibitem [{\citenamefont {Batalin}\ and\ \citenamefont
  {Shabad}(1968)}]{batalin_preprint_1968}%
  \BibitemOpen
  \bibfield  {author} {\bibinfo {author} {\bibfnamefont {I.~A.}\ \bibnamefont
  {Batalin}}\ and\ \bibinfo {author} {\bibfnamefont {A.~E.}\ \bibnamefont
  {Shabad}},\ }\href@noop {} {\bibfield  {journal} {\bibinfo  {journal} {FIAN
  Preprint}\ }\textbf {\bibinfo {volume} {166}} (\bibinfo {year}
  {1968})}\BibitemShut {NoStop}%
\bibitem [{\citenamefont {Batalin}\ and\ \citenamefont
  {Shabad}(1971)}]{batalin_greens_1971}%
  \BibitemOpen
  \bibfield  {author} {\bibinfo {author} {\bibfnamefont {I.~A.}\ \bibnamefont
  {Batalin}}\ and\ \bibinfo {author} {\bibfnamefont {A.~E.}\ \bibnamefont
  {Shabad}},\ }\href@noop {} {\bibfield  {journal} {\bibinfo  {journal} {Sov.
  Phys. {JETP}}\ }\textbf {\bibinfo {volume} {33}},\ \bibinfo {pages} {483}
  (\bibinfo {year} {1971})}\BibitemShut {NoStop}%
\bibitem [{\citenamefont {Olver}\ \emph {et~al.}(2010)\citenamefont {Olver},
  \citenamefont {Lozier}, \citenamefont {Boisvert},\ and\ \citenamefont
  {Clark}}]{olver_nist_2010}%
  \BibitemOpen
  \bibinfo {editor} {\bibfnamefont {F.~W.~J.}\ \bibnamefont {Olver}}, \bibinfo
  {editor} {\bibfnamefont {D.~W.}\ \bibnamefont {Lozier}}, \bibinfo {editor}
  {\bibfnamefont {R.~F.}\ \bibnamefont {Boisvert}}, \ and\ \bibinfo {editor}
  {\bibfnamefont {C.~W.}\ \bibnamefont {Clark}},\ eds.,\ \href@noop {} {\emph
  {\bibinfo {title} {{{N}{I}{S}{T}} {H}andbook of {M}athematical
  {F}unctions}}}\ (\bibinfo  {publisher} {Cambridge University Press},\
  \bibinfo {year} {2010})\BibitemShut {NoStop}%
\bibitem [{\citenamefont {Dirac}(1949)}]{dirac_forms_1949}%
  \BibitemOpen
  \bibfield  {author} {\bibinfo {author} {\bibfnamefont {P.~A.~M.}\
  \bibnamefont {Dirac}},\ }\href {\doibase 10.1103/RevModPhys.21.392}
  {\bibfield  {journal} {\bibinfo  {journal} {Rev. Mod. Phys.}\ }\textbf
  {\bibinfo {volume} {21}},\ \bibinfo {pages} {392} (\bibinfo {year}
  {1949})}\BibitemShut {NoStop}%
\bibitem [{\citenamefont {Neville}\ and\ \citenamefont
  {Rohrlich}(1971)}]{neville_quantum_1971}%
  \BibitemOpen
  \bibfield  {author} {\bibinfo {author} {\bibfnamefont {R.~A.}\ \bibnamefont
  {Neville}}\ and\ \bibinfo {author} {\bibfnamefont {F.}~\bibnamefont
  {Rohrlich}},\ }\href {\doibase 10.1103/PhysRevD.3.1692} {\bibfield  {journal}
  {\bibinfo  {journal} {Phys. Rev. D}\ }\textbf {\bibinfo {volume} {3}},\
  \bibinfo {pages} {1692} (\bibinfo {year} {1971})}\BibitemShut {NoStop}%
\bibitem [{\citenamefont {Leader}(2001)}]{leader_spin_2001}%
  \BibitemOpen
  \bibfield  {author} {\bibinfo {author} {\bibfnamefont {E.}~\bibnamefont
  {Leader}},\ }\href@noop {} {\emph {\bibinfo {title} {{S}pin in {P}article
  {P}hysics}}}\ (\bibinfo  {publisher} {Cambridge University Press},\ \bibinfo
  {year} {2001})\BibitemShut {NoStop}%
\end{thebibliography}
\end{document}